\documentclass[prd,aps,
tightenlines,
superscriptaddress,showpacs,nofootinbib
,eqsecnum,amsfonts,amsmath
,twocolumn
]{revtex4}
\usepackage{bm,graphics,graphicx,epsfig,color
,subfigure,rotating,latexsym
}
\usepackage{amsmath}
\usepackage{amsthm}
\usepackage{amssymb}
\usepackage{url}
\usepackage{fancyhdr}
\usepackage{alltt}
\usepackage{array}
\usepackage{hyperref}

\def\be{\begin{equation}}
\def\ee{\end{equation}}

\def\bea{\begin{eqnarray}}
\def\eea{\end{eqnarray}}

\def\Gpc{\textrm{Gpc}}
\def\mHz{\textrm{mHz}}
\def\Hz{\textrm{Hz}}
\def\m{\textrm{m}}
\def\s{\textrm{s}}

\newcommand{\Ms}{M_{\odot}}

\newcommand{\df}{{\rm d}f}
\newcommand{\etal}{\textit{et al. }}

 \def\gsim{\mathrel{
 \rlap{\raise 0.511ex \hbox{$>$}}{\lower 0.511ex
 \hbox{$\sim$}}}}
 \def\lsim{\mathrel{
 \rlap{\raise 0.511ex \hbox{$<$}}{\lower 0.511ex
 \hbox{$\sim$}}}}

\begin{document}

\title{Weak lensing effects in the measurement of the dark energy equation of state with LISA}

\author{Chris Van Den Broeck}
\email{vdbroeck@nikhef.nl}
\affiliation{Nikhef, National Institute for Subatomic Physics, Science Park 105, 1098 XG Amsterdam, 
The Netherlands}
\affiliation{School of Physics and Astronomy, Cardiff University, 
Queen's Buildings, The Parade, Cardiff, CF24 3AA, United Kingdom}
\author{M.\ Trias} 
\email{miquel.trias@uib.es} 
\affiliation{Departament de F\'{\i}sica, Universitat de les Illes 
Balears, Cra.~Valldemossa Km.~7.5, E-07122 Palma de Mallorca, Spain}
\author{B.\ S.\ Sathyaprakash} 
\email{B.Sathyaprakash@astro.cf.ac.uk} 
\affiliation{School of Physics and Astronomy, Cardiff University, 
Queen's Buildings, The Parade, Cardiff, CF24 3AA, United Kingdom}
\author{A.\ M.\ Sintes} 
\email{alicia.sintes@uib.es} 
\affiliation{Departament de F\'{\i}sica, Universitat de les Illes 
Balears, Cra.~Valldemossa Km.~7.5, E-07122 Palma de Mallorca, Spain}

\begin{abstract}
The Laser Interferometer Space Antenna's (LISA's) observation of supermassive
binary black holes (SMBBH) could provide a new tool for precision cosmography. 
Inclusion of sub-dominant signal harmonics in the inspiral signal allows for high-accuracy
sky localization, dramatically improving the chances of finding the host galaxy and 
obtaining its redshift. 
A SMBBH merger can potentially have component masses from a wide range ($10^5 - 10^8\,\Ms$)
over which parameter accuracies vary considerably. We perform an in-depth study in order to
understand (i) what fraction of possible SMBBH mergers allow for sky localization, depending on the
parameters of the source, and (ii) how accurately $w$ can be measured when the host galaxy can be 
identified. We also investigate how accuracies on all parameters improve when 
a knowledge of the sky position can be folded into the estimation of errors. We find that  
$w$ can be measured to within a few percent in most cases, if the only error 
in measuring the luminosity distance is due to LISA's instrumental noise and the confusion
background from Galactic binaries. However, weak lensing-induced 
errors will severely degrade the accuracy with which $w$ can be obtained, emphasizing that 
methods to mitigate weak lensing effects would be required to take advantage of LISA's 
full potential.
\end{abstract}

\maketitle

\section{Introduction}
The Laser Interferometer Space Antenna (LISA) is a space-based gravitational-wave
detector that can observe the merger of supermassive black holes in a binary
with signal-to-noise ratios (SNRs) of hundreds to thousands. So far, it has
not been possible to predict the rate of mergers in the Universe, and the event
rate for LISA, with any precision. The initial seed black holes could  
have masses anywhere in the range $\sim  100$-$10^5\,M_\odot$ depending on
the formation scenario. Collapse of metal-free massive stars at $z\gsim 20$
could lead to {\em light} seeds, with masses of a few hundred solar masses 
\cite{Volonteri:2002vz}.
On the other hand, gravitational instability of massive proto-galactic disks 
could lead to the formation at $z \gsim 10$ of {\em heavy} seeds with masses 
$\sim 10^5M_\odot$ \cite{Begelman:2006db}. These black hole seeds might grow by {\em prolonged} 
accretion whereby in-falling matter has a constant angular momentum direction 
and spins-up the black hole \cite{Bardeen70,Thorne74}. Alternatively, the seeds 
might grow by chaotic accretion associated with sporadic in-fall of small amounts of matter 
from a fragmented disc \cite{KingPringle}. The merger history in LISA's past light-cone depends
on how black holes formed and evolved. If seed black holes were heavy, then
one expects several tens of mergers per year and LISA will observe all of them.
If the seeds were small then the rate in LISA's past light cone could be 
two to three times larger, but LISA's sensitivity to smaller black holes
will be poorer and the number of mergers observed is again a few tens per year.
Of these, a handful of events might be close enough to be useful for cosmography
\cite{LISAPE_paper}.

LISA may open a new era for precision cosmography since black hole binaries 
are {\em self-calibrating standard sirens} \cite{Schutz:1986gp,HolzHugh03,HolzHugh05}. 
In astronomy, a standard candle is a 
source whose absolute luminosity can be deduced from certain observed 
properties such as the time-variability of its light-curve, spectral 
characteristics, etc. Analogously, the term {\em standard siren} is used for binary black 
holes, since the way gravitational waves interact with matter is more akin 
to sound waves, although, of course, they are transverse waves traveling at
the speed of light. Black holes are also {\em self-calibrating} since they 
don't need other measures of distance to calibrate their luminosity. This is 
because the luminosity of a binary black hole depends only on its chirp mass\footnote{The
chirp mass $\cal M$ of a binary of total mass $M$ and reduced mass $\mu$ is
${\cal M} = \mu^{3/5} M^{2/5}$.} and its luminosity distance from LISA. 
For a chirping binary, i.e., a binary whose gravitational-wave frequency 
changes by an observable amount during the course of observation, one can 
deduce the chirp mass and measure its amplitude and thereby infer its luminosity
distance. Therefore, binary black holes can provide a new calibration for the
high redshift Universe avoiding all the lower rungs of the cosmic distance
ladder.

Supermassive binary black hole (SMBBH) mergers are, therefore, potential tools for cosmology.
Indeed, a single SMBBH observation by LISA might already significantly constrain
the dark energy equation-of-state parameter $w$ \cite{HolzHugh03,HolzHugh05,Arun:2007hu,AIMSSV}. From 
the gravitational waveform, one obtains the luminosity distance, $D_{\rm L}$. Due
to LISA's orbital motion over the observation time, one can also obtain an approximate
sky position. If this allows for localization of the host galaxy, then redshift $z$ can
be measured. Since the relationship between $D_{\rm L}$ and $z$ depends sensitively on
$w$ (among other cosmological parameters), the latter can then be constrained.
One of the hurdles in 
availing of LISA for cosmology was that LISA's angular resolution might not be good enough to
localize the host galaxy -- a step that is crucial for obtaining the redshift
of the source. However, more recent work has mitigated this hurdle by showing that
the use of the full signal waveform, which contains not only the dominant harmonic
at twice the orbital frequency $2f_{\rm orb},$ but also other harmonics 
$k f_{\rm orb},$ $k=1,3,4, \ldots,$  in parameter estimation can improve the angular resolution to
a level that enables the localization of the source for a large fraction of 
systems in LISA's band \cite{Arun:2007hu,AIMSSV,Trias:2007fp,Trias:2008pu,Porter:2008kn}. 
The sub-dominant harmonics, therefore, are essential for precision cosmology with LISA.

The above conclusions have mainly been drawn by computing the covariance matrix
of the intrinsic and extrinsic parameters associated with an SMBBH. A binary
consisting of non-spinning black holes on a quasi-circular orbit is characterized 
by nine parameters: the chirp mass $\mathcal{M}$ and reduced mass $\mu$, 
a radial vector $(D_{\rm L}, \theta_{\rm N}, \varphi_{\rm N})$ giving the location of the
source, the orientation of the angular momentum $(\theta_{\rm L}, \varphi_{\rm L})$
at a fiducial time, the epoch of merger $t_{\rm C}$, 
and the signal's phase $\varphi_{\rm C}$ at that epoch. The information 
matrix (inverse of the covariance matrix) tends to be rather ill-conditioned and 
it is necessary to exercise a lot of care in its computation and inversion. Several
groups have independently confirmed the measurement accuracies and it is now
widely believed that sub-dominant harmonics truly bring about a dramatic improvement 
in the estimation of parameters \cite{Sintes:1999ch,Sintes:1999cg,Moore:1999zw,Arun:2007hu,AIMSSV,Trias:2007fp,Trias:2008pu,Porter:2008kn,LISAPE_paper}. 
One of the implications is that LISA will be able to measure the masses of the component 
black holes very accurately and obtain the mass function of (seed) black holes.

Localizing the source well enough for its host galaxy to be identified is
crucial for measuring $w$, since a knowledge of the redshift is needed. One may
then ask how parameter estimation in general benefits from a knowledge of the 
sky position. In the full 9-dimensional parameter space, $\theta_{\rm N}$
and $\phi_{\rm N}$ are partially correlated with the other 7 parameters. 
The associated degeneracies in parameter estimation get broken if $(\theta_{\rm N},\phi_{\rm N})$
are exactly known. The covariance matrix then gets reduced from a $9 \times 9$ to a 
$7 \times 7$ matrix, leading to smaller uncertainties \cite{Schutz:2009a,AIMSSV}.

In Ref.\ \cite{Arun:2007hu,AIMSSV}, it was shown how the inclusion of sub-dominant signal harmonics 
allows for a sufficiently good localizability of the source in the sky and measurement
of the luminosity distance that inference of $w$ becomes possible. However, in those papers
only a limited number of possible parameter values were considered. The event rates mentioned above
are integrated over a large range of masses, and it is important to know how accurately 
LISA can measure $w$ depending on which binaries it observes during its lifetime; indeed,
the quality of parameter estimation varies widely depending on the properties of the sources
\cite{Trias:2007fp,Trias:2008pu,Porter:2008kn}. 
Here we report on an exhaustive study of the relevant part of parameter space. We look at 
15 mass pairs with (observed) component masses
roughly in the range $10^5 - 10^8\,\Ms$, at a fixed redshift. For each of these, a 
Monte-Carlo simulation is performed with 5000 instances of sky position and orientation of the 
orbital plane. We then compute in what percentage of these cases one would be able to estimate
sky position well enough that host identification should be possible. Using only these 
instances, we calculate the distribution of the uncertainties in $w$. Our Monte-Carlo results
are in such a form that they allow for easy rescaling to different redshifts $z$; we study what
happens at $z = 0.55$, $z = 0.7$, and $z = 1$. LISA's instrumental noise will not be the
only restriction in the measurement of $w$; weak lensing will affect the determination
of luminosity distance. To assess weak lensing effects, our distance errors are combined
in quadrature with a 4\% additional error due to weak lensing \cite{Kocsisetal06}.

This paper is structured as follows. In Section \ref{sec:signal model} 
we give a brief description of the signal model used in our study. 
Our choice of systems is explained in Section \ref{sec:systems} and we study the impact of source
localizability on parameter estimation in Section \ref{sec:source_localizability}. In Section 
\ref{sec:cosmology} we describe how we go about determining the dark 
energy parameter $w$, including our criteria for localizability of the source. 
In Section \ref{sec:results} we discuss the results of our study, giving the
fraction of systems for which the host galaxy can be localized, and the 
level at which $w$ can be estimated depending on what kind of system is observed. 
We will show that weak lensing-induced errors in $D_{\rm L}$ will severely limit LISA's 
ability to measure $w$. We conclude in Section \ref{sec:conclusions} with the 
message that future studies should focus on correcting the effect of weak 
lensing.

Throughout this paper we set $G = c = 1$ unless stated otherwise.

\section{Signal model and LISA configuration}
\label{sec:signal model}

The coalescence of black hole binary systems is commonly divided into 
three successive epochs: inspiral, merger, and ringdown. During the 
inspiral, the two black holes are well separated, and the radial inspiral 
timescale is much larger than the orbital timescale. As a consequence, 
the gravitational wave signal emitted during this regime is
well-understood analytically, in terms of the post-Newtonian 
(PN) approximation of general relativity. The latter is a perturbative 
approach whereby the amplitudes and phases of gravitational waveforms 
are expressed in terms of a characteristic velocity $v$ (see \cite{BlanchetLRR} and
the extensive references therein). The dynamics 
of the binary system will be modeled very well in this way for many orbits, 
but eventually there comes a point where the PN approximation fails; 
after that a numerical solution to the full Einstein equations is called 
for. This happens when the \emph{innermost stable circular 
orbit} (ISCO) is reached, after which the black holes plunge 
towards each other to form a single black hole; this is referred to as the 
merger phase of the coalescence. The resulting black hole then undergoes
``ringdown" as it gradually settles down to a quiescent Kerr black hole. 

Although most of the signal-to-noise ratio (SNR) is accumulated during 
the final stages of inspiral and merger,
disentangling the parameters that characterize the system and 
extracting physical information relies critically
on longer observation times, so that it is important to carefully study 
the inspiral phase. In this work we focus on LISA parameter estimation from 
observations of just the inspiral process. By doing so we 
can make use of analytical expressions for the detector response and its derivatives with 
respect to the different parameters which are needed to perform parameter 
estimation. This allows us to implement fast algorithms and explore the 
parameter space in a comprehensive way, unlike in numerical simulations 
where one can only consider a single choice of parameters (masses, spins, 
$\ldots$) at a time. In this way we are able to carry out an 
extensive study of LISA's performance, in particular in measuring the dark energy 
equation of state. Nevertheless, by not including the merger and ringdown 
we are missing a fraction of the total SNR that LISA would observe 
(especially for the higher mass systems), which would improve parameter 
estimation. While there has been recent work on parameter 
estimation with LISA using numerical waveforms that include merger and 
ringdown \cite{Babak:2008bu,Thorpe:2008wh}, some more understanding 
may be needed before embarking on extensive studies (but see the 
semi-analytic approach of \cite{McWilliamsetal09}). As far as inspiral itself is
concerned, Stavridis et al.~\cite{Stavridisetal09} studied the effect of
spin-induced precession of the orbital plane on our ability to measure $w$, though
without inclusion of higher signal harmonics in the waveform. The work presented
here is complementary, in that it assumes zero spins but does include higher
harmonics in the analysis. 

%
%

The inspiral PN waveforms in the two polarizations $h_+$ and $h_\times$ 
take the general form 
\begin{widetext}
\begin{equation}
h_{+ , \times} = \frac{2 M \nu}{D_{\rm L}} (M \omega)^{2/3} 
\{H_{+ , \times}^{(0)} + x^{1/2} H_{+ , \times}^{(1/2)} 
+ x H_{+ , \times}^{(1)} + x^{3/2} H_{+ , \times}^{(3/2)} 
+ x^2 H_{+ , \times}^{(2)} + x^{5/2} H_{+ , \times}^{(5/2)}
+ x^3 H_{+, \times}^{(3)} \}.
\end{equation} 
\end{widetext}
Here $x(t) \equiv [2\pi M F(t)]^{2/3}$ is the post-Newtonian expansion 
parameter, with $F(t)$ the instantaneous
orbital frequency; $\mathcal{O}(x^q)$ is referred to as 
$q$th PN order. For observed component masses $m_1$ and $m_2$, $M = m_1 + m_2$ and 
$\nu = m_1 m_2/M^2$ are, respectively, the observed total mass and the symmetric 
mass ratio, and $D_{\rm L}$ is the luminosity 
distance to the source. The explicit expressions for $H_{+ , \times}^{(i/2)}$ 
up to 3PN can be found in \cite{Polarizations3PN}. We neglect the contribution of 
spins, so that the gravitational waveform can be parametrized by $9$ 
parameters: luminosity distance, $D_{\rm L}$; two angles 
$(\phi_{\rm N}, \theta_{\rm N})$ defining the 
source position; another two angles $(\theta_{\rm L}, \phi_{\rm L})$ specifying 
the orientation of the orbital angular momentum; two mass parameters; the
phase at coalescence, $\varphi_{\rm C}$; and the time of coalescence, $t_{\rm C}$.
The sky position and orientation angles are defined with respect to a solar system 
barycentric frame, as in \cite{Cutler:1997ta}.

For LISA observation of supermassive black hole binary inspirals, most 
of the SNR accumulates at frequencies below $10\,\mHz$, in which case it 
is appropriate to use the low-frequency approximation to the LISA 
response function \cite{Cutler:1997ta}. In this approximation the 
detector can be regarded as two independent Michelson interferometers, 
and the measured strain in each of these separately can be written as
\begin{equation}
h^{(i)} (t) = \frac{\sqrt{3}}{2} [F_+ ^{(i)}(t) h_+ (t) 
+ F_\times ^{(i)}(t) h_\times(t)] \; ,
\end{equation}
where $i = I,II$ labels the two independent interferometers. 
The response functions $F_+^{(i)}$ and $F_\times^{(i)}$ depend on 
time through the sky position and orientation of the source with respect to
LISA, which vary over the observation time because of LISA's orbital motion.
The factor $\sqrt{3}/2$ is due to the 
$60${\mbox{$^{\circ}$}} angle between the interferometers' arms.

It is convenient to express the waveform in the Fourier domain using the 
stationary phase approximation \cite{Sathyaprakash:1991mt}. The Fourier 
transform $\tilde{h}^{(i)}(f)$ of the response of detector $i$ then takes 
the form \cite{Arun:2007hu,Trias:2007fp}:
\begin{widetext}
\begin {equation}
\tilde{h}^{(i)}(f) = \frac{\sqrt{3}}{2} \frac{2M\nu}{D_{\rm L}}
\,\sum_{k=1}^{8}\,\sum_{n=0}^6\,
\frac{A^{(i)}_{(k,n/2)}(t(f_k))
\,x^{\frac{n}{2}+1}(t(f_k))\,e^{-i\phi^{(i)}_{(k,n/2)}(t(f_k))}}
{2\sqrt{k\dot{F}(t(f_k))}}\,
\exp\left[i\,\psi_{f,k}(t(f_k))\right],
\label {FT}
\end {equation}
\end{widetext}
where $f_k\equiv f/k$, an overdot denotes the derivative with respect to time, 
and $\psi_{f,k}(t(f_k))$ is given by
\begin {equation}
\psi_{f,k}(t(f_k)) = 2\pi f\,t(f_k) -
k\,\Psi(t(f_k))-k\,\phi_{\rm D}(t(f_k)) - \pi/4.
\label {phase}
\end {equation}
The waveform is a superposition 
of harmonics of the orbital frequency (labeled by the index $k$), and each 
harmonic has PN contributions to the amplitude (labeled by $n$; note that 
currently one can only go up to $n=6$, as no amplitude corrections are explicitly 
known beyond 3PN \cite{Polarizations3PN}). As the PN order in amplitude is increased, more and 
more harmonics appear; at 3PN order there are eight, which is why the 
index $k$ only runs up to $k=8$.  Quantities in Eqs.~(\ref{FT}) and 
(\ref{phase}) with the argument $t(f_k)$ denote their values at the time 
when the instantaneous orbital frequency $F(t)$ sweeps past the value $f/k$. 
$A^{(i)}_{(k,n/2)}(t)$ 
and $\phi^{(i)}_{(k,n/2)}(t)$ are the polarization amplitudes and phases of 
the $k$th harmonic appearing at $n/2$th PN order.  $\Psi(t)$ is the 
orbital phase of the binary and $\phi_{\rm D}(t)$ is a time-dependent 
term representing Doppler modulation due to LISA's motion around the Sun. Explicit expressions for 
$A^{(i)}_{(k,n/2)}$ and $\phi^{(i)}_{(k,n/2)}$ can be found in 
\cite{VanDenBroeck:2006ar,VanDenBroeck:2006qu}; time dependence of these 
quantities arises through the beam pattern functions due to the varying 
sky position and orientation of the source relative to the detector 
\cite{Cutler:1997ta}. The expression for $\phi_{\rm D}(t)$ is given in 
\cite{Cutler:1997ta}. For the PN expansions for $t(f)$, 
$\Psi(t)$, $F(t)$, and $\dot{F}(t)$  we refer to \cite{Blanchet:2001ax}. 

Each harmonic in $\tilde h_{(i)}(f)$ is taken to be zero outside a certain 
frequency range. The upper cut-off frequencies are dictated by the ISCO, 
beyond which the PN approximation breaks down. For simplicity we assume 
that this occurs when the orbital frequency $F(t)$ reaches $F_{\rm ISCO}$,  
the orbital frequency at ISCO of a test particle in Schwarzschild 
geometry\footnote{Note that the cut-off is placed on the orbital
frequency of the binary, not on the dominant harmonic in the gravitational
wave signal, hence the extra factor of 2 in the denominator of the expression for 
$F_{\rm ISCO}$ compared to what one often finds in other literature.}: 
$F_{\rm ISCO} = (6^{3/2} 2\pi M)^{-1}$.  Consequently, in the frequency 
domain, the contribution to $\tilde{h}^{(i)}(f)$ from the $k$th harmonic is 
set to zero for frequencies above $k F_{\rm ISCO}$. Thus, the $k$th 
harmonic ends at a frequency
\begin{equation}
F^{(k)}_{\rm ISCO} = 2.198 \times 10^{-3} \,k\,\left(\frac{10^6\,M_\odot}{M}\right)\,\mbox{Hz}.
\label{eq:fisco}
\end{equation}
In determining the lower cut-off frequencies we assume that the source 
is observed for at most one year, and the $k$th harmonic is truncated 
below a frequency $k F_{\rm in}$, where $F_{\rm in}$ is the value of 
the orbital frequency one year before ISCO is reached \cite{Arun:2007hu}:
\begin{equation}
F_{\rm in} = F(t_{\rm ISCO} - \Delta t_{\rm obs}) = 
\frac{F_{\rm ISCO}}{\left(1 + \frac{256\nu}{5 M}
\Delta t_{\rm obs} v_{\rm ISCO}^8\right)^{3/8}}.
\end{equation}
For simplicity, $F_{\rm in}$ was computed using the quadrupole formula. In the above, $t_{\rm ISCO}$ 
and $v_{\rm ISCO} = 1/\sqrt{6}$ are, respectively, the time and orbital 
velocity at the last stable orbit, and $\Delta t_{\rm obs} = 1$ yr.
However, LISA's sensitivity becomes poorer and poorer as one goes to lower frequencies, 
and current estimates normally assume a ``noise wall" at a frequency no lower than 
$f_{\rm s} = 10^{-5}$ Hz. Thus, we take the lower cut-off frequency of the $k$th harmonic to be the 
maximum of $f_{\rm s}$ and $k F_{\rm in}$. 
 
As has been shown by several groups, both for Earth-based detectors 
\cite{VanDenBroeck:2006qu, VanDenBroeck:2006ar} and for LISA 
\cite{Sintes:1999ch, Sintes:1999cg, Moore:1999zw, Arun:2007qv, Arun:2007hu, 
Trias:2007fp, Trias:2008pu, Porter:2008kn}, taking into account all the 
harmonics significantly improves the parameter estimation, and at the same 
time it extends the mass reach to higher mass systems. However, in our computer 
code we have restricted ourselves to 2PN order in both 
amplitude and phase. We emphasize that there is no technical difficulty in 
going to higher orders, but as shown in \cite{VanDenBroeck:2006ar,Arun:2007hu,AIMSSV,Trias:2007fp,Trias:2008pu}, 
the main improvement in parameter estimation occurs 
in going from 0PN to 0.5PN order in amplitude, and 2PN order will be more 
than sufficient for our purposes.  

Due to the large SNR values that will be measured by LISA in observing SMBBH 
events, the Fisher information matrix formalism can be used to perform the 
parameter estimation. In the limit of high SNR, the probability density
distribution of the true parameters near the measured value can be approximated
by a multivariate Gaussian distribution 
whose covariance matrix is given by the inverse of the Fisher 
information matrix \cite{Finn:1992wt}. For each of the interferometers $i = I, II$, the Fisher 
matrix takes the form
\begin{eqnarray} \label{eq.gamma_ij}
\Gamma^{(i)}_{\alpha\beta} & \equiv & (\partial_\alpha h^{(i)} | 
\partial_\beta h^{(i)}) \nonumber \\ & = & 2 \int_0^{\infty} 
\frac{\partial_\alpha \tilde{h}^{(i)*}(f) \partial_\beta 
\tilde{h}^{(i)}(f) + \partial_\alpha \tilde{h}^{(i)}(f) 
\partial_\beta \tilde{h}^{(i)*}(f)} {S_n (f)} \df, \nonumber \\
\end{eqnarray}
where $\partial_\alpha \equiv \partial / \partial \lambda^\alpha$,
with $\lambda^\alpha$ the parameters to be estimated. Specifically, 
we take these to be
\begin{equation}
\bar{\lambda} = (t_{\rm C}, \phi_{\rm C}, \cos(\theta_{\rm N}), \phi_{\rm N}, \cos(\theta_{\rm L}), \phi_{\rm L}, \ln D_{\rm L}, \mathcal{M}, \mu).
\end{equation}
The Fisher matrix for LISA as a whole is then
\begin{equation}
\Gamma_{\alpha\beta} = \Gamma_{\alpha\beta}^I + \Gamma_{\alpha\beta}^{II}.
\label{eq.gamma_tot}
\end{equation}
The covariance matrix is $\Sigma^{\alpha\beta} = (\Gamma^{-1})^{\alpha\beta}$,
which gives the covariances between parameters,
\begin{equation}
\langle \delta \lambda^\alpha \delta \lambda^\beta \rangle = \Sigma^{\alpha\beta},
\end{equation} 
and hence also the 1-$\sigma$ uncertainties, 
\begin{equation}
\Delta\lambda^\alpha \equiv \left[\langle (\delta \lambda^\alpha)^2 \rangle\right]^{1/2} =  \sqrt{\Sigma^{\alpha\alpha}}, 
\end{equation}
where no summation over repeated indices is assumed.

As mentioned above, we use the stationary phase approximation to the 
gravitational waveform in the frequency domain, which provides a way 
of getting analytical expressions for the observed 
signals, $\tilde{h}^{(i)}(f)$, and for their derivatives with respect to 
the parameters, $\partial_\alpha \tilde{h}^{(i)}(f)$ \cite{Arun:2007hu,Trias:2007fp}. \
Only the final integral over $f$ in Eq.~(\ref{eq.gamma_ij}) 
needs to be done numerically.

The code we use to generate our results has been validated by the LISA 
Parameter Estimation (LISA PE) Taskforce \cite{LISAPE_web} through 
cross-checking of the output with different codes from other groups 
\cite{LISAPE_paper}. 

In Eq.~(\ref{eq.gamma_ij}), $S_n(f)$ is the one-sided noise power 
spectral density, which is a combination of instrumental and galactic 
confusion noise. We take our noise curve to be the one that was used by 
all the members of the LISA PE Taskforce in \cite{LISAPE_paper}, which 
also corresponds to the noise curve from the second round of the Mock 
LISA Data Challenges \cite{Arnaud:2007jy, Babak:2007zd}. The 
sky-averaged instrumental noise is defined by
%
\bea
S_{\mathrm{inst}} (f) & = & \frac{1}{L^2} \left\lbrace \left[ 1+\frac{1}{2} \left( \frac{f}{f_\ast}\right)^2 \right] S_p  \right. \nonumber \\
& & + \left. \left[ 1 + \left (\frac{0.1\,\rm mHz}{f} \right)^2 \right] \frac{4 S_a}{(2\pi f)^4} \right\rbrace \; ,
\eea 
where $f$ is in Hz, $L = 5 \times 10^9\,\m$ is the armlength, \quad \quad \quad \quad \quad
$S_p = 4 \times 10^{-22}~\m^2~\Hz^{-1}$ is the (white) position noise level,
$S_a = 9 \times 10^{-30}~\m^2~\s^{-4}~\Hz^{-1}$ is the white acceleration noise level and
$f_\ast = c/(2\pi L)$ is the LISA arm transfer frequency (see Ref.\, \cite{LISAPE_paper} for further comments 
and details).

The galactic confusion noise is estimated by simulating the 
population synthesis of galactic binaries with periods shorter than 
$2 \times 10^{-4}$ s \cite{Nelemans:2001,Nelemans:2003ha}. 
The confusion noise at the output can be fitted as \cite{Cornish:2007if}
\begin{equation}
S_{\mathrm{conf}} (f) = \left\lbrace \begin{array}{ll}
10^{-44.62}~f^{-2/3},  &   f \leq 10^{-3}{\rm Hz}   \\
10^{-50.92}~f^{-4.4},  &   10^{-3}{\rm Hz}  < f \leq 10^{-2.7}{\rm Hz}   \\
10^{-62.8}~f^{-8.8},   &   10^{-2.7}{\rm Hz}  < f \leq 10^{-2.4}{\rm Hz}   \\
10^{-89.68}~f^{-20},   &   10^{-2.4}{\rm Hz}  < f \leq 10^{-2}{\rm Hz}   \\
0,                     &   f > 10^{-2}{\rm Hz} \end{array} \right. \; 
\end{equation}
The total noise curve is the sum of instrumental and confusion noise, 
\begin{equation}
S_n(f) = S_{\mathrm{inst}}(f) + S_{\mathrm{conf}}(f).
\end{equation}
Finally, as in \cite{LISAPE_paper}, we also apply a lower frequency 
cut-off at $10^{-5}$ Hz.


\section{Choice of systems studied}
\label{sec:systems}

\begin{figure*}[t]
\begin{center}
\includegraphics[width = 3in]{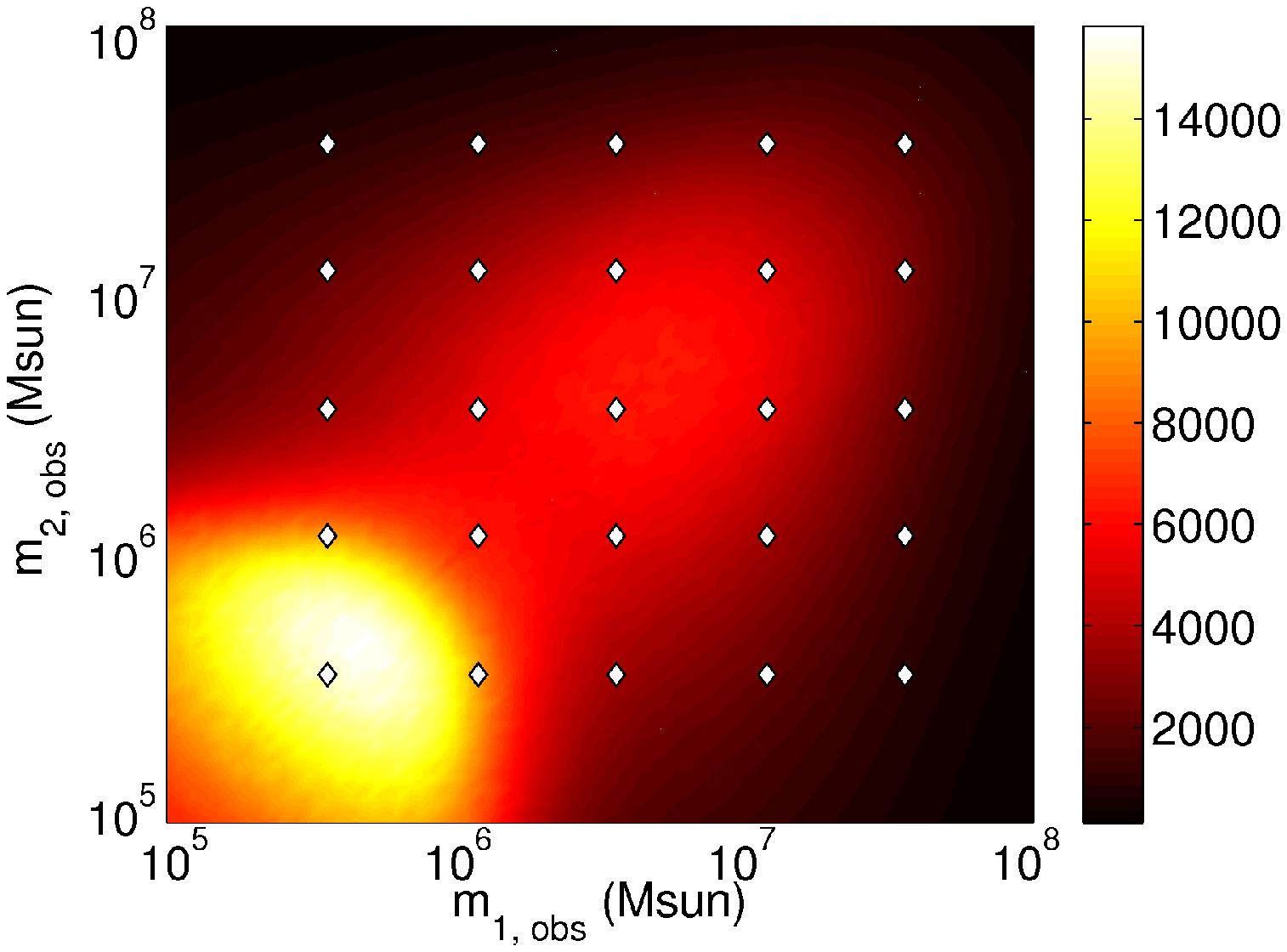}
\includegraphics[width = 3in]{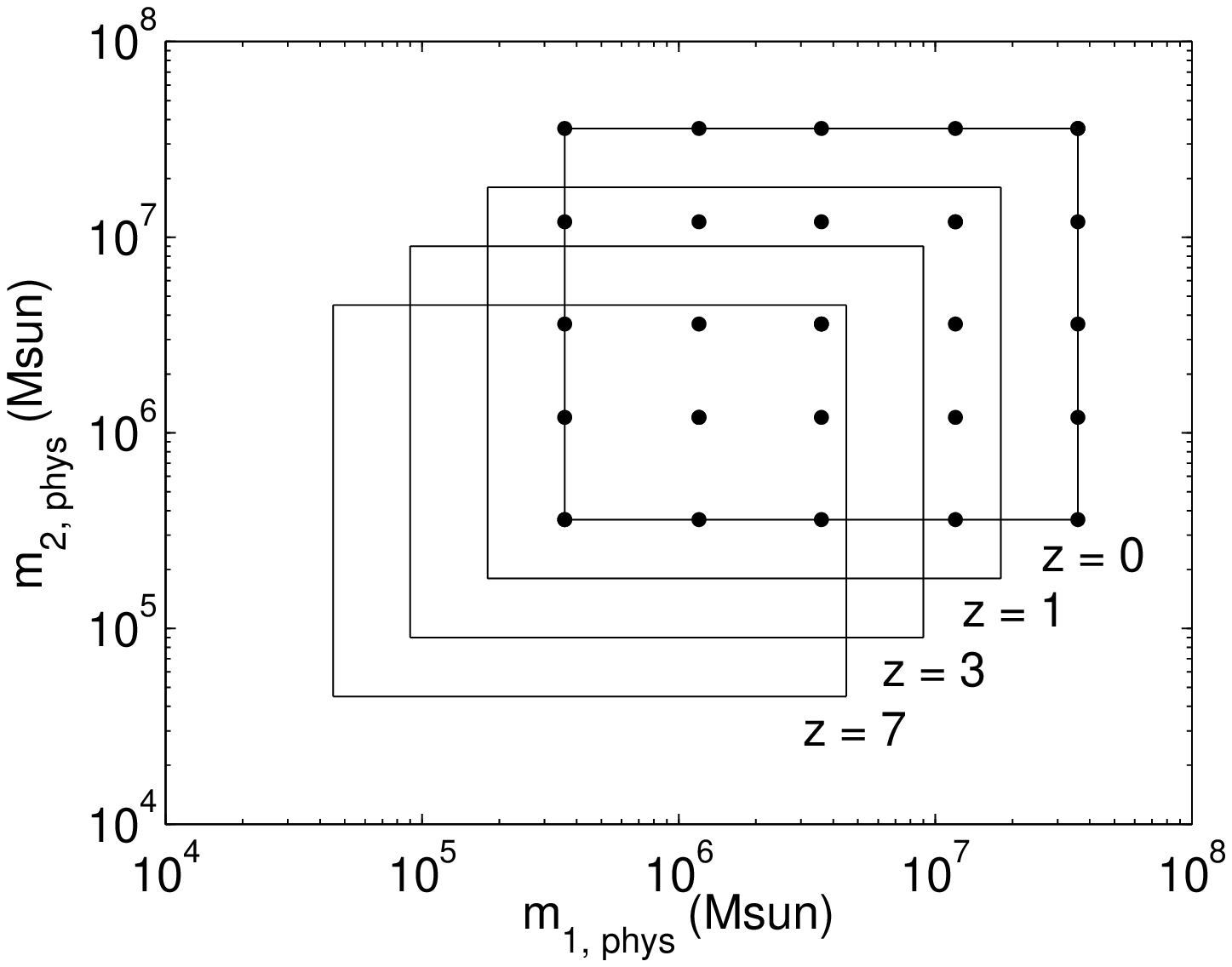}
\end{center}
\caption{Left: Median values of the measured signal-to-noise ratio 
($\mbox{SNR}_{1\,\Gpc}$) from the inspiral phase of SMBBH systems at 
$1\,\Gpc$, as a function of the observed 
masses of the individual black holes, obtained from $1000$ Monte-Carlo 
simulations over the sky location and orientation of the source. The 
quantity plotted is related to the observed SNR at any luminosity 
distance $D_{\rm L}$ as $\mbox{SNR}_{\rm D_{\rm L}} = \mbox{SNR}_{1\,\Gpc} 
\times \frac{\Gpc}{D_{\rm L}}$. The superimposed diamonds represent the 
grid of observed mass cases considered to study parameter estimation and 
its implications in measuring the dark energy equation of state.
Right: Representation of how the grid of observed masses in 
Fig.~\ref{Fig.SNR_vs_masses} (here as dots) translates into physical 
masses as we increase the redshift, $z$, of the source.}
\label{Fig.SNR_vs_masses}
\end{figure*}


LISA's sensitivity band stretches from $10^{-5}$ Hz to $0.1$ Hz, which 
means (see Eq.~(\ref{eq:fisco})) that LISA will be able to see the coalescence of 
SMBBH systems of (observed) total mass from $\sim 10^5\,\Ms$ to 
$\sim 10^8\,\Ms$ with signal-to-noise ratios of several hundreds to thousands (see 
Fig.~\ref{Fig.SNR_vs_masses}) almost anywhere in the observable Universe. Signals 
from higher mass systems will not significantly enter the frequency band, 
and systems with a total mass lower than $\sim 10^5\,\Ms$ will have signal 
amplitudes in LISA's band that quickly become too low to be observable.

How many SMBBH events is LISA expected to see in a year, and of what kind? 
There are several possible SMBBH formation scenarios that are able to reproduce 
the measured optical luminosity function of Active Galactic Nuclei in the redshift range 
$1 \lesssim z \lesssim 6$, but they differ in (i) the formation 
mechanism and masses of the ``seed" black holes, as well as in (ii) the 
details of how accretion causes black holes to grow in time. For instance, 
Volonteri \etal \cite{Volonteri:2002vz,Sesana:2007sh, Sesana:2008ur} consider a scenario where light 
``seed" black holes (of a few hundred $\Ms$) were produced as remnants of 
metal-free stars at $z \gtrsim 20$. Alternatively, gravitational instability
of massive proto-galactic disks at $z \gtrsim 10$ could have led to the 
formation of much heavier seeds with masses $\sim 10^5\,\Ms$ \cite{Begelman:2006db}. 
Regardless of seed masses, the seeds may have grown by an accretion process 
in which infalling matter has a constant angular momentum direction, spinning up the black holes 
\cite{Bardeen70,Thorne74}, or by chaotic accretion from a fragmented disc, causing
much smaller spins \cite{KingPringle}.  

The implications for LISA of these various scenarios were recently assessed by
the LISA PE Taskforce \cite{LISAPE_paper}. Generally, scenarios with heavy seeds
lead to several tens of mergers per year in LISA's past light cone, all of which
will be detectable. If the seeds were smaller then the rate in the past light cone 
will be a factor of several larger, but LISA's sensitivity to light black hole 
coalescences is smaller; here too the number of \emph{observed} mergers is a few tens
per year. Whatever scenario, in the course of its lifetime LISA
may see a few merger events that are close enough ($z \lesssim 2$) to be useful as 
standard sirens, but their expected masses differ depending on the scenario. 

For this reason, in this work we did not focus attention on any particular part of 
the mass range and considered a uniform sampling of systems within LISA's mass reach. 
In particular, we have analyzed $15$ different pairs of masses forming an almost 
uniform grid in the $\log m_1 - \log m_2$ plane 
that covers most\footnote{The lower mass systems, despite their low 
amplitude, remain in the LISA band for many cycles, which makes the 
parameter estimation highly expensive in terms of computational time. For 
this reason we restricted ourselves to systems with masses higher than 
$3.6 \times 10^5 \Ms$.} of the region of systems observable by LISA, as 
shown in Fig.~\ref{Fig.SNR_vs_masses}: we consider all possible 
combinations of observed component masses $m_{1,2} = \{ 3.6 \times 10^5, \, 
1.2 \times 10^6, \, 3.6 \times 10^6, \, 1.2 \times 10^7, \, 3.6 \times 10^7 \} 
\Ms$. As pointed out by a number of authors, e.g. 
\cite{Trias:2007fp,Arun:2007qv, Arun:2007hu}, the observed SNRs and 
parameter estimation depend sensitively on the sky location and orientation 
of the source, so that Monte-Carlo (MC) simulations are called for if one
wants to draw general conclusions about LISA's performance. Thus, in this work
for each of our $15$ pairs of masses, we have carried out $5000$ MCs 
over $\{ \cos \theta_{\rm N}, \phi_{\rm N}, \cos \theta_{\rm L},  \phi_{\rm L} \}$
drawn from a uniform distribution, computing in every case SNRs and parameter errors.

The luminosity distance, $D_{\rm L}$, and redshift, $z$ are extrinsic 
parameters in the problem. Since SNR, parameter uncertainties, and 
masses scale with these in simple ways, one can obtain a variety 
of results from calculations that were done with \emph{fixed} values of $D_{\rm L}$
and $z$.\footnote{Note that the output of a simulation is obtained without 
making any assumptions about the relationship between $z$ and $D_{\rm L}$.} 

Let us discuss the dependence of the results (SNRs and parameter 
errors) on the luminosity distance and redshift. First note that:
\begin{enumerate}
\item From conservation of energy, the amplitude of the gravitational wave
signal is inversely proportional 
to $D_{\rm L}$.
\item The frequency is redshifted because of the expansion of the Universe, 
which causes a blue shift in the observed masses relative to the physical
ones: $m_{\rm obs} = (1+z) m_{\rm phys}$, where $m_{\rm phys}$ stands for
any intrinsic parameter with dimensions of mass.
\end{enumerate}
Since the SNR and the Fisher information matrix 
are proportional to the signal amplitude and its square, respectively, 
we have\footnote{The error in the luminosity distance 
behaves differently because we are using the derivative with respect to 
the parameter we are varying.} 
\begin{equation} \label{eq.props2DL}
\mbox{SNR} \propto \frac{1}{D_{\rm L}}; \quad
\Delta D_{\rm L} \propto D_{\rm L}^2; \quad
\Delta\lambda \propto D_{\rm L},
\end{equation}
where $\lambda$ is any parameter different from $D_{\rm L}$. 
Thus, the quantities 
$\mbox{SNR} \times D_{\rm L}$, $\Delta D_{\rm L} / D_{\rm L}^2$, 
and $\Delta\lambda / D_{\rm L}$ are independent of $D_{\rm L}$. 

On the other hand, the redshift experienced by any signal due to 
the expansion of the Universe  translates into a shift in the physical 
masses. Thus, a single simulation made for some pair of observed masses, 
say $\{ m_{1, \, \rm obs} \, , \, m_{2, \, \rm obs}  \},$ will be 
representative of an infinite set of systems at different 
redshifts, $z$, with physical masses 
\begin{equation}
m_{1, \, \rm phys} = \frac{m_{1, \,\rm obs}}{1+z}, \, \, \, \, \, \, \,
m_{2, \, \rm phys} = \frac{m_{2, \,\rm obs}}{1+z}. 
\end{equation}
In Fig.~\ref{Fig.SNR_vs_masses}, the panel on the 
right illustrates how the observed masses in the left panel correspond to 
progressively lower physical masses as we consider sources at successively 
higher redshifts.


\section{Impact of source localizability on parameter estimation}
\label{sec:source_localizability}

\begin{figure*}[t]
\begin{tabular}{ccc}
$m_{phys} : ( 2.32 ~,~ 2.32 ) \times 10^5 \Ms$ & 
$m_{phys} : ( 2.32 ~,~ 0.77 ) \times 10^6 \Ms$ & 
$m_{phys} : ( 7.74 ~,~ 7.74 ) \times 10^6 \Ms$ \\
\includegraphics[width = 6cm]{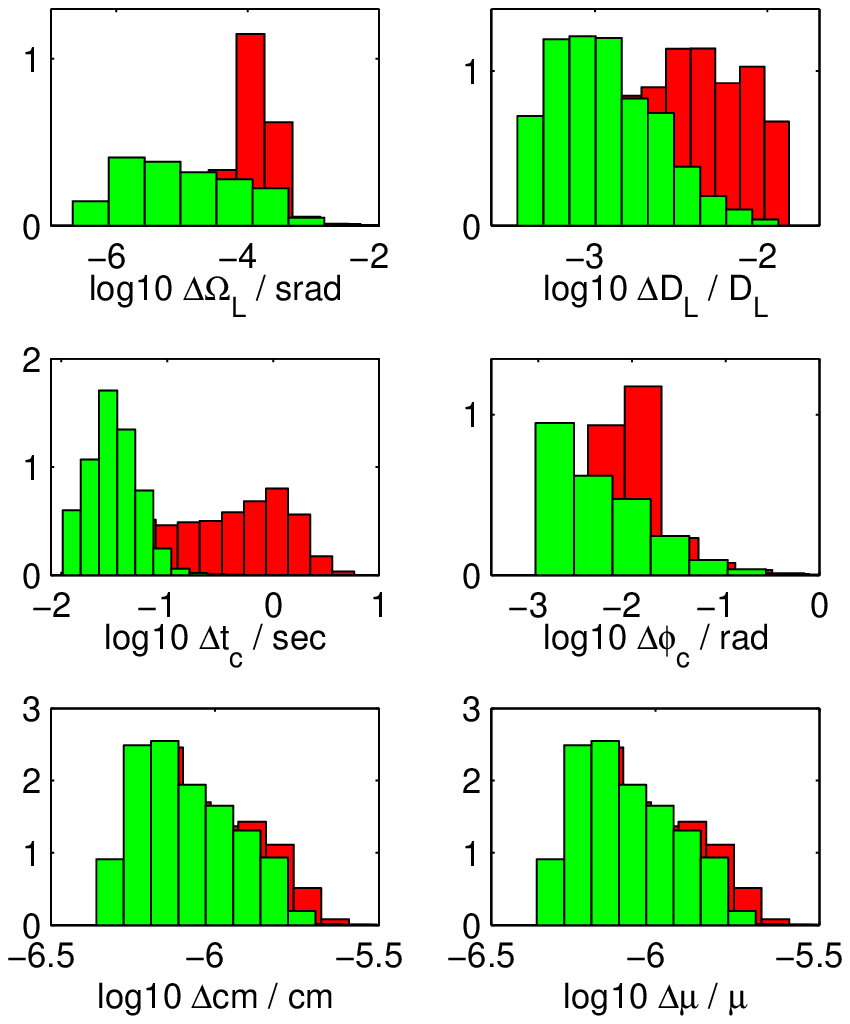} & 
\includegraphics[width = 6cm]{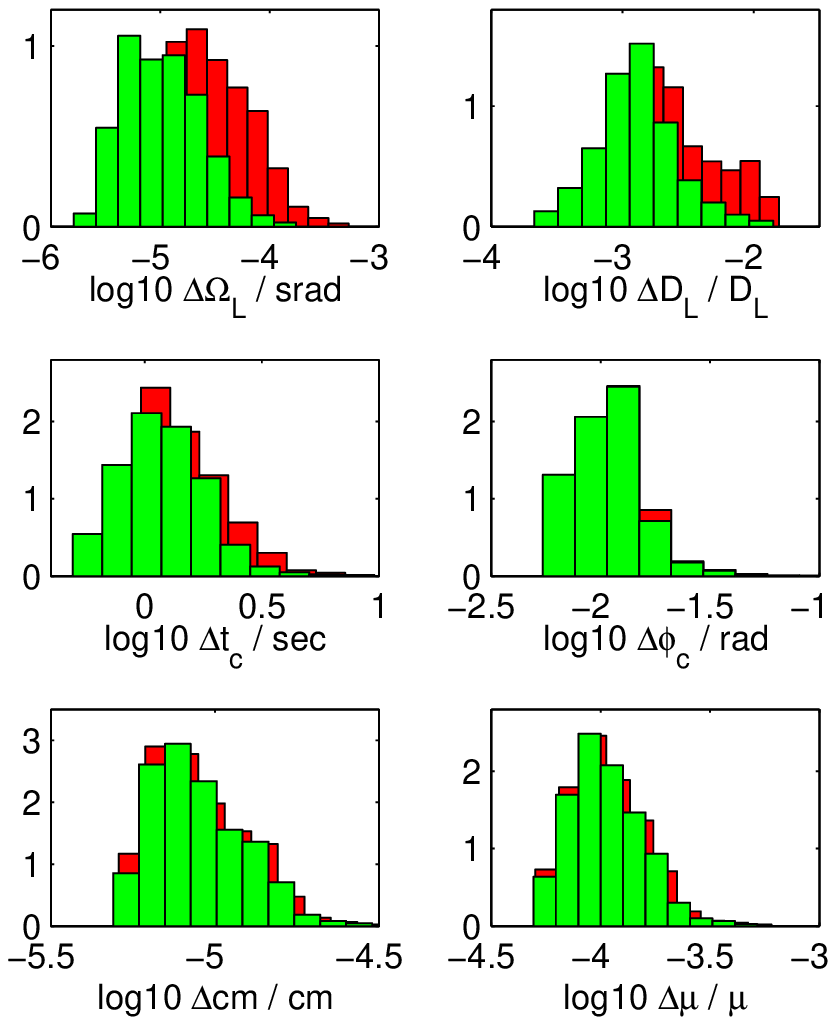} & 
\includegraphics[width = 6cm]{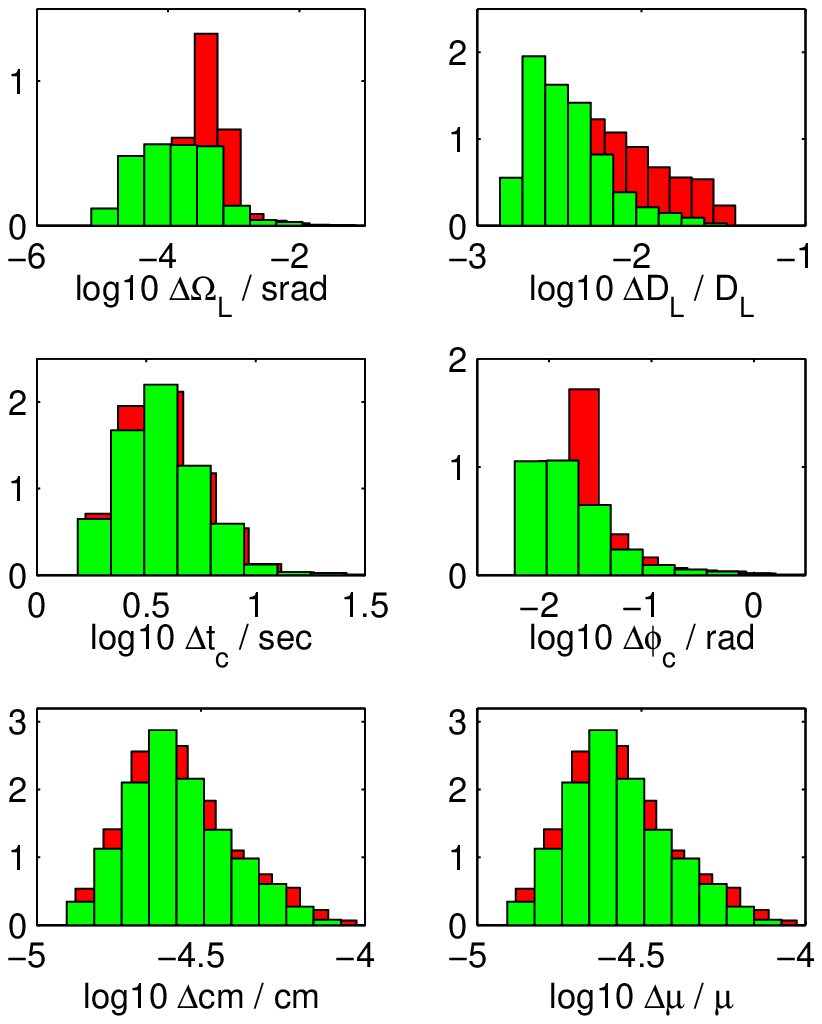}
\end{tabular}
\caption{Examples of distribution of errors on different parameters
before (histograms in the back) and after (front histograms) fixing the
sky location angles for three different combinations of masses:
light and symmetric (left), intermediate and asymmetric (middle), and
heavy and symmetric (right). $\Delta\Omega_{\rm L}$, the error on the
orientation of the orbital plane, is defined analogously to
$\Delta\Omega_{\rm N}$ (Eq.~(\ref{skyellipse})). 
We are assuming a fiducial luminosity distance of $D_{\rm L} = 3~\Gpc$ ($z_0 = 0.55$), and
we only consider the ``localizable'' systems for this case. 
}
\label{Fig.beforeafter}
\end{figure*}

\begin{table*}[t]
\begin{tabular}{crcrc}
\begin{tabular}{ccccc}
\multicolumn{5}{c}{\boldmath $\Delta\Omega_{\rm L}$} \\
           &         &         &         &  1.533 \\
           &         &         &  3.471  &  2.179 \\
           &         &  4.846  &  2.947  &  2.173 \\
           &  3.581  &  2.906  &  2.655  &  1.722 \\
   14.871  &  5.937  &  2.952  &  2.135  &  1.279
\end{tabular}
& \hspace{1cm} &
\begin{tabular}{ccccc}
\multicolumn{5}{c}{\boldmath $\Delta D_{\rm L}/D_{\rm L}$} \\
           &         &         &         &  1.342 \\
           &         &         &  1.980  &  1.432 \\
           &         &  2.428  &  1.723  &  1.507 \\
           &  1.960  &  1.753  &  1.565  &  1.411 \\
    3.644  &  2.380  &  1.750  &  1.356  &  1.202
\end{tabular}
& \hspace{1cm} &
\begin{tabular}{ccccc}
\multicolumn{5}{c}{\boldmath $\Delta t_{\rm C}$} \\
           &         &         &         &  1.000 \\
           &         &         &  1.019  &  1.138 \\
           &         &  1.495  &  1.038  &  1.005 \\
           &  2.963  &  1.146  &  1.004  &  1.002 \\
   15.564  &  1.982  &  1.150  &  1.005  &  1.005
\end{tabular}

\\ \\ \\

\begin{tabular}{ccccc}
\multicolumn{5}{c}{\boldmath $\Delta\phi_{\rm C}$} \\
           &         &         &         &  1.055 \\
           &         &         &  1.399  &  1.153 \\
           &         &  1.756  &  1.024  &  1.004 \\
           &  1.344  &  1.026  &  1.002  &  1.002 \\
    2.830  &  1.008  &  1.010  &  1.004  &  1.007
\end{tabular}
& \hspace{1cm} &
\begin{tabular}{ccccc}
\multicolumn{5}{c}{\boldmath \textbf{$\Delta\mathcal{M}/\mathcal{M}$}} \\
           &         &         &         &  1.000 \\
           &         &         &  1.006  &  1.099 \\
           &         &  1.021  &  1.031  &  1.004 \\
           &  1.029  &  1.011  &  1.006  &  1.002 \\
    1.055  &  1.015  &  1.010  &  1.005  &  1.006
\end{tabular}
& \hspace{1cm} &
\begin{tabular}{ccccc}
\multicolumn{5}{c}{\boldmath \textbf{$\Delta\mu/\mu$}} \\
           &         &         &         &  1.000 \\
           &         &         &  1.006  &  1.141 \\
           &         &  1.021  &  1.033  &  1.004 \\
           &  1.029  &  1.009  &  1.002  &  1.003 \\
    1.055  &  1.013  &  1.003  &  1.005  &  1.006
\end{tabular}
\end{tabular}
\caption{Improvement factors of the errors on different parameters after fixing the sky location
angles
for the ``localizable'' systems at $z_0 = 0.55$.
In particular, the quoted numbers represent the ratios between the median values of the 
error distributions before and after fixing the sky location. 
Note that these are independent of $D_{\rm L}$. We provide the results for $6$ physical parameters 
and 
15 choices of observed component mass pairs. The way the values are arranged corresponds to the
location
of the analyzed systems in the  $\log(m_1) - \log(m_2)$
plane as in Fig.~\ref{Fig.frac_median}. 
}
\label{Tab.beforeafter}
\end{table*}


Much of the literature on parameter estimation with LISA has used the 
restricted post-Newtonian waveform, which suggested that the position 
uncertainty would be too large and 
it would not be possible to find the host galaxy. But as pointed out 
by Sintes and Vecchio \cite{Sintes:1999ch,Sintes:1999cg} and Hellings 
and Moore \cite{Moore:1999zw}, and recently studied more thoroughly by 
Arun \etal \cite{Arun:2007hu}, Trias and Sintes \cite{Trias:2007fp,Trias:2008pu},  
and Porter and Cornish \cite{Porter:2008kn}, the higher harmonics in 
the orbital frequency that will also be present in the signal, carry a 
significant amount of information, and including them in search 
templates can vastly improve parameter estimation. (This is also the 
case for ground-based detectors; see \cite{VanDenBroeck:2006ar}.) 
In particular, with the inclusion of sub-dominant signal harmonics, 
in many cases the uncertainty in sky position decreases dramatically, 
so that host identification becomes possible, allowing for accurate 
measurement of $w$, as shown in \cite{Arun:2007hu,AIMSSV}. However, 
in the latter papers only a small number of example systems were 
considered. In the present work we aim to sample the parameter space 
far more thoroughly, enabling a much more detailed assessment of what 
might be possible.

If the host galaxy of an inspiral event can be found, then the sky position 
will be known with essentially no error. One would then match-filter the 
signal against a template family with fixed values for $\theta_{\rm N}$ and 
$\phi_{\rm N}$. This will help in removing the correlation between 
$(\theta_{\rm N}, \phi_{\rm N})$ and $D_{\rm L}$, 
resulting in a smaller uncertainty in the estimation of the luminosity 
distance than before \cite{AIMSSV}. The distance error $\Delta D_{\rm L}$ 
resulting from this smaller Fisher matrix is what determines the error 
on the dark energy equation-of-state parameter $w$ \cite{Schutz:2009a}. 

\begin{figure*}[t]
\begin{center}
\includegraphics[height=5cm,width = 6 cm]{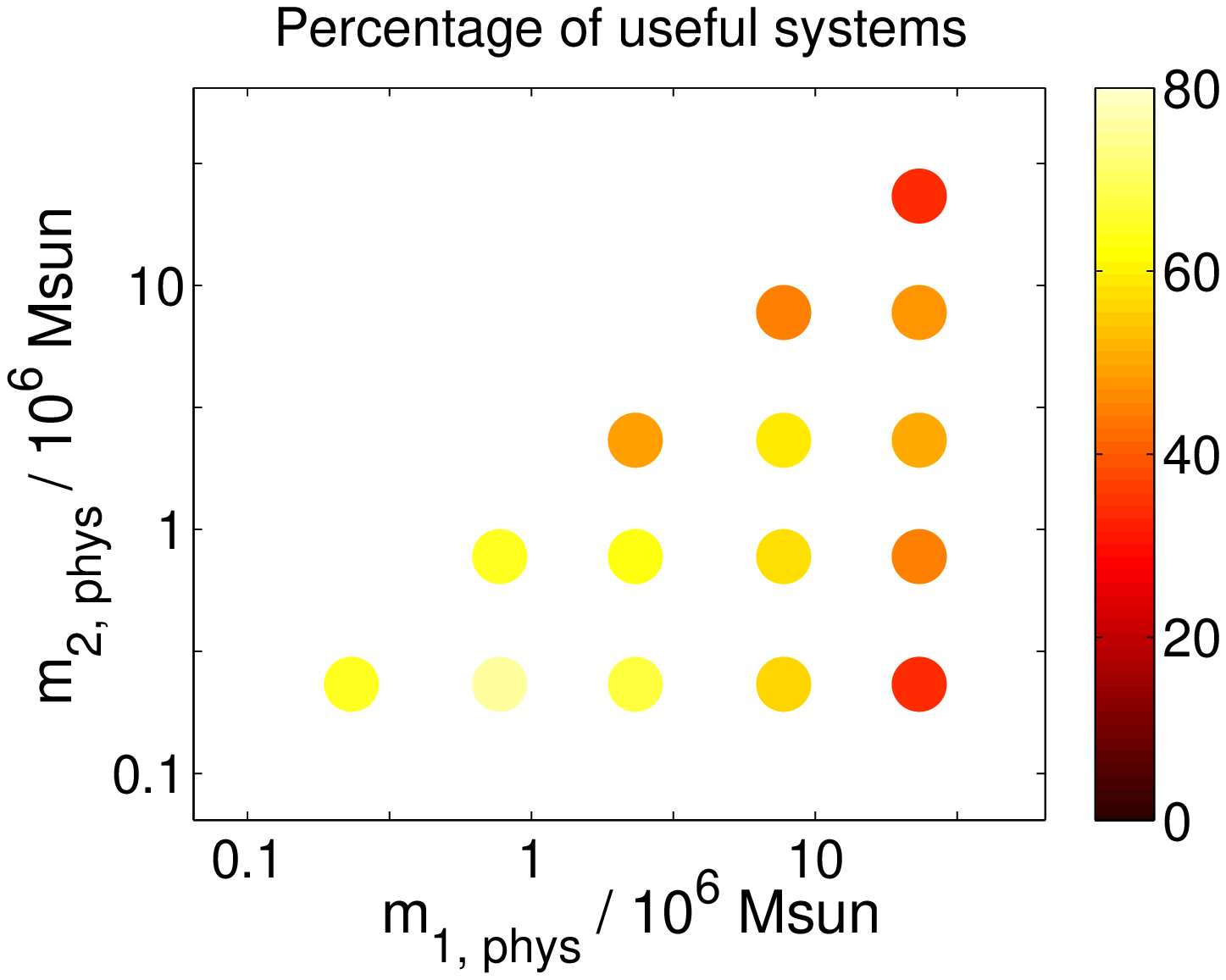}
\includegraphics[height=5cm,width = 6 cm]{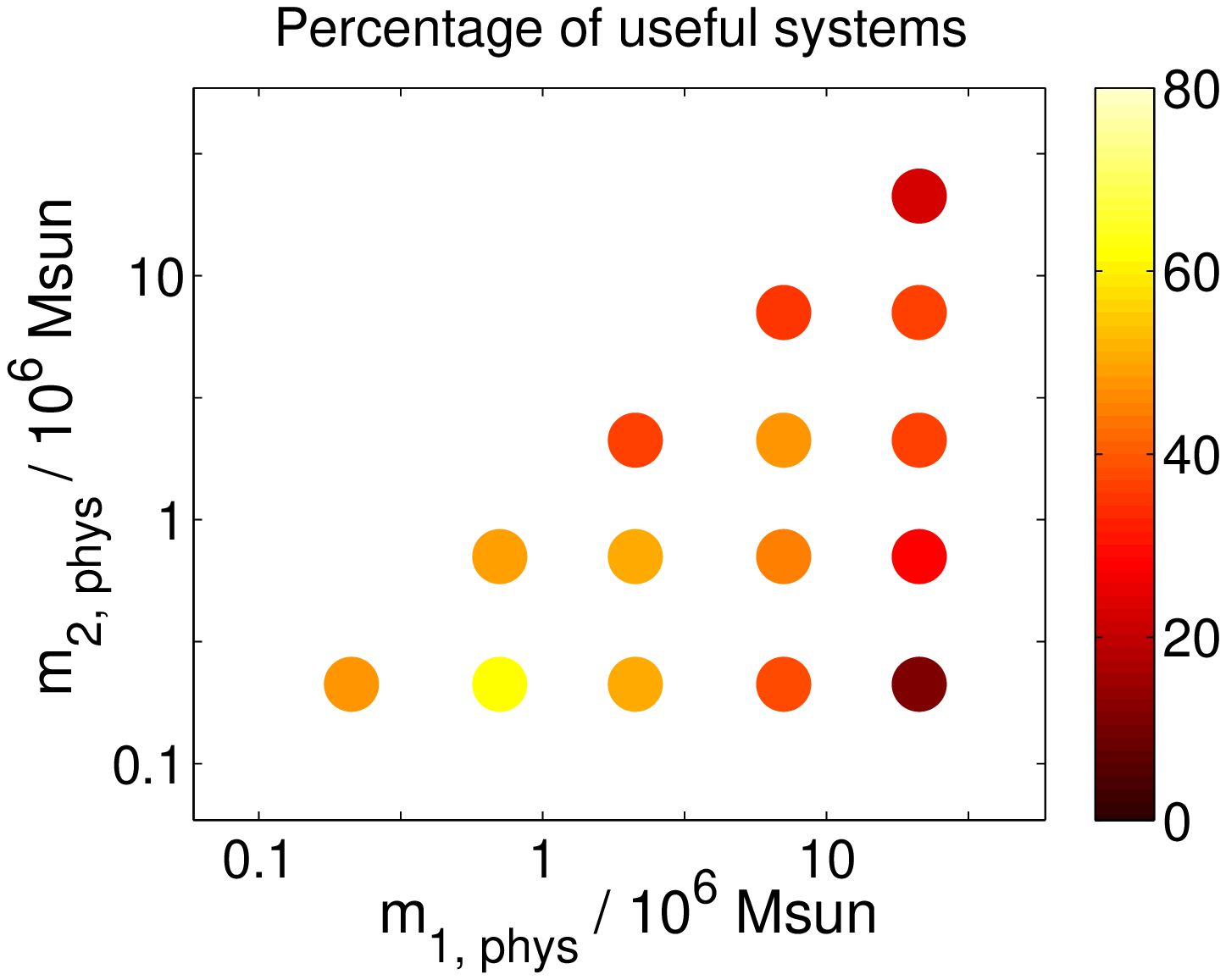}
\includegraphics[height=5cm,width = 6 cm]{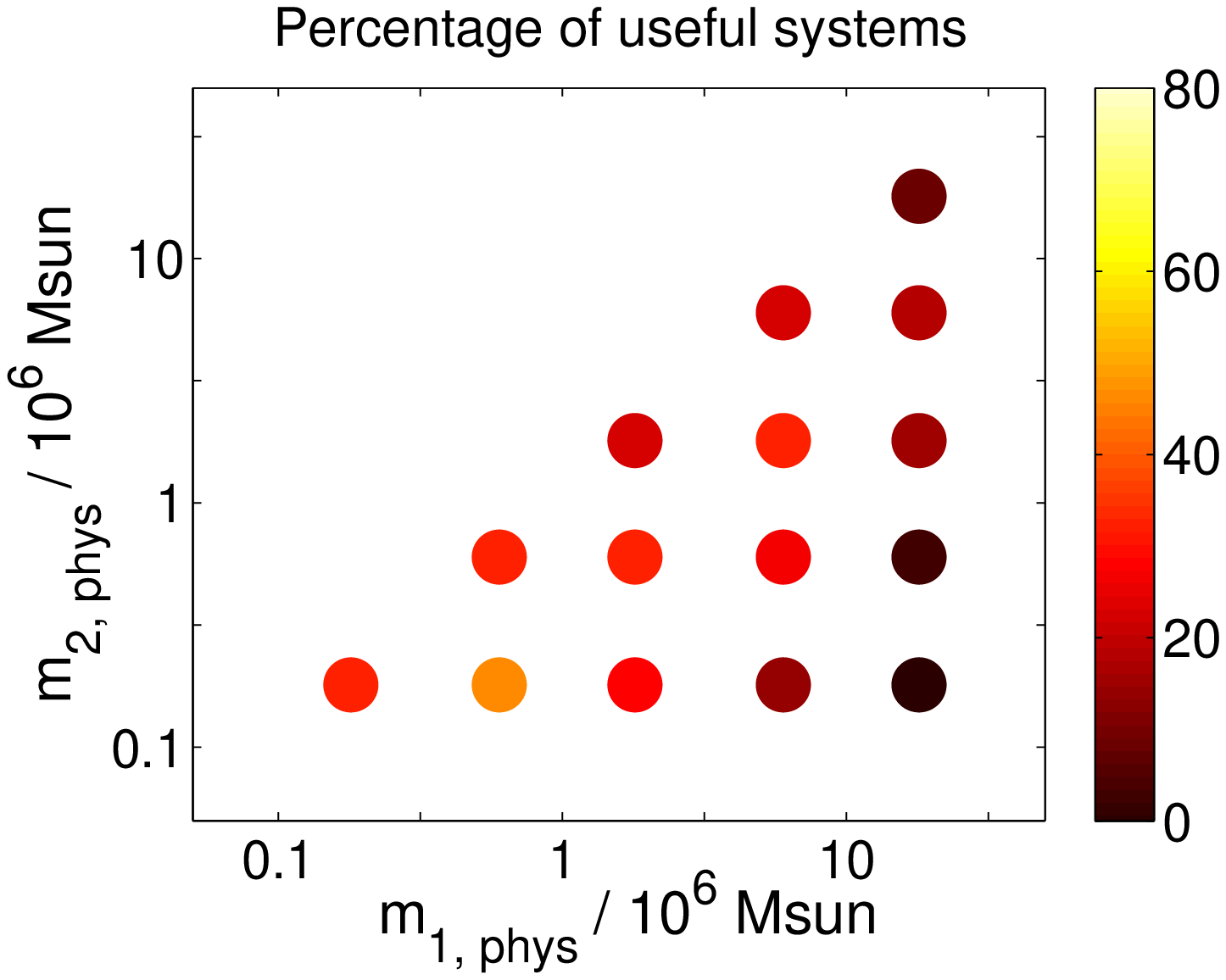}
\vskip 0.5 cm
\includegraphics[height=5cm,width = 6 cm]{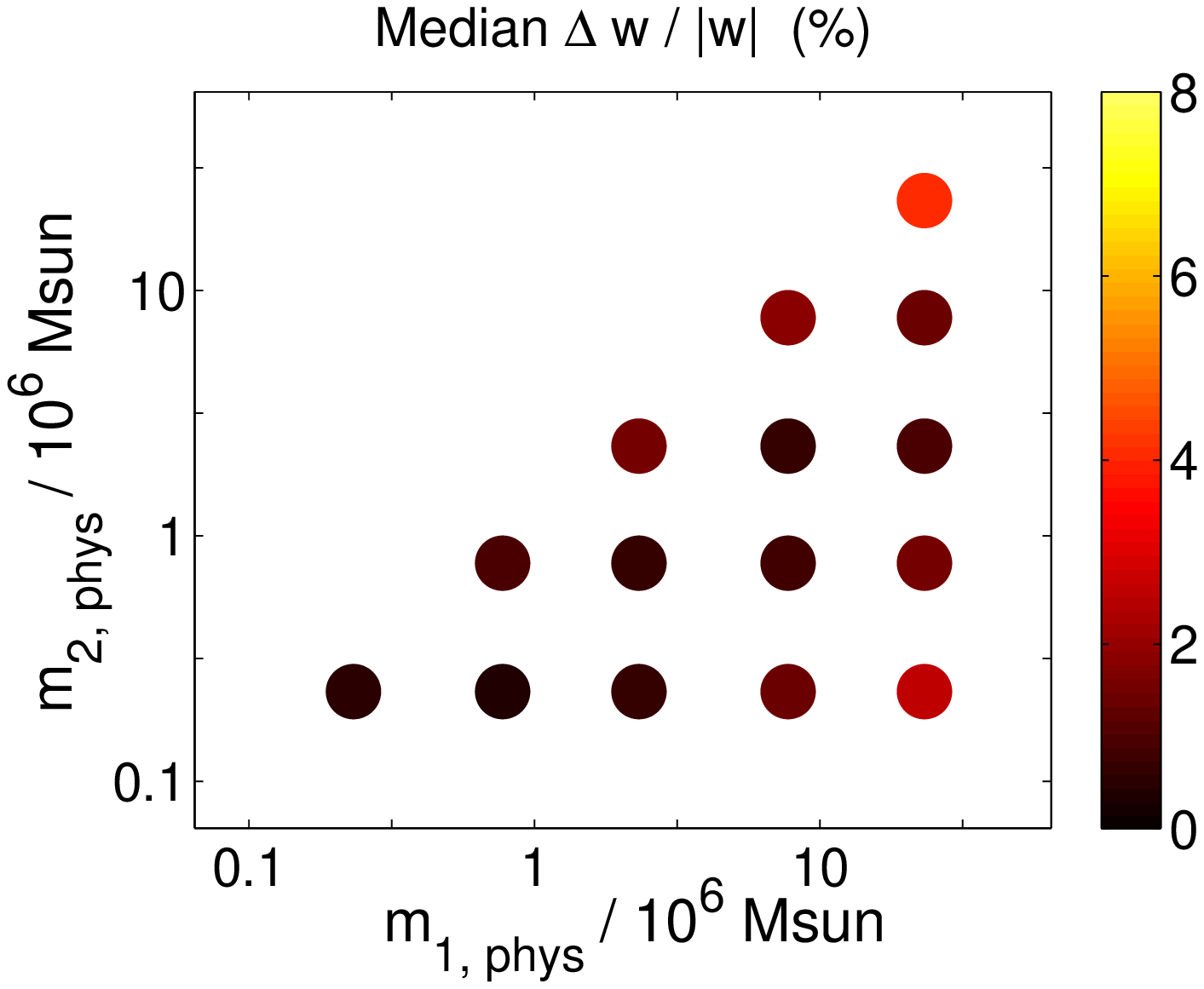}
\includegraphics[height=5cm,width = 6 cm]{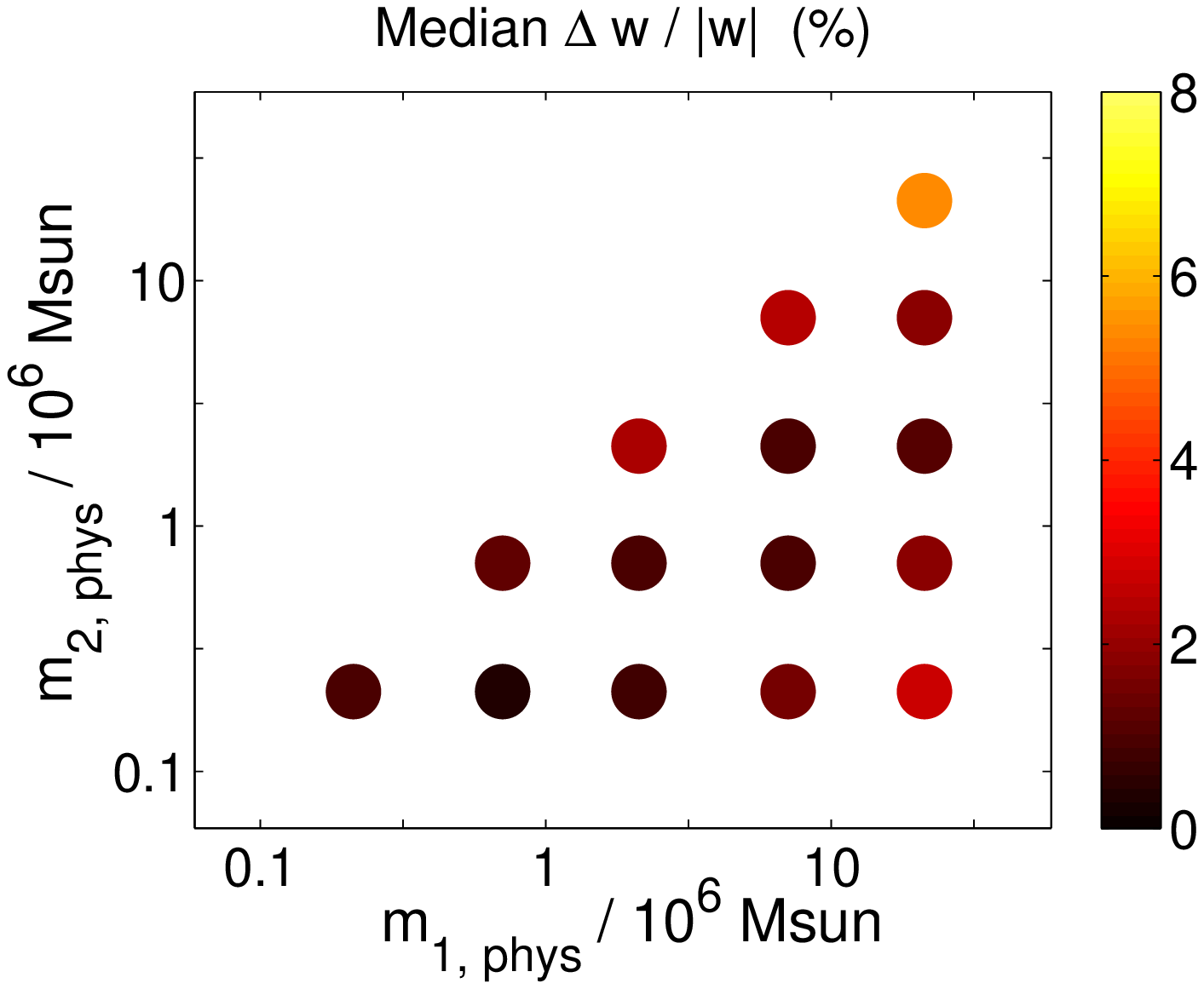}
\includegraphics[height=5cm,width = 6 cm]{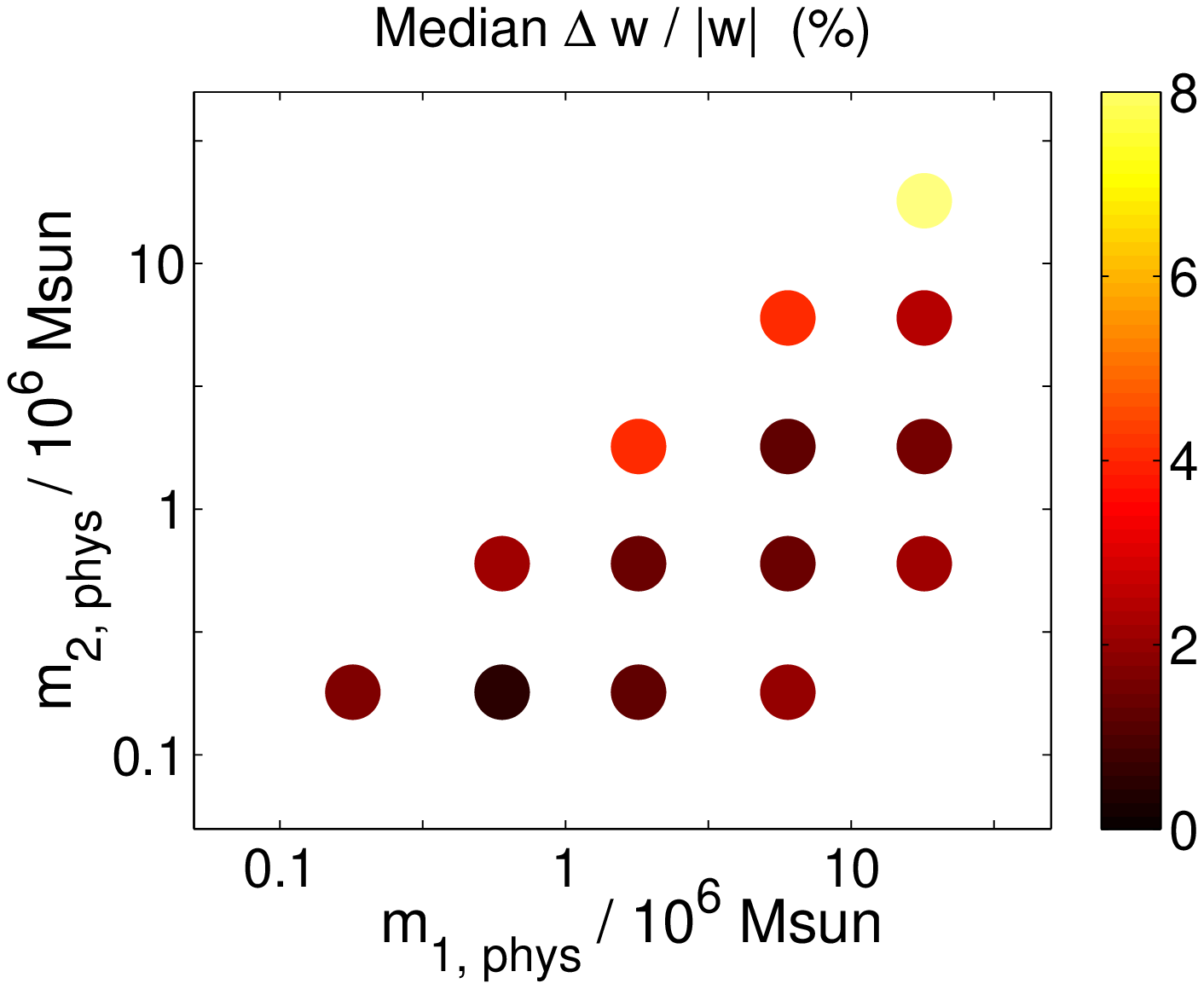}

\end{center}
\caption{Results for three choices of redshift: $z = 0.55$ (left column),
$z= 0.7$ (middle column), and $z= 1$ (right column). In each column, the top
plots show the fraction of inspiral events for which the host galaxy can be 
identified (by our criterion $N_{\rm clusters} < 3$) for the 
various choices of component masses we considered. The bottom plots show
the median uncertainties in the determination of $w$ for the systems that
allow for host determination and hence measurement of redshift. The masses
are the physical ones.}
\label{Fig.frac_median}
\end{figure*}

The source localizability criterion that we consider in this paper is explained in next section, but
first we are interested in the impact of the localization on the other parameters associated with
inspiral events. Figure~\ref{Fig.beforeafter} and Table \ref{Tab.beforeafter} show how
knowledge of sky position improves
the estimation of the unit normal to the inspiral plane (where
$\Delta\Omega_{\rm L}$ is the stereal angle subtended by the 
two-dimensional uncertainty ellipse on the unit sphere), the luminosity distance 
$D_{\rm L}$, the coalescence time $t_{\rm C}$, the phase at
coalescence $\phi_{\rm C}$, the chirp mass $\mathcal{M}$, and the
reduced mass $\mu$. In these figures, we only consider ``localizable'' systems, consisting in a
certain fraction (represented in Fig.~\ref{Fig.frac_median}) of the total.

We see the following trends:
\begin{enumerate} 
\item Knowing sky position has a bigger effect on light systems than on
heavy ones. Heavier systems deposit less power into LISA's band and have a smaller SNR. 
In that case our localizability requirement can only be fulfilled by systems where the correlations
between parameters were already relatively small to begin with, so that parameter estimation
would already have been good beforehand. Adding the information on sky position then will not lead
to significantly more improvement.
\item Symmetric systems show more improvement in parameter
estimation when sky location is known than asymmetric ones. Indeed, the odd harmonics are
all proportional to the difference between component masses $(m_1 - m_2)$,
so that they are absent for symmetric systems. For asymmetric systems, the presence
of the odd harmonics helps break degeneracies, and adding sky position information again
does not lead to great improvement. 
\end{enumerate}
The improvements in the estimation of chirp mass and reduced mass are modest; depending on
the system, the gains are between a fraction of a percent and 10\% (for $\mathcal{M}$) or
14\% (for $\mu$). Much greater improvements can be seen in the measurement of the luminosity
distance
and the orientation of the inspiral plane [i.e., the unit vector determined by $(\theta_{\rm L},
\phi_{\rm L})$],
as these are much more strongly correlated with sky position. The great accuracy in the
determination of 
$D_{\rm L}$ (typically a fraction of a percent even for quite massive systems) will translate into
excellent estimation
of $w$, if weak lensing can be subtracted, as we shall see in the next sections.


\section{Cosmology and measurement of dark energy}
\label{sec:cosmology}

As argued in \cite{HolzHugh05,DaHHJ06}, by treating SMBBH as
{\em standard sirens}, LISA could play 
a role in studying the physical nature of dark energy. From the 
gravitational wave signal itself one can measure the luminosity 
distance $D_{\rm L}$ with good accuracy, but not the redshift. 
However, the amplitude and phase modulations induced in the observed 
gravitational waveform due to LISA's motion around the Sun allow 
for a determination of the source's position in the sky. If the error ellipse 
associated with the sky position measurement is small enough that it 
contains a sufficiently small number of galaxies or galaxy clusters, 
then it may be feasible to identify the host galaxy, possibly with 
the help of an electromagnetic counterpart to the inspiral event. In 
that case a redshift $z$ can be obtained. Now, the relationship 
between $D_{\rm L}$ and $z$ depends sensitively on cosmological 
parameters such as $H_0$, $\Omega_{\rm M}$, $\Omega_{\rm DE}$, and 
$w$ -- respectively, the Hubble parameter at the current epoch, the 
matter and dark energy density (normalized by the critical density), and 
the dark energy equation-of-state parameter. Hence, separate 
measurements of the distances and redshifts to four or more sources
would constrain these parameters. 

For the purposes of this paper we assume a spatially flat 
Friedman-Lema\^itre-Robertson-Walker (FLRW) Universe 
with constant $w$. In that case, the relationship between the luminosity 
distance $D_{\rm L}$ and redshift $z$ is given by
\begin{equation}
D_{\rm L}(z) = \frac{(1+z)}{H_0}\,\int_0^z \frac{dz'}{\left[ \Omega_{\rm M}(1+z')^3 + \Omega_{\rm DE} (1+z')^{3(1+w)} \right]^{1/2}}.
\label{DLgeneral}
\end{equation} 
In principle, a measurement of the cosmological parameters could proceed 
as follows. Imagine that a number of SMBBH inspiral events have 
been found in LISA data, and that their host galaxy has been identified. The 
redshifts $z$ can then be determined with negligible error. From the 
gravitational wave signals themselves, the luminosity distances could be 
extracted. A fit of $D_{\rm L}$ as a function of $z$ using the expression 
(\ref{DLgeneral}) would then allow us to deduce the values of $H_0$, 
$\Omega_{\rm M}$, $\Omega_{\rm DE}$, and $w$. In practice, however, there 
may not be a large enough number of sources for which the sky position can 
be determined sufficiently well to allow the identification of the host 
galaxy, in which case it will be impracticable to constrain all four 
cosmological parameters at the same time.

In this paper we consider an illustrative example. We will assume that we only have access to a single
inspiral event, which will be used to estimate one cosmological parameter, $w$ in our case; 
the other parameters will be considered known with negligible errors. Such an
observation would complement the estimation of $(H_0, 
\Omega_{\rm M}, \Omega_{\rm DE}, w)$ obtained by other means, e.g., via gravitational-wave
observations of other LISA sources such as the extreme mass ratio inspirals (EMRIs)
\cite{MacLeod:2008ab}, or via the observations of stellar mass compact binaries with ground-based detectors like 
the Advanced LIGO \cite{Nissanke:2009ab} or the Einstein Telescope \cite{Sathyaprakash:2009ab}. The
latter may see as many as 500 (stellar mass) inspiral events per year with identifiable
electromagnetic counterparts, giving several thousands over a period of five years. From 
each of the signals a luminosity distance could be extracted, and the 
electromagnetic counterpart would allow us to find the host and obtain a 
value for redshift. Fitting the function $D_{\rm L}(z)$ would then 
severely constrain at least a subset of the unknowns $(H_0, \Omega_{\rm M}, 
\Omega_{\rm DE}, w)$ \cite{Sathyaprakash:2009ab}. Even without these other gravitational wave measurements, by the time
LISA is operational, all of the parameters (including $w$) may already have been measured
with good accuracy through electromagnetic means, by continued studies of the Cosmic 
Microwave Background, baryon acoustic oscillations, gravitational lensing, and a larger 
population of Type Ia supernovae \cite{DE_taskforce}. 
What LISA can add, even if only one parameter is measured, is an important \emph{consistency check}: 
gravitational wave astronomy brings the unique benefit that cosmological parameters can be 
constrained without reference to a cosmic distance ladder. It is then natural not to make any
a priori assumptions on $w$.

As explained in the previous section, we simulated a large number of 
instances of SMBBH inspirals, with different masses, sky positions, 
and orientations of the orbital plane relative to the observer. 
We aim to answer two questions: 
\begin{enumerate}
\item What fraction of these instances allows for identification of the 
host galaxy?\\
\item For each of the events where the host can be found, how accurately 
can we measure $w$?
\end{enumerate}

First we need a criterion to discriminate between cases where the host can 
be identified and cases where it can not. In order to do this, we define a 
fiducial cosmological model which we will use as a reference; say, a 
spatially flat FLRW Universe with $H_0 = 75\,\mbox{km}\,\mbox{s}^{-1}\mbox{Mpc}^{-1}$, 
$\Omega_{\rm M} = 0.27$, $\Omega_{\rm DE} = 0.73$, and $w = -1$. Using this 
fiducial model, to the measured value of $D_{\rm L}$ we can associate 
a fiducial redshift value $z_0$. Deciding whether or not the host galaxy 
can be found will involve counting the number of galaxies or galaxy 
clusters in some volume error box around the approximate sky position 
and distance of the inspiral event. The error ellipse in the sky, 
$\Delta \Omega_{\rm N}$, will provide one constraint in determining such 
an error box; one has
\begin{equation}
\Delta\Omega_{\rm N} = 2\pi \sqrt{ ( \Delta\cos(\theta_{\rm N}) \Delta\phi_{\rm N} )^2 
- \langle \delta\cos(\theta_{\rm N}) \delta\phi_{\rm N} \rangle^2}. 
\label{skyellipse}
\end{equation}
However, we are not allowed to also use the error in the 
determination of \emph{distance}, as measuring $w$ requires an 
\emph{independent} determination of the luminosity distance and redshift. 
Instead, we will count how many galaxies or clusters there are in an 
error box determined by the sky position error ellipse $\Delta\Omega_{\rm N}$, 
and a large redshift interval centered on the fiducial redshift $z_0$. For 
concreteness we take this interval to be $[0.8 z_0, 1.2 z_0]$. This is a 
generous choice: given the above values of $(H_0, \Omega_{\rm M}, \Omega_{\rm DE})$, 
to reconcile the measured $D_{\rm L}$ with a redshift that differs by 10\% 
from our fiducial $z_0$ will typically require picking a value of $w$ that 
lies far outside the existing bounds from WMAP and supernovae studies 
\cite{Arun:2007hu,AIMSSV}. The sky position error $\Delta \Omega_{\rm N}$ 
together with the redshift interval $[0.8 z_0, 1.2 z_0]$ will, through 
our fiducial cosmological model, imply a comoving volume $\Delta V_{\rm C}$ 
in which to search for host galaxies:
\begin{widetext}
\begin{equation}
\Delta V_{\rm C} = \int_{0.8 z_0}^{1.2 z_0} dz' 
\frac{\Delta \Omega_{\rm N}}{H_0} \frac{D_{\rm L}^2(z')}{(1+z')^2}
\frac{1}{\sqrt{\Omega_{\rm M}(1+z')^3 + \Omega_{\rm DE}(1+z')^{3(1+w)}}}.
\end{equation} 
\end{widetext}
To arrive at an actual number of galaxies or galaxy clusters within the 
volume $\Delta V_{\rm C}$ we need an estimate for the density of 
clusters\footnote{We note that at redshifts $z \sim 1$ and beyond,  
galaxy clusters become increasingly ill-defined; we will merely use the 
number of clusters as a quantitative way of judging whether the host of 
an inspiral event will be identifiable.}. This density is not known very 
well; here we follow Bahcall et al.~\cite{Bahcalletal}, who give 
$\rho_{\rm clusters} \sim 2 \times 10^{-5} h^3 \mbox{Mpc}^{-3}$, with 
$h$ the Hubble parameter at the current epoch in units of $100\, \mbox{km}\, 
\mbox{s}^{-1} \mbox{Mpc}^{-1}$. The number of clusters within our volume 
error box can then be estimated as $N_{\rm clusters} \simeq 
\rho_{\rm clusters} \Delta V_{\rm C}$. If for a particular inspiral event 
$N_{\rm clusters}$ turns out to be of order 1 then the host cluster can 
be found and a redshift value can be obtained. It could be that the host 
can be identified even when $N_{\rm clusters} \gg 1$, as the binary SMBBH 
merger might be accompanied by a distinctive electromagnetic counterpart 
which could be found by future large survey instruments though electromagnetic counterparts \cite{LangHughes08,
Kocsisetal08}. Even so, we will take $N_{\rm clusters} < 3$ 
to be our localizability criterion. Issues related to finite cluster size
and identification of the actual host galaxy will be discussed below. At first
instance one wants to identify the host cluster, and for that it will typically
not be problematic if there are several clusters within the volume box. As an example, 
consider an inspiral at 3 Gpc, which in our fiducial cosmological model corresponds to $z = 0.55$. Suppose the
volume box contains a few clusters with redshifts differing by 10\%, e.g., 
imagine there is a potential host cluster at $z = 0.6$. Then in order to reconcile
this slightly larger redshift with the measured distance, for the same values
of $H_0$, $\Omega_{\rm M}$, and $\Omega_{\rm DE}$ one would arrive at $w = -0.47$, a value that
is strongly excluded by WMAP and supernovae studies.

When the host galaxy or galaxy cluster of an inspiral event can be found, we can assume that the
sky position will be known with essentially no error,\footnote{For systems at $z=1$, LISA's sky
resolution will be at best $10^{-4}$~srad \cite{Trias:2007fp}, whereas the solid angle
subtended by a galaxy cluster at the same redshift is at least two orders of magnitude smaller.
Hence, for the purposes of Fisher matrix calculations, we can assume that the sky position
has negligible error when the host cluster can be identified.}
which will allow us to recompute a reduced Fisher matrix removing the correlations between
$(\theta_{\rm N}, \phi_{\rm N})$ and the other parameters. In particular, this will translate into
an improvement in the estimation of 
$D_{\rm L}$ (see Table \ref{Tab.beforeafter}), which is what determines the error on the
equation-of-state parameter $w$ \cite{Schutz:2009a}. 

\begin{figure*}[t]
\begin{tabular}{ccccc}
 & & & & 
\includegraphics[width = 3.5 cm]{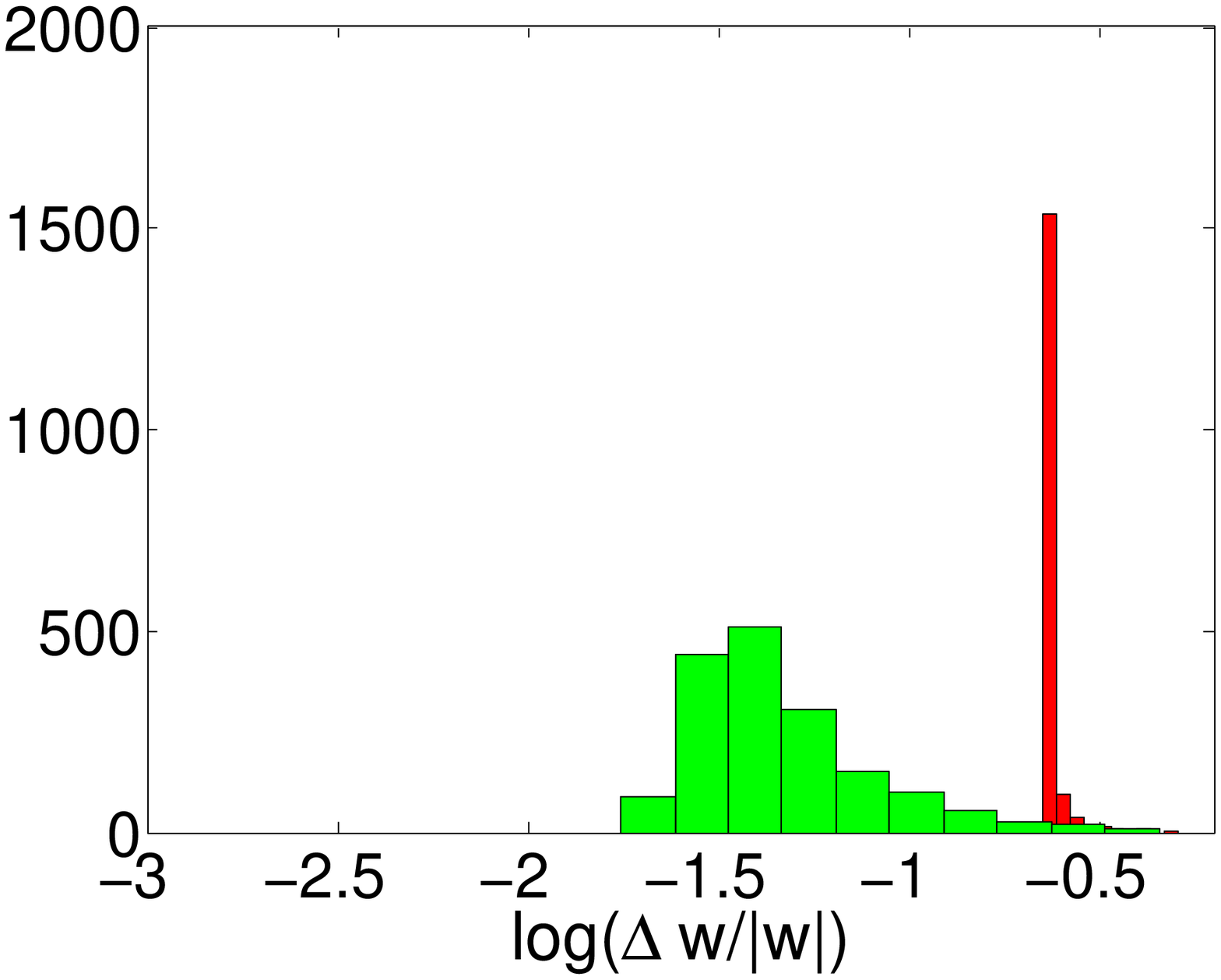}
\\
 & & &
\includegraphics[width = 3.5 cm]{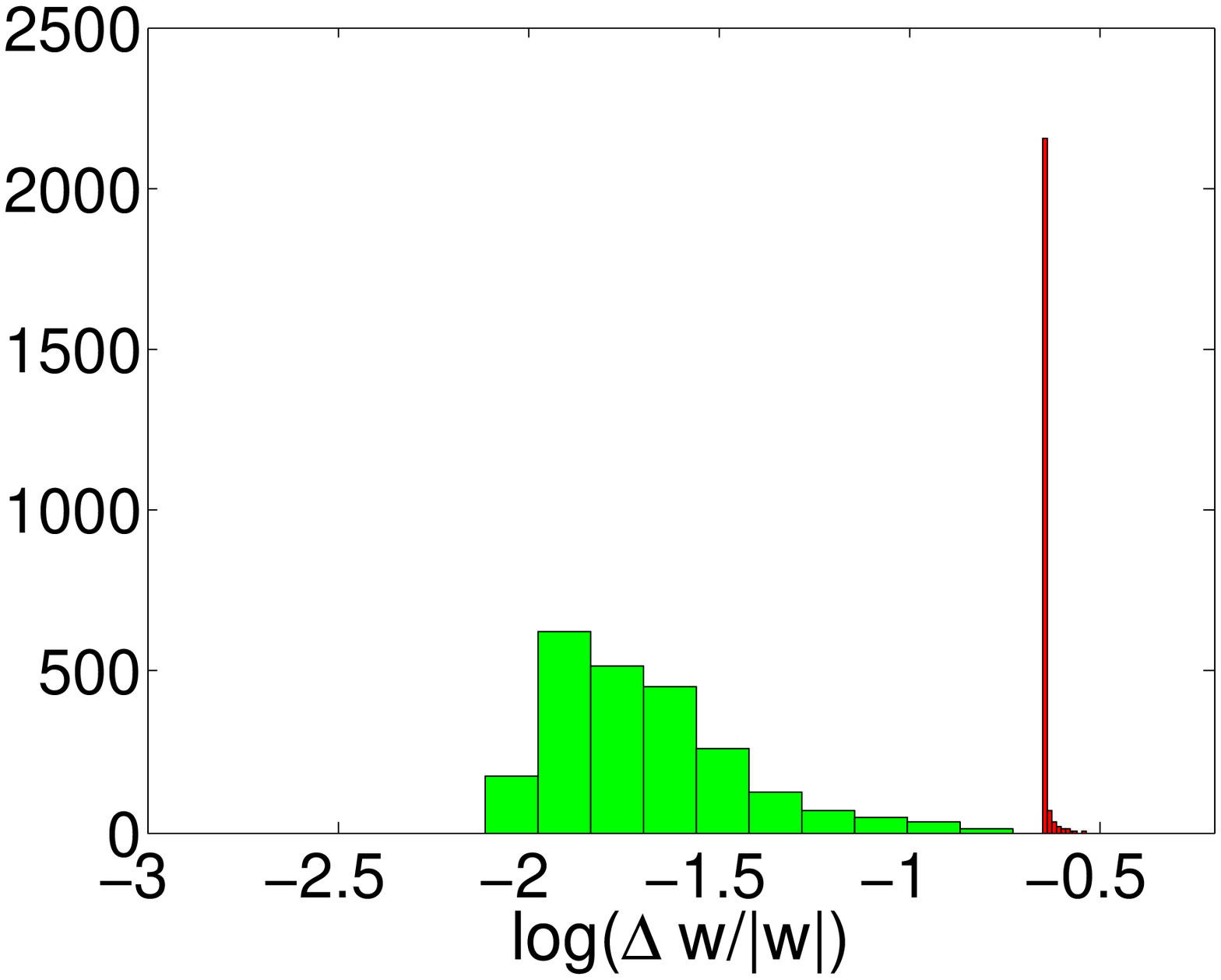} &
\includegraphics[width = 3.5 cm]{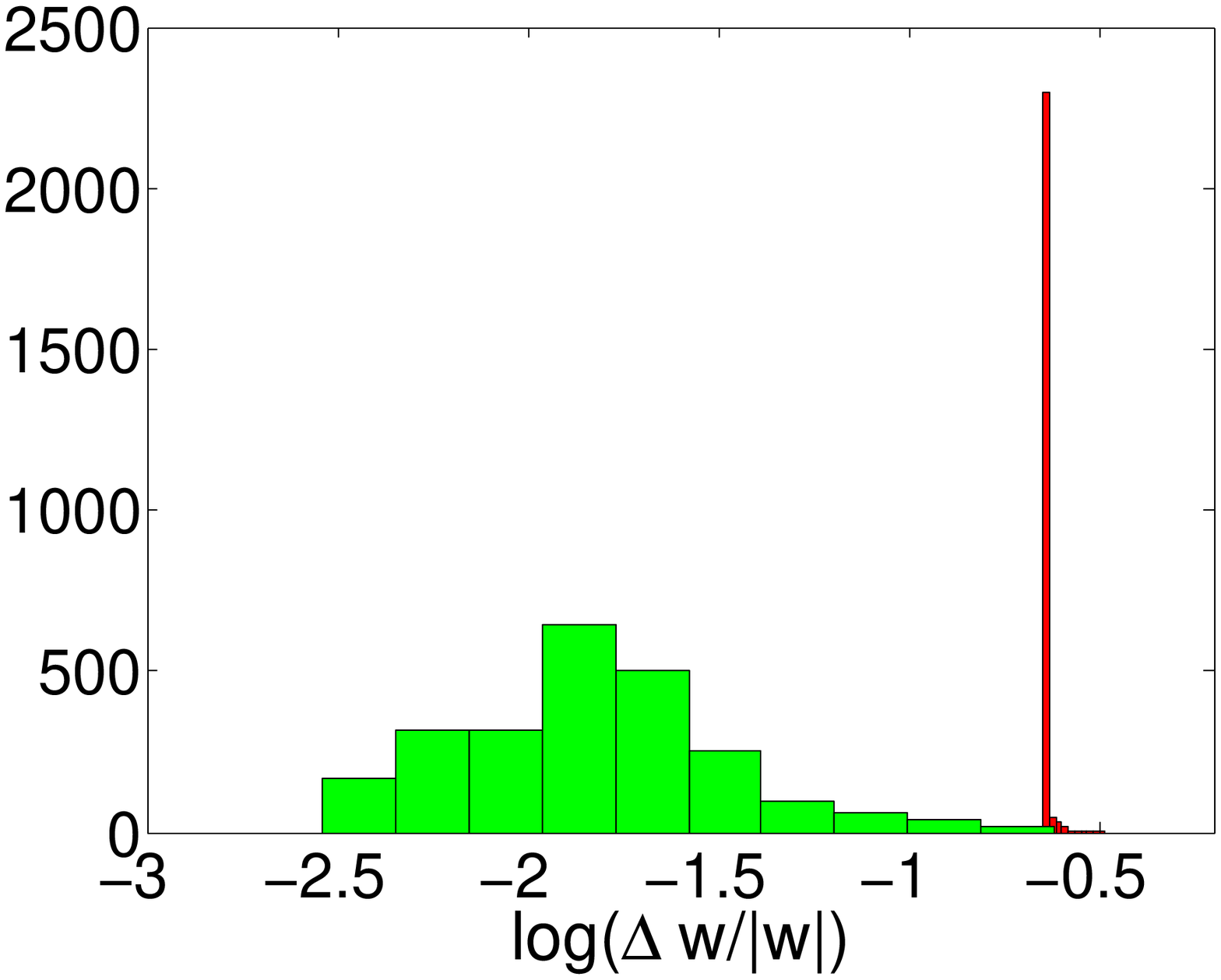}
\\
 & &
\includegraphics[width = 3.5 cm]{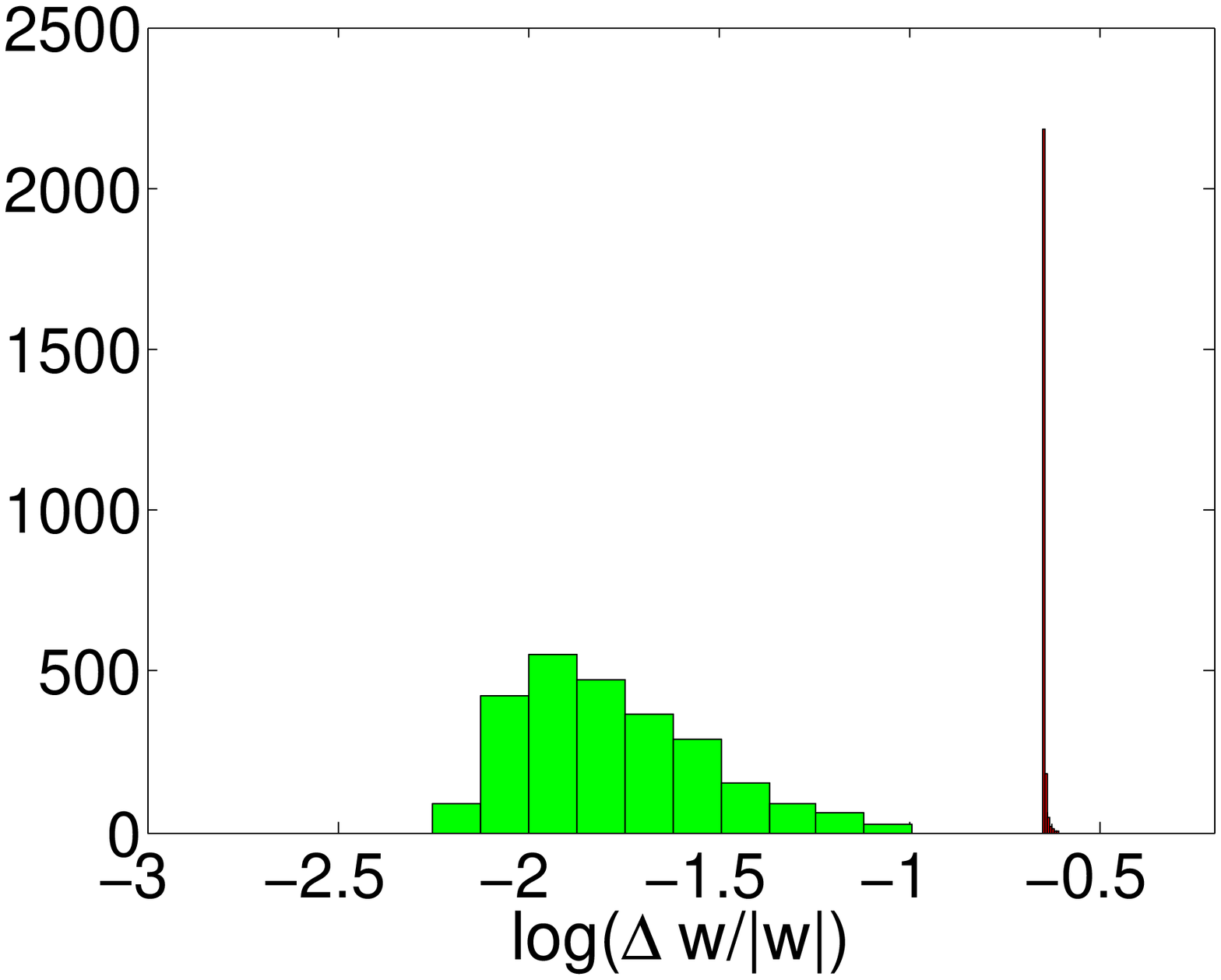} &
\includegraphics[width = 3.5 cm]{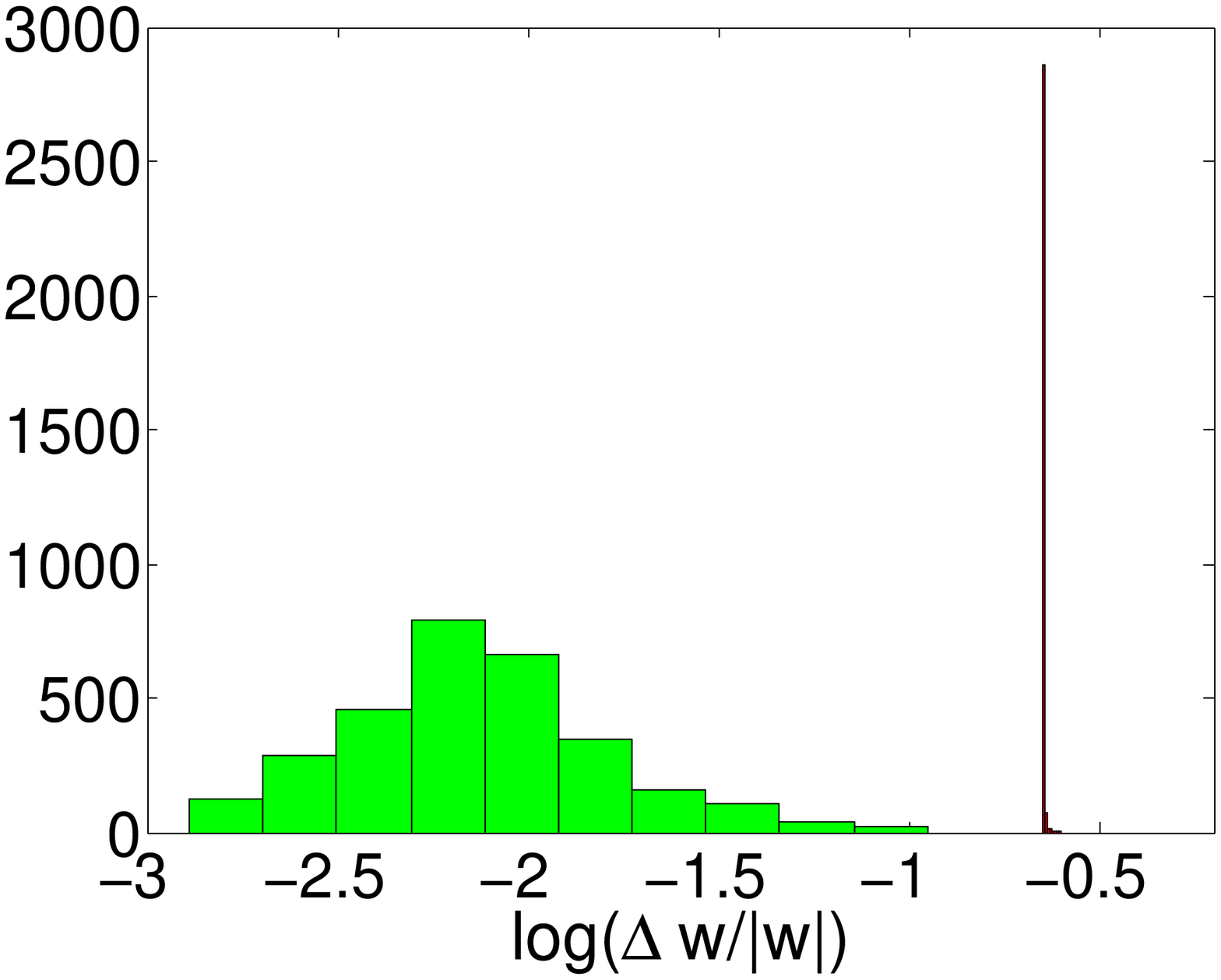} &
\includegraphics[width = 3.5 cm]{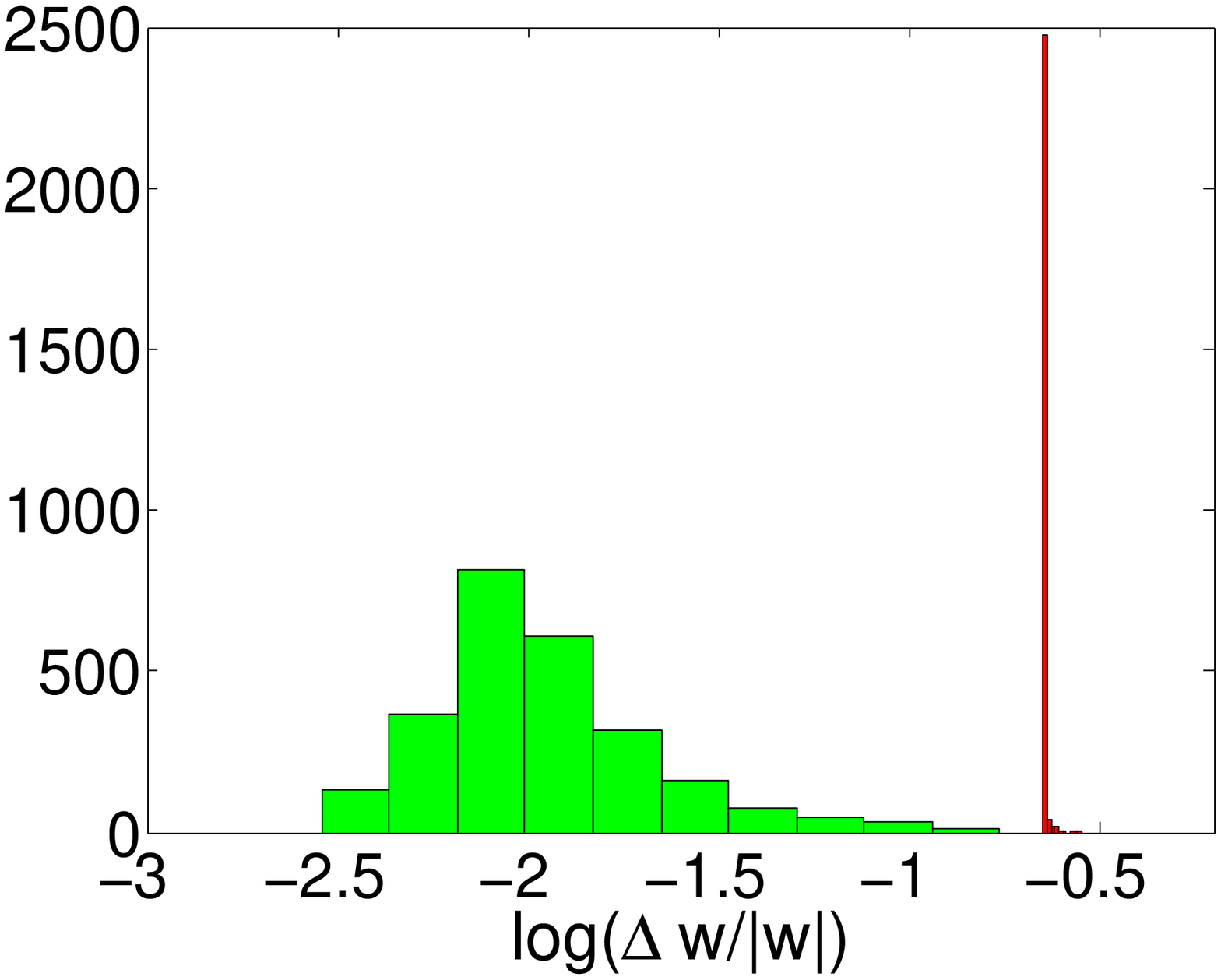} 
\\
 & 
\includegraphics[width = 3.5 cm]{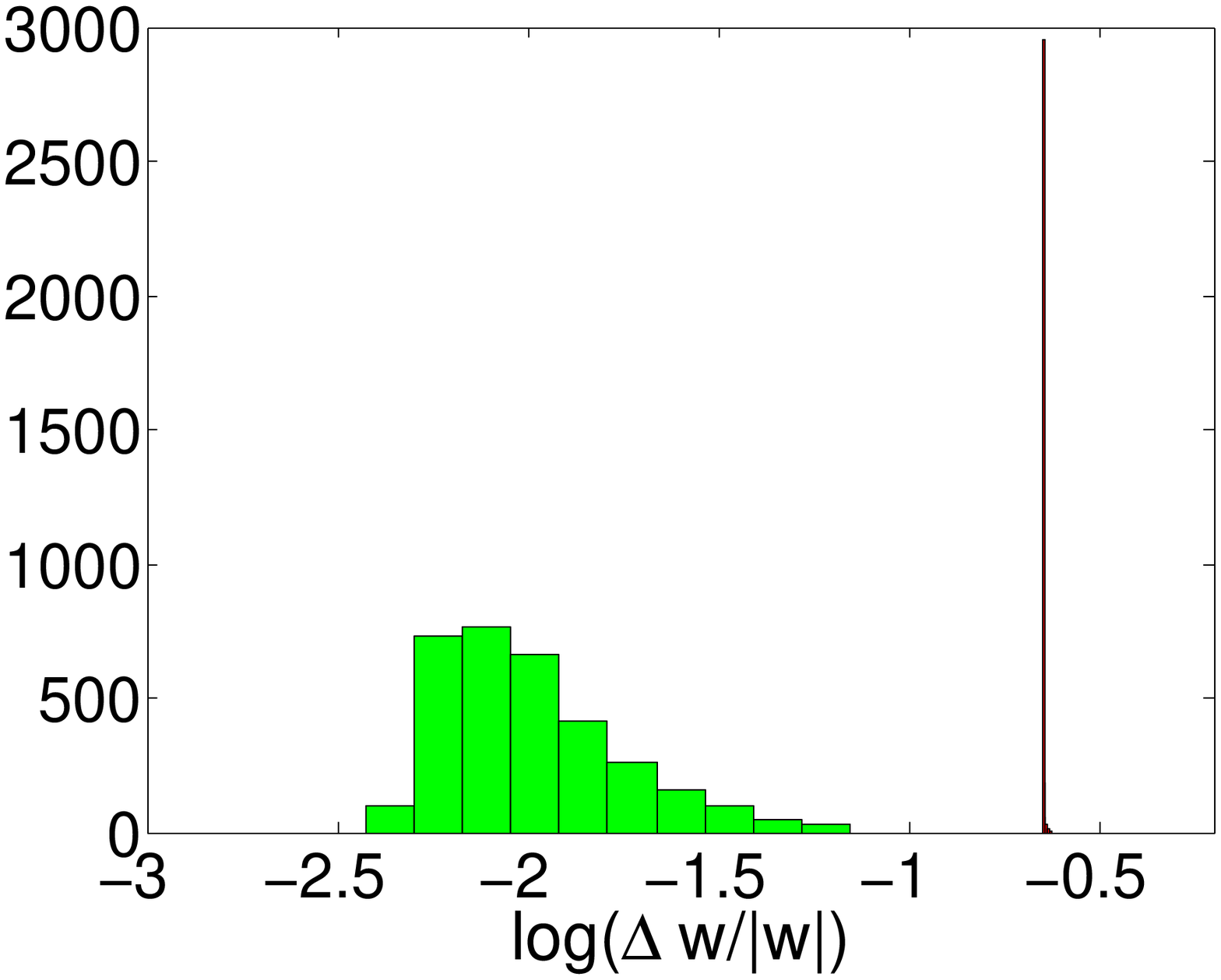} &
\includegraphics[width = 3.5 cm]{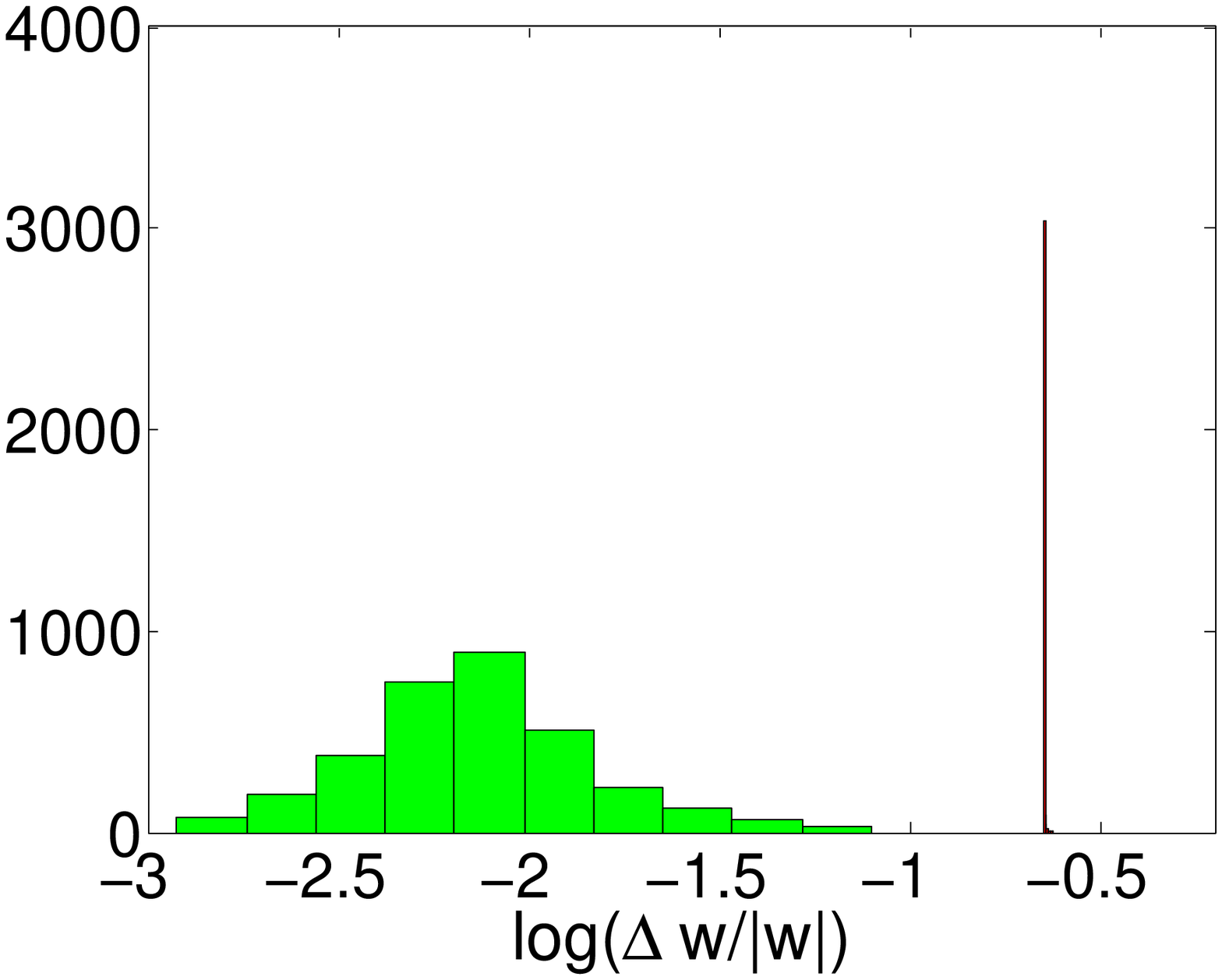} &
\includegraphics[width = 3.5 cm]{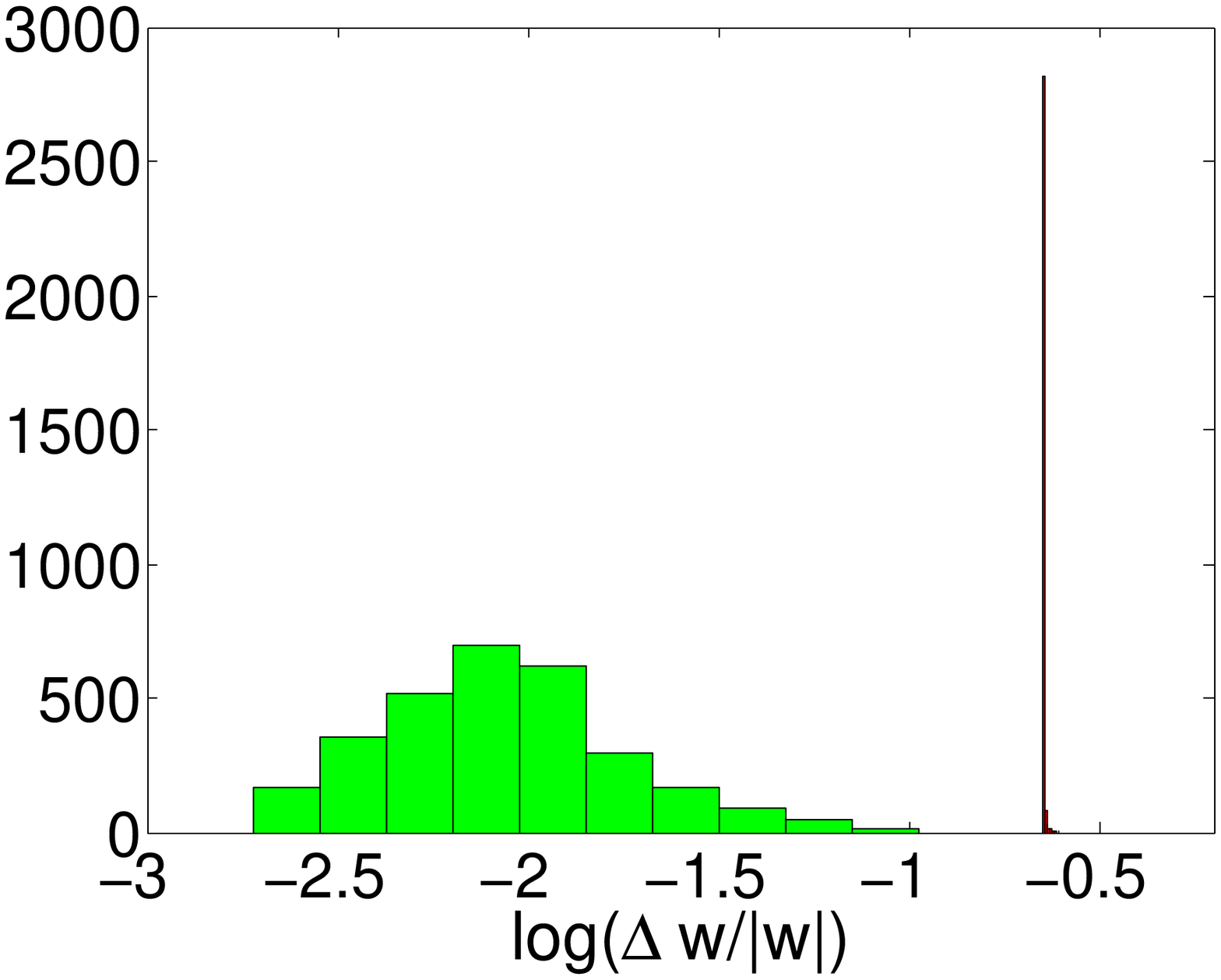} &
\includegraphics[width = 3.5 cm]{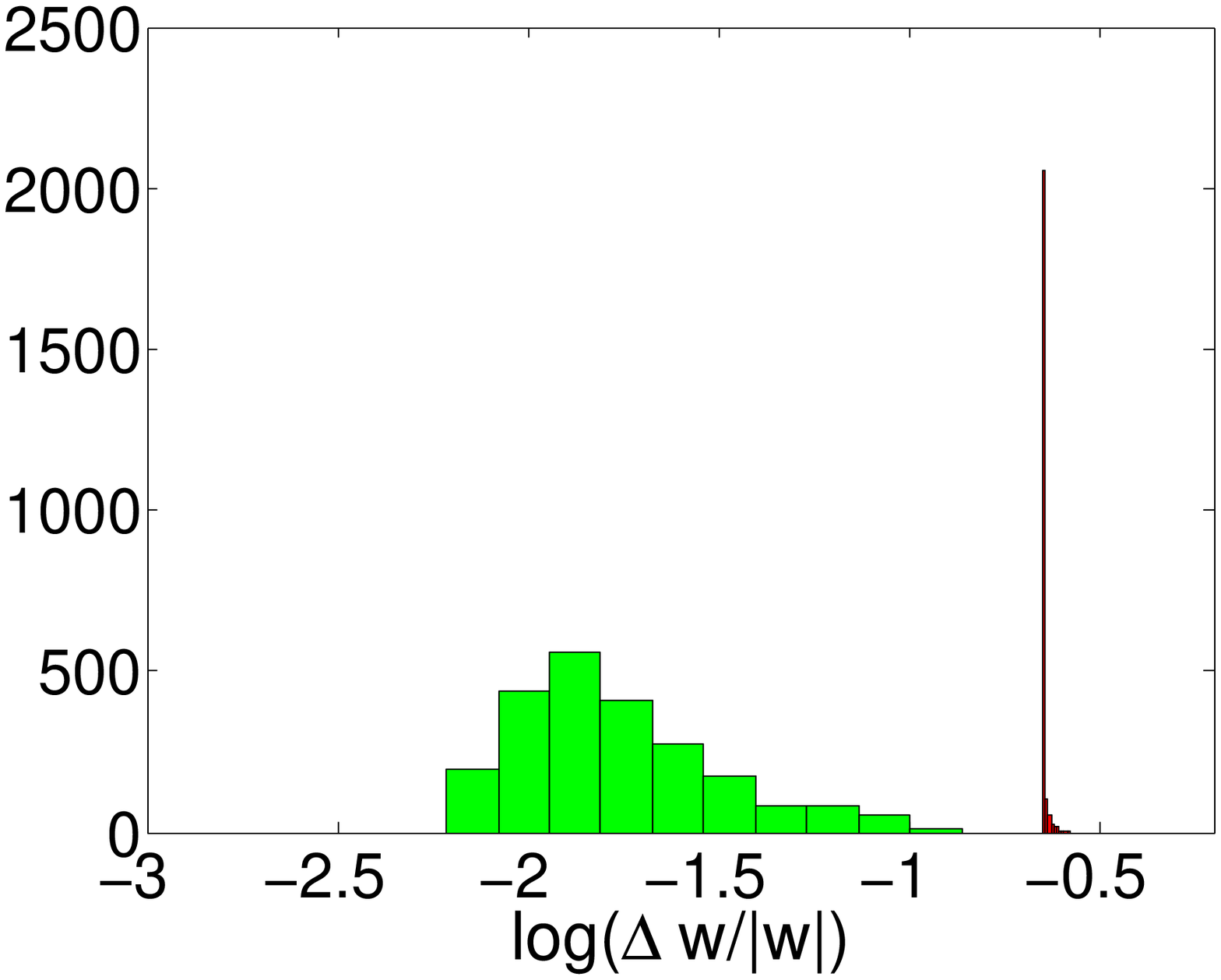}
\\
\includegraphics[width = 3.5 cm]{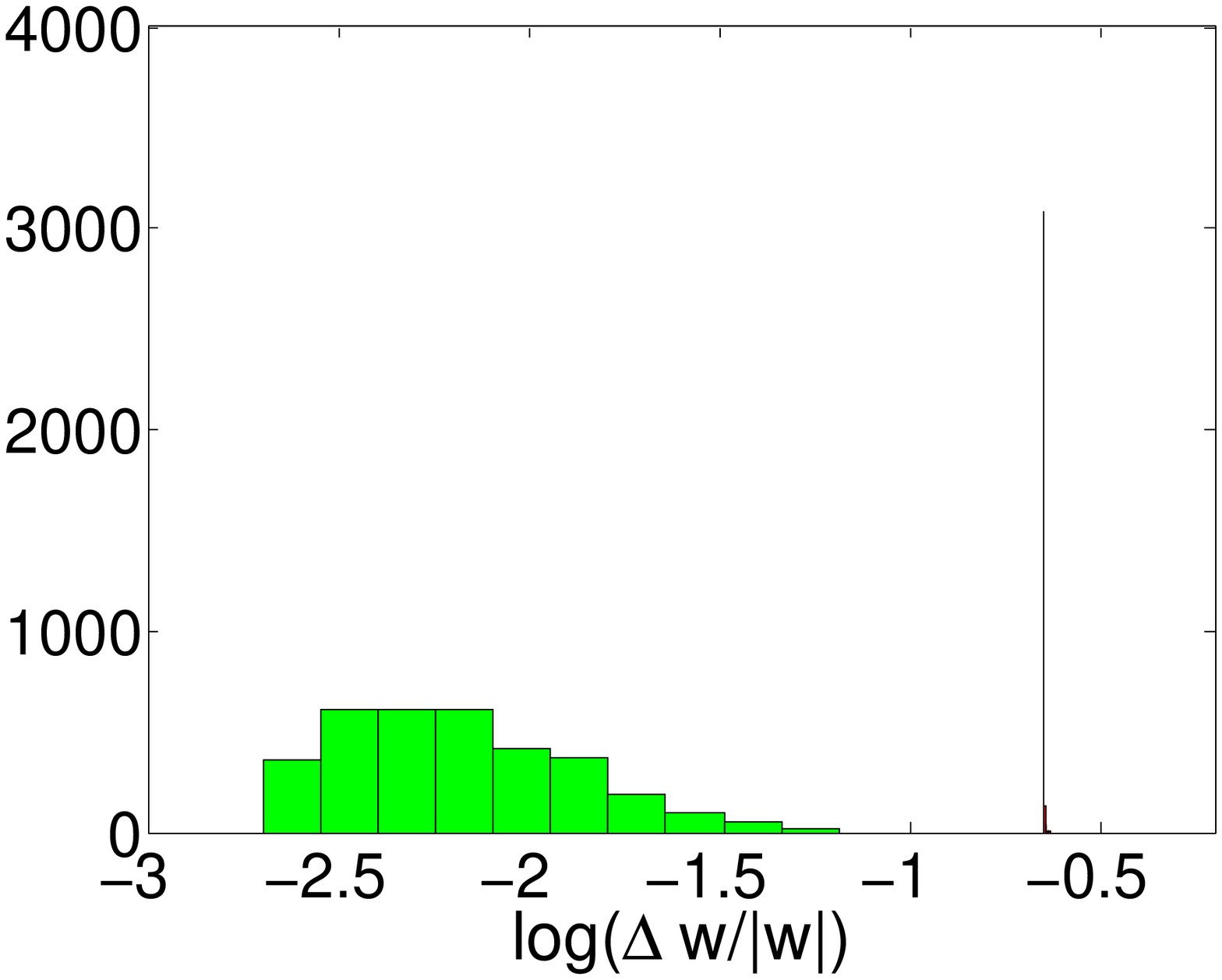} &
\includegraphics[width = 3.5 cm]{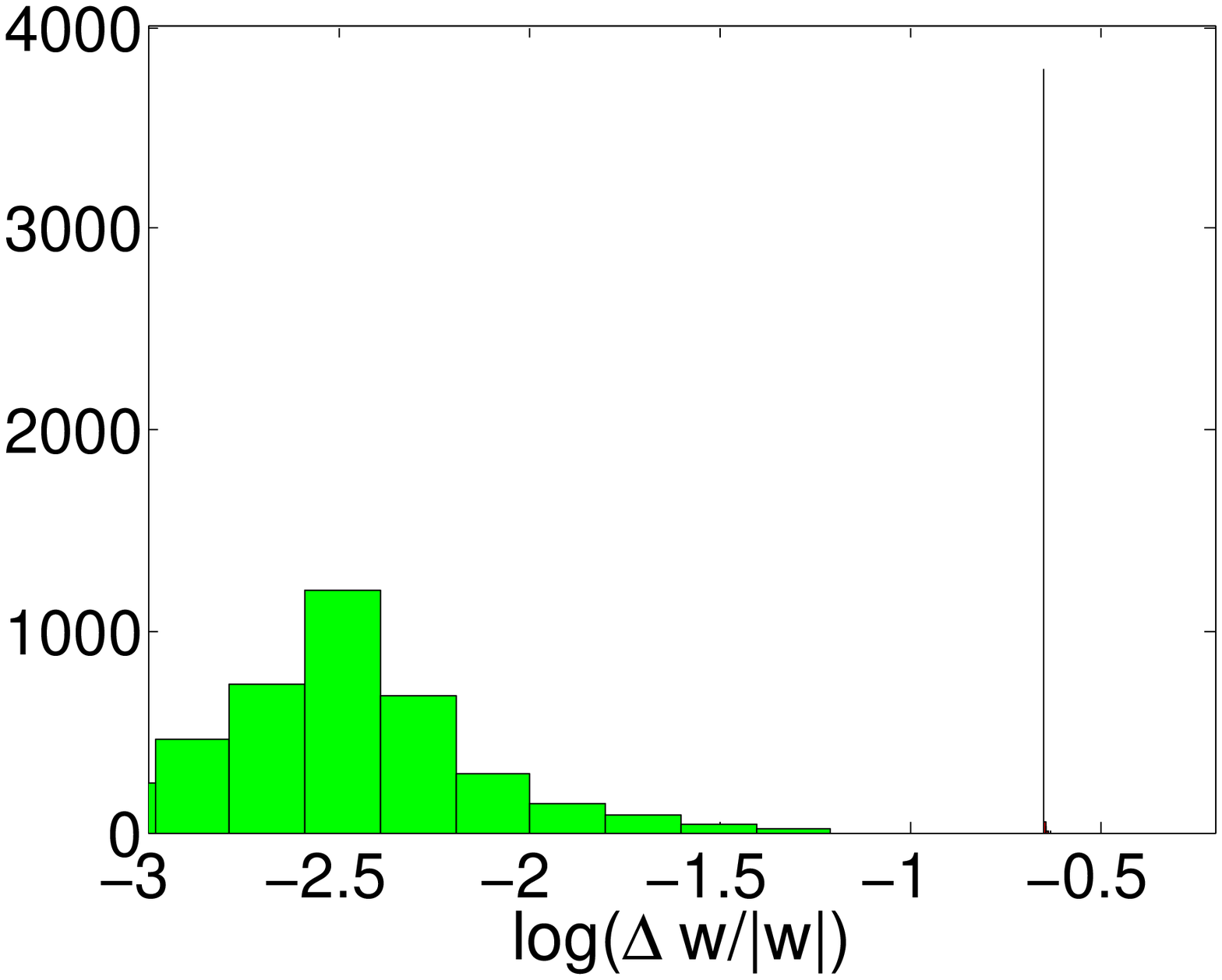} &
\includegraphics[width = 3.5 cm]{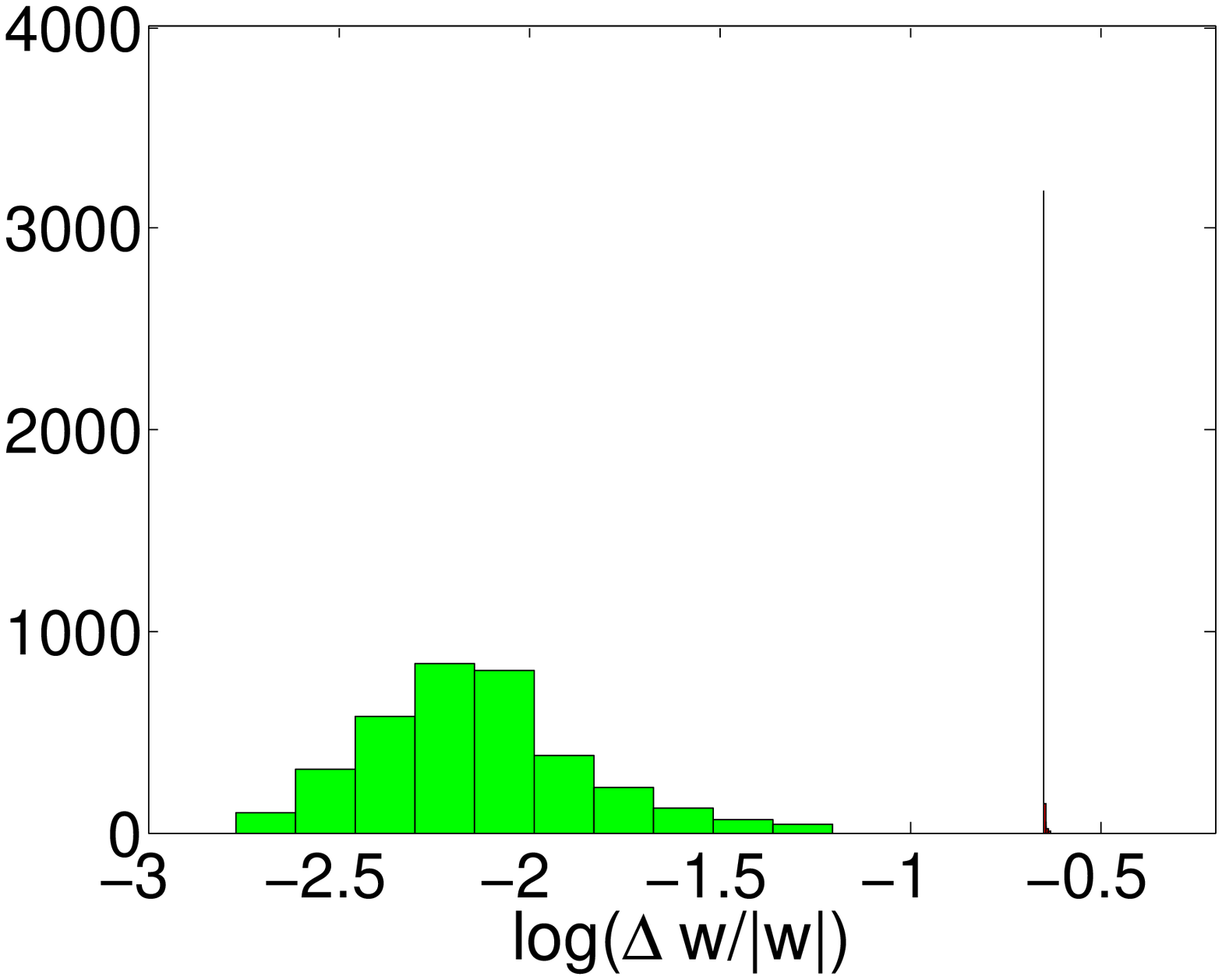} &
\includegraphics[width = 3.5 cm]{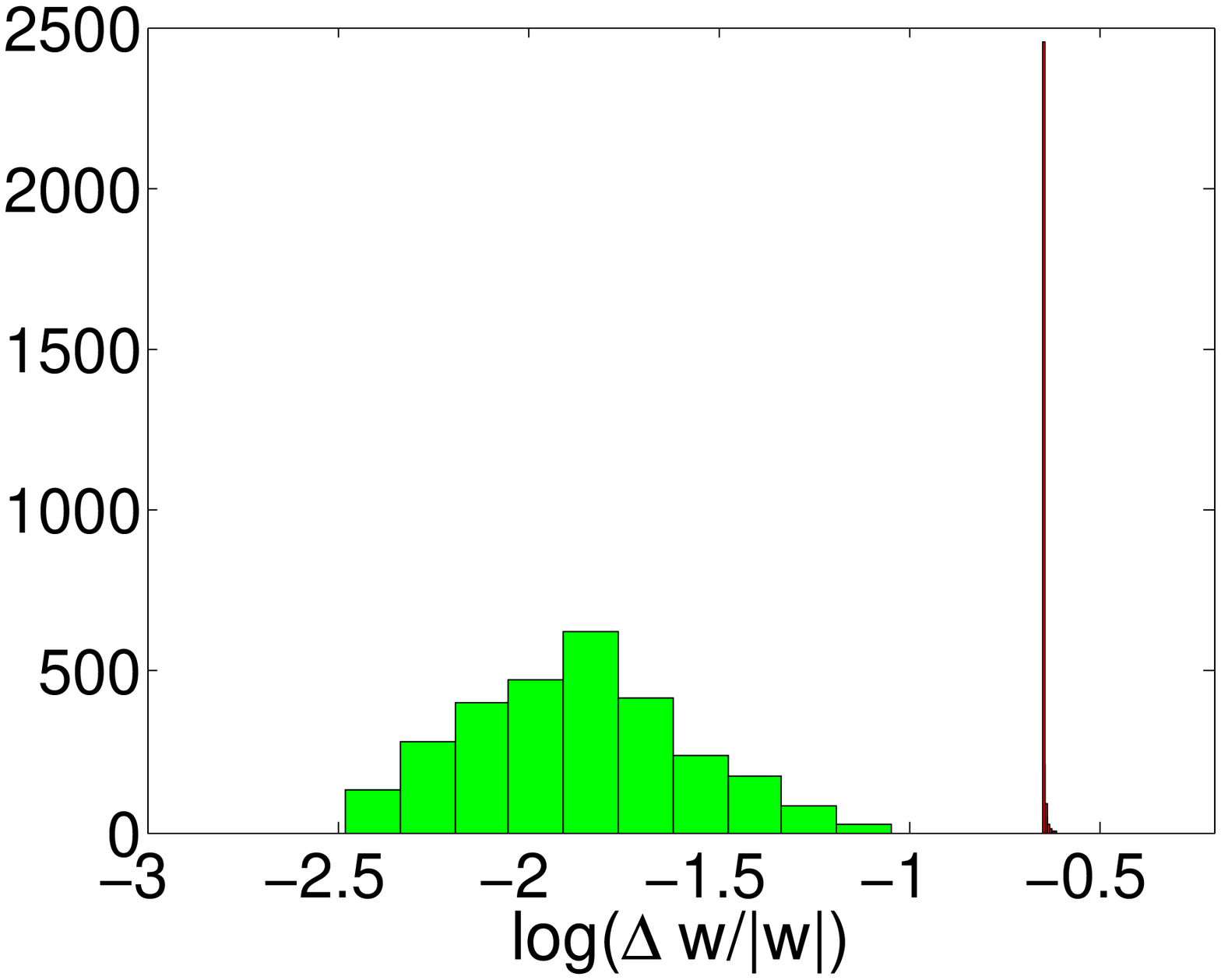} &
\includegraphics[width = 3.5 cm]{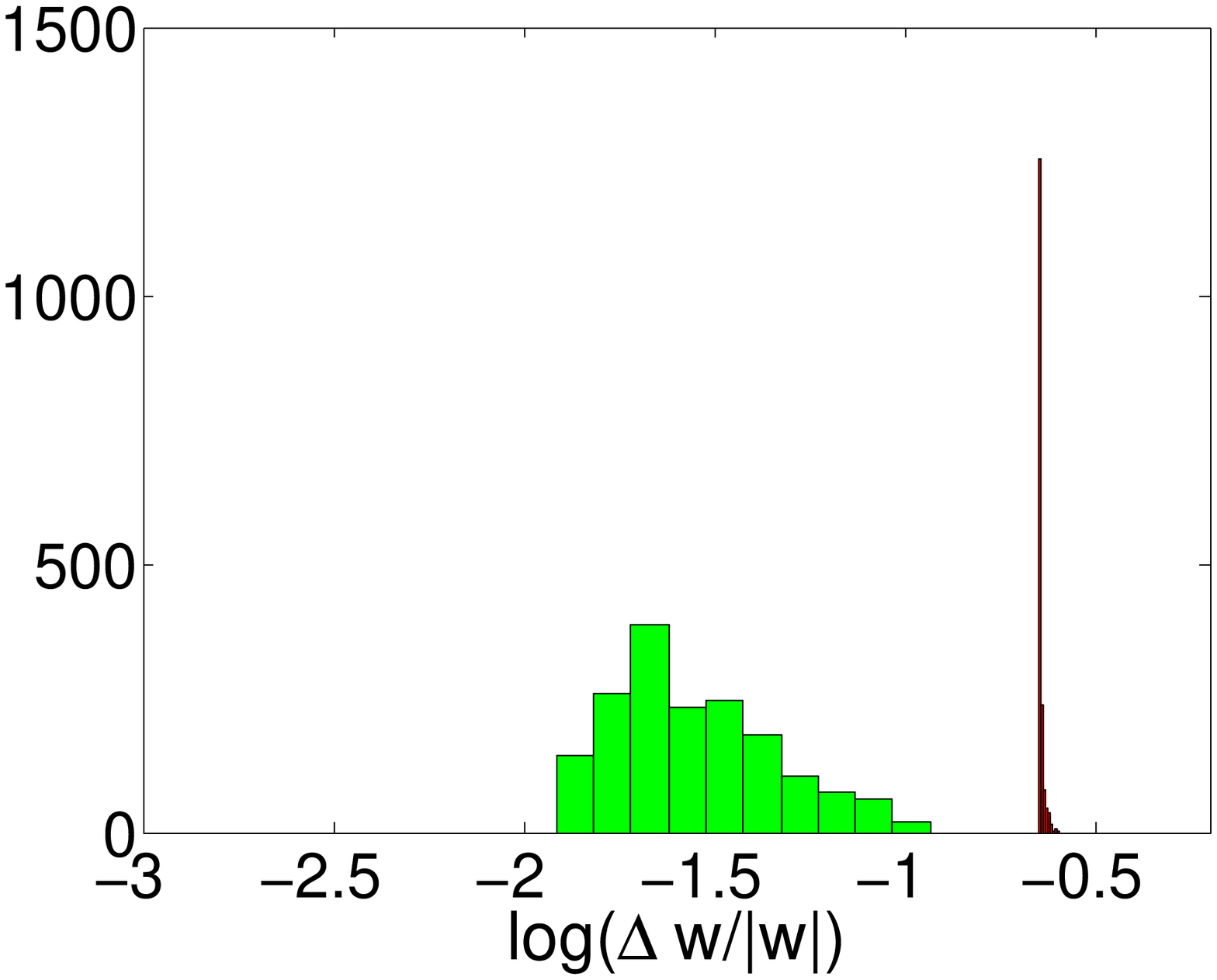}
\end{tabular}
\caption{Accuracies on the estimation of $w$ for $z_0 = 0.55$ ($D_{\rm L} = 3~\Gpc$,
according to our fiducial model). The way the plots are 
arranged corresponds to the location of systems in the $\log(m_1) - \log(m_2)$
plane as in Fig.~\ref{Fig.frac_median}. The light (green) distributions are without 
weak lensing; in the dark (red) distributions a 4\% error in distance 
estimation has been folded in, as a heuristic way to account for weak lensing. Clearly, 
if the effects of weak lensing are not removed then they will 
dominate the uncertainty on $w$.}
\label{Fig.cases}
\end{figure*} 

\begin{figure*}[t]
\begin{tabular}{ccccc}
 & & & &
\includegraphics[width = 3.5 cm]{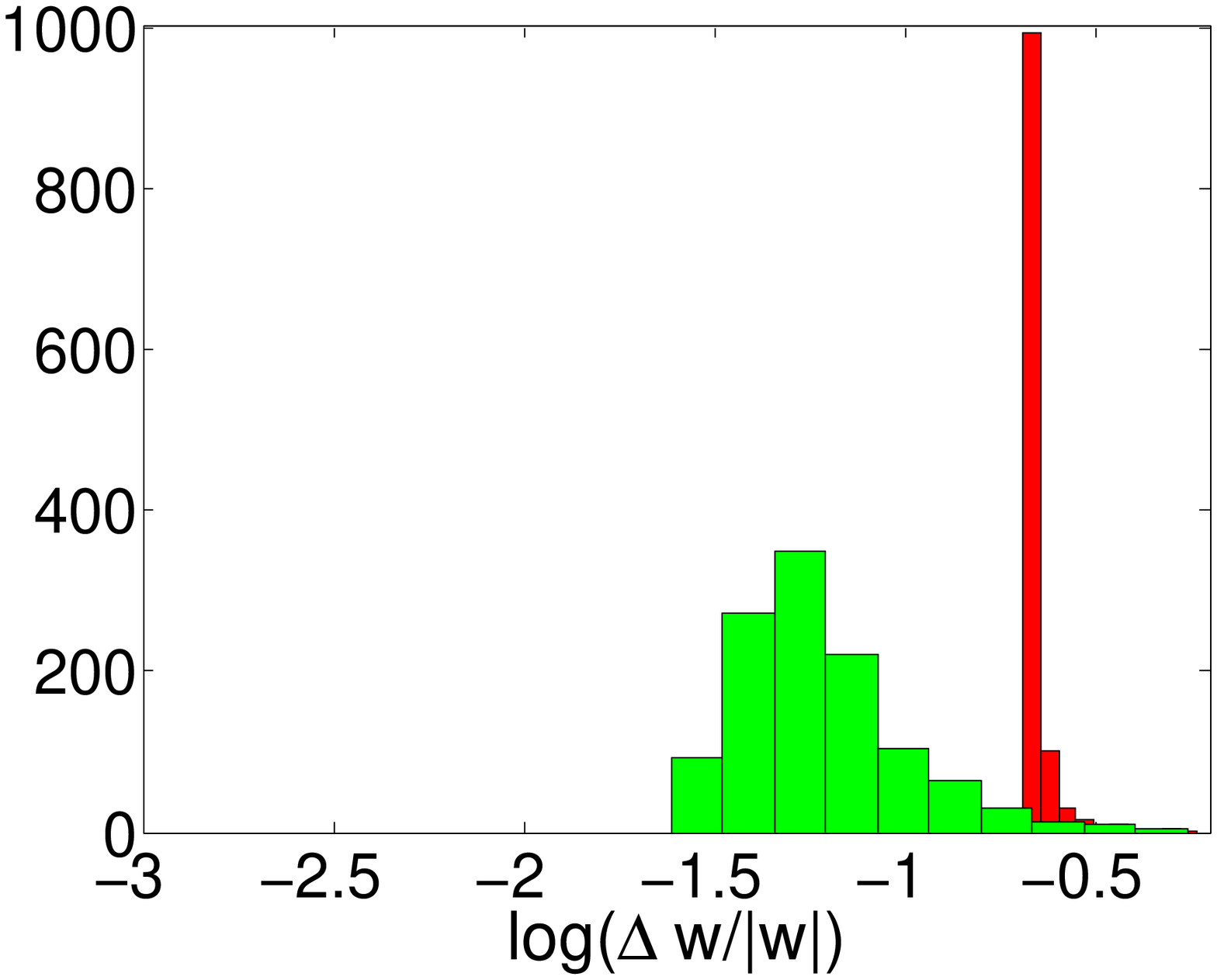}
\\
 & & &
\includegraphics[width = 3.5 cm]{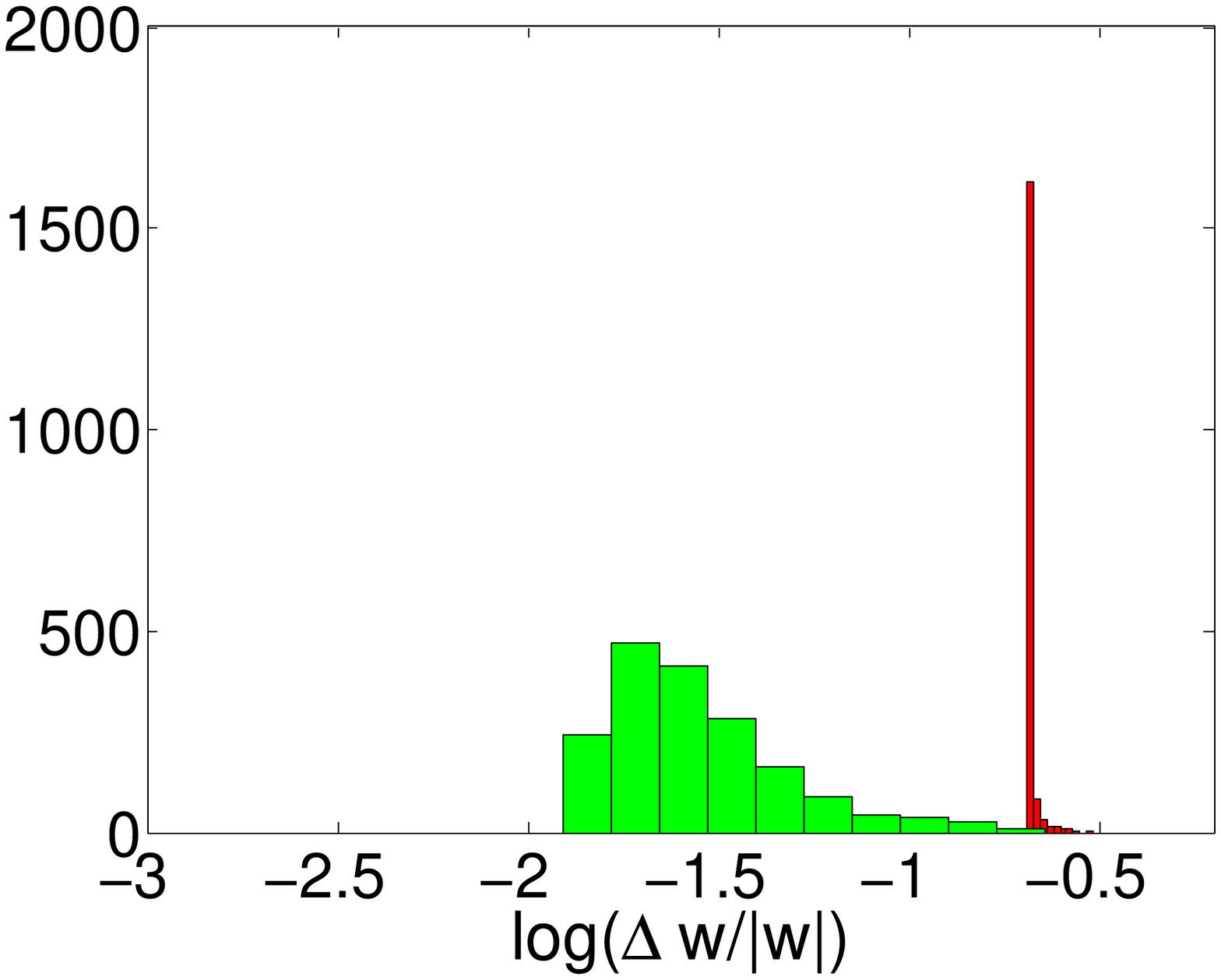} &
\includegraphics[width = 3.5 cm]{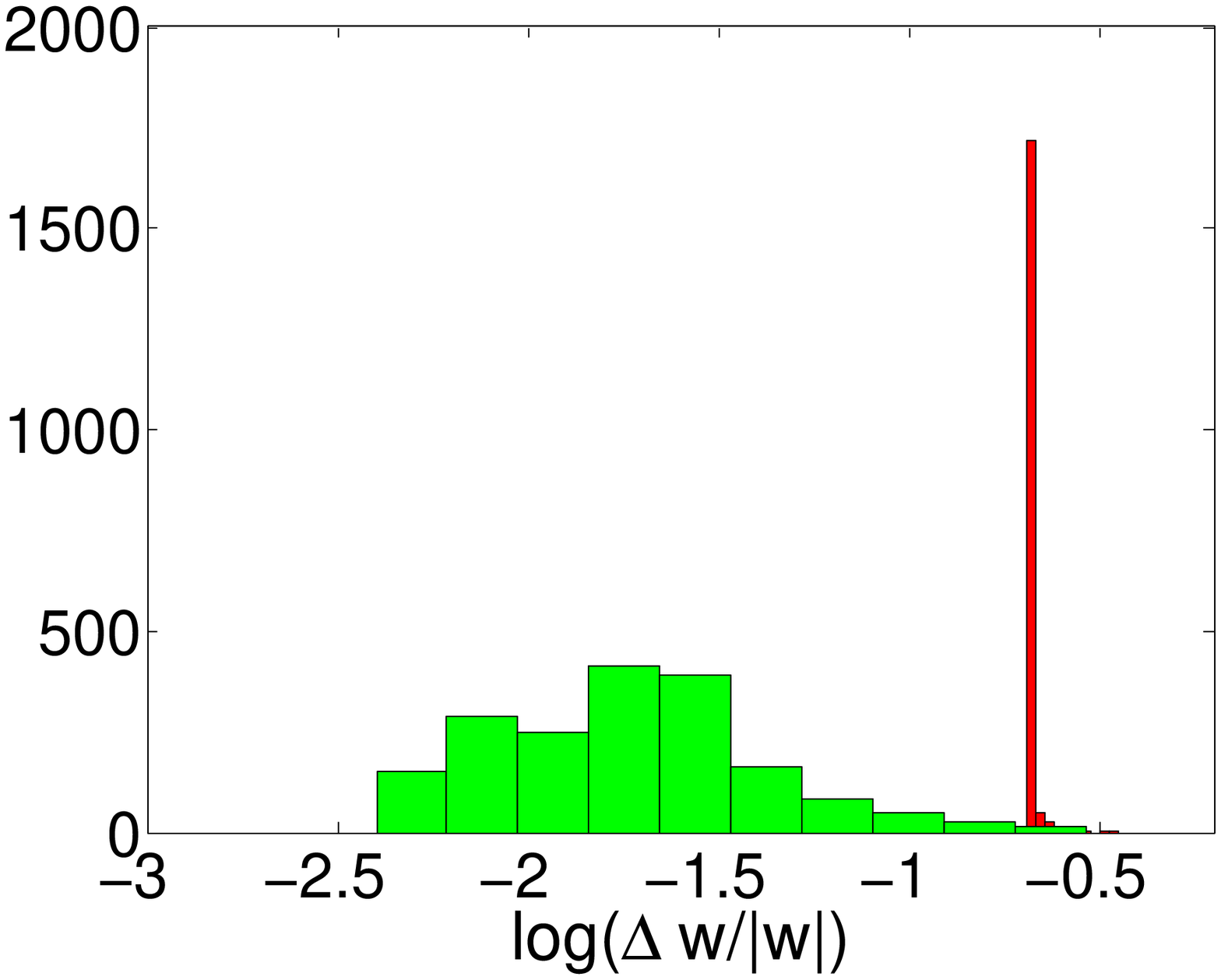}
\\
 & &
\includegraphics[width = 3.5 cm]{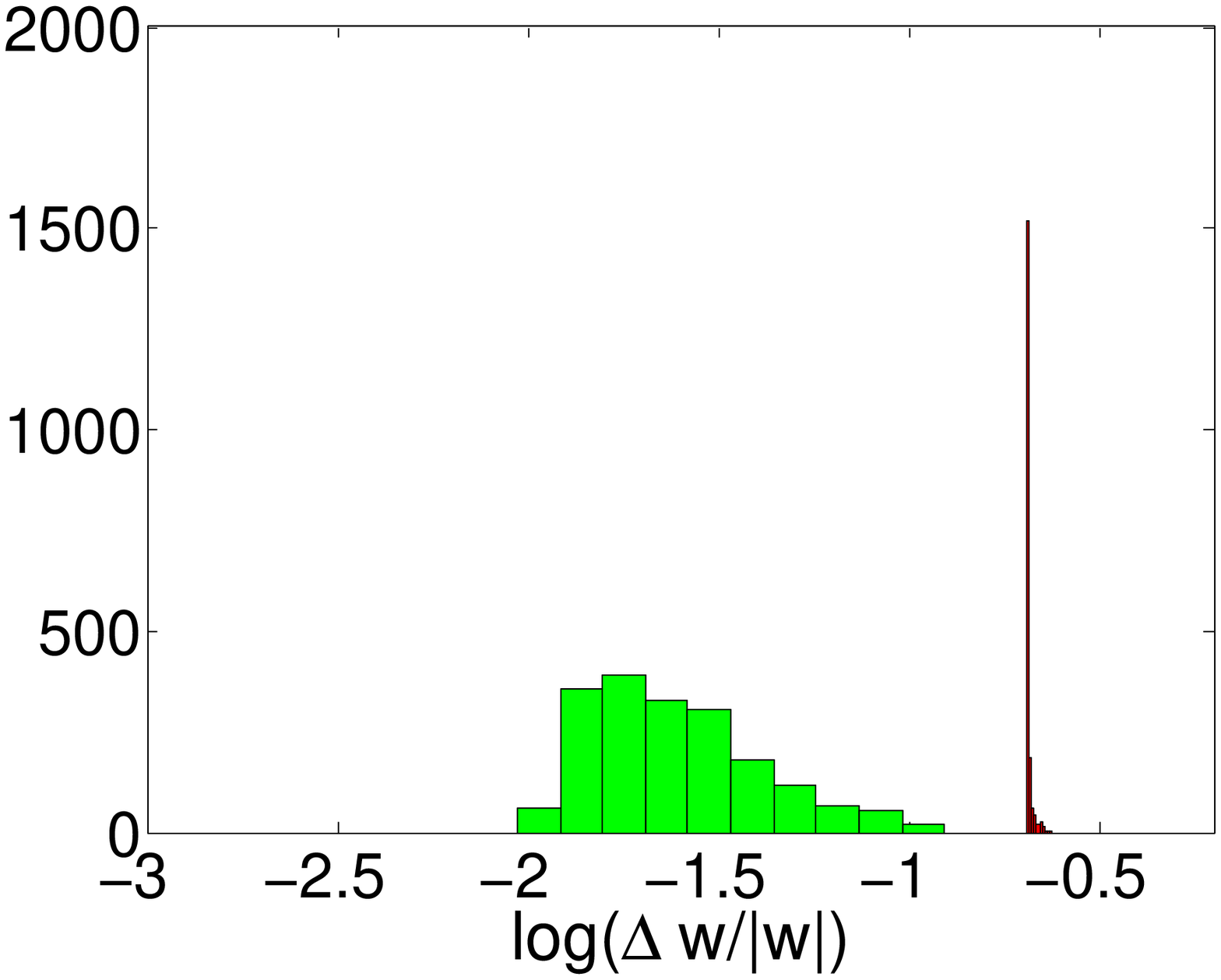} &
\includegraphics[width = 3.5 cm]{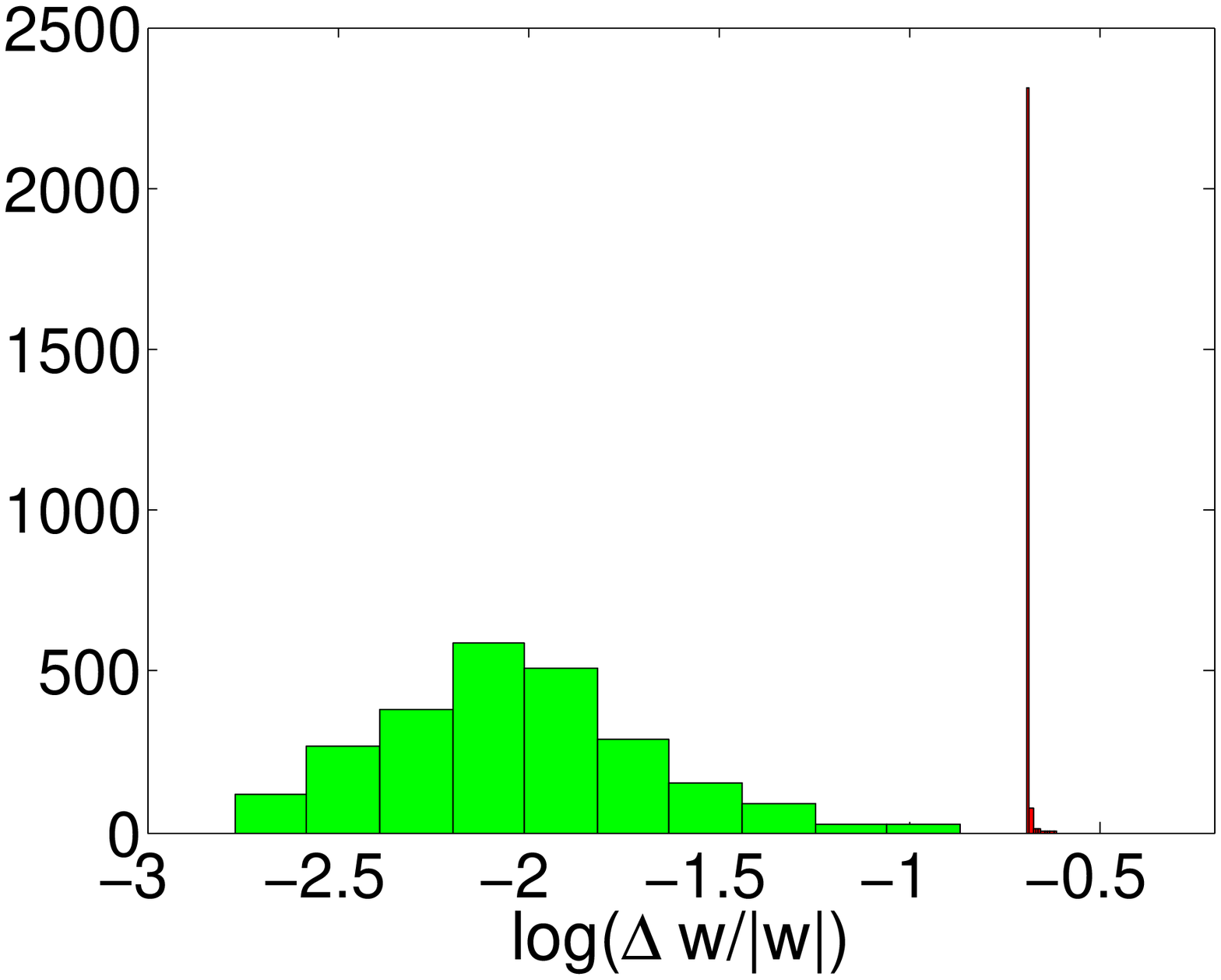} &
\includegraphics[width = 3.5 cm]{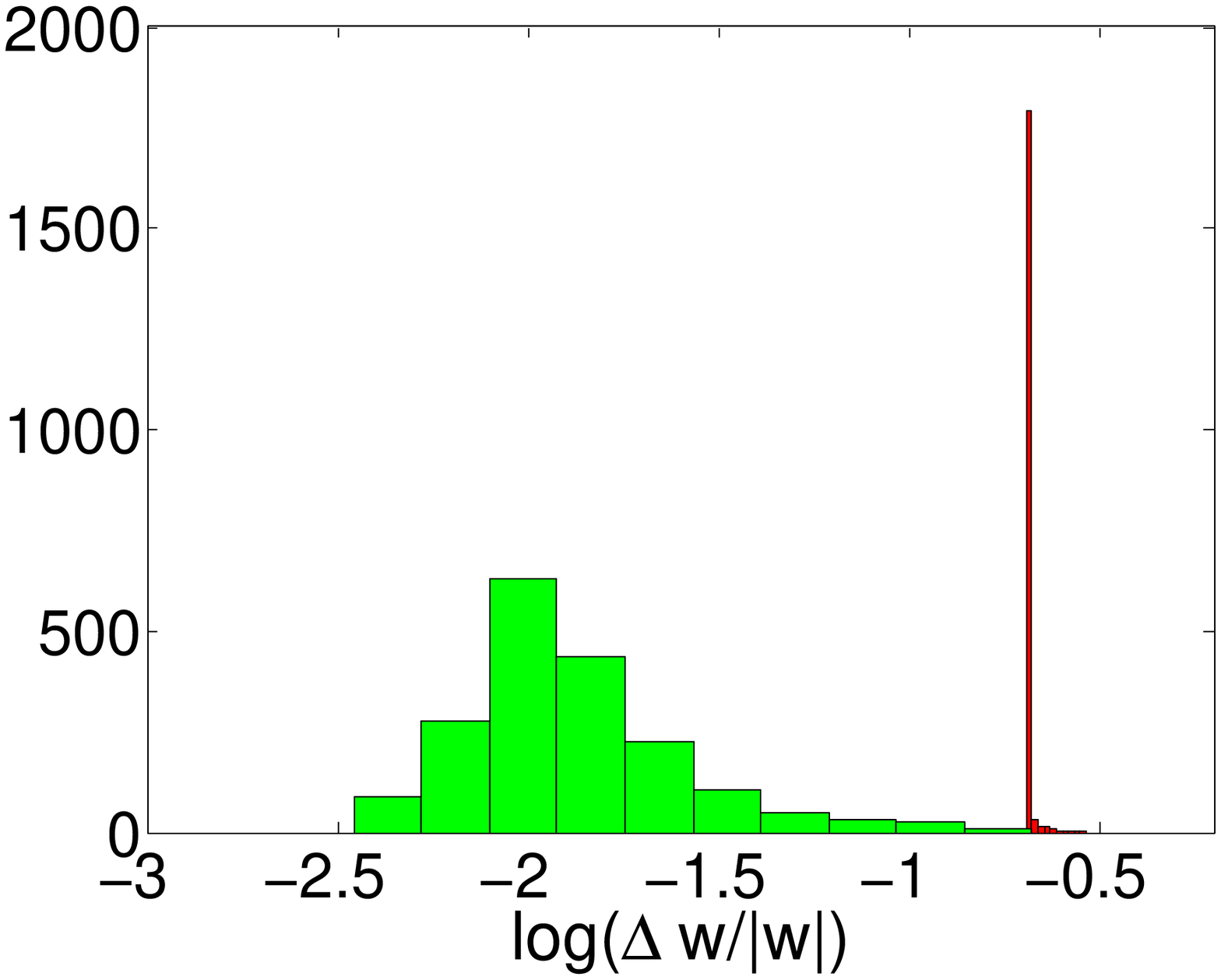}
\\
 &
\includegraphics[width = 3.5 cm]{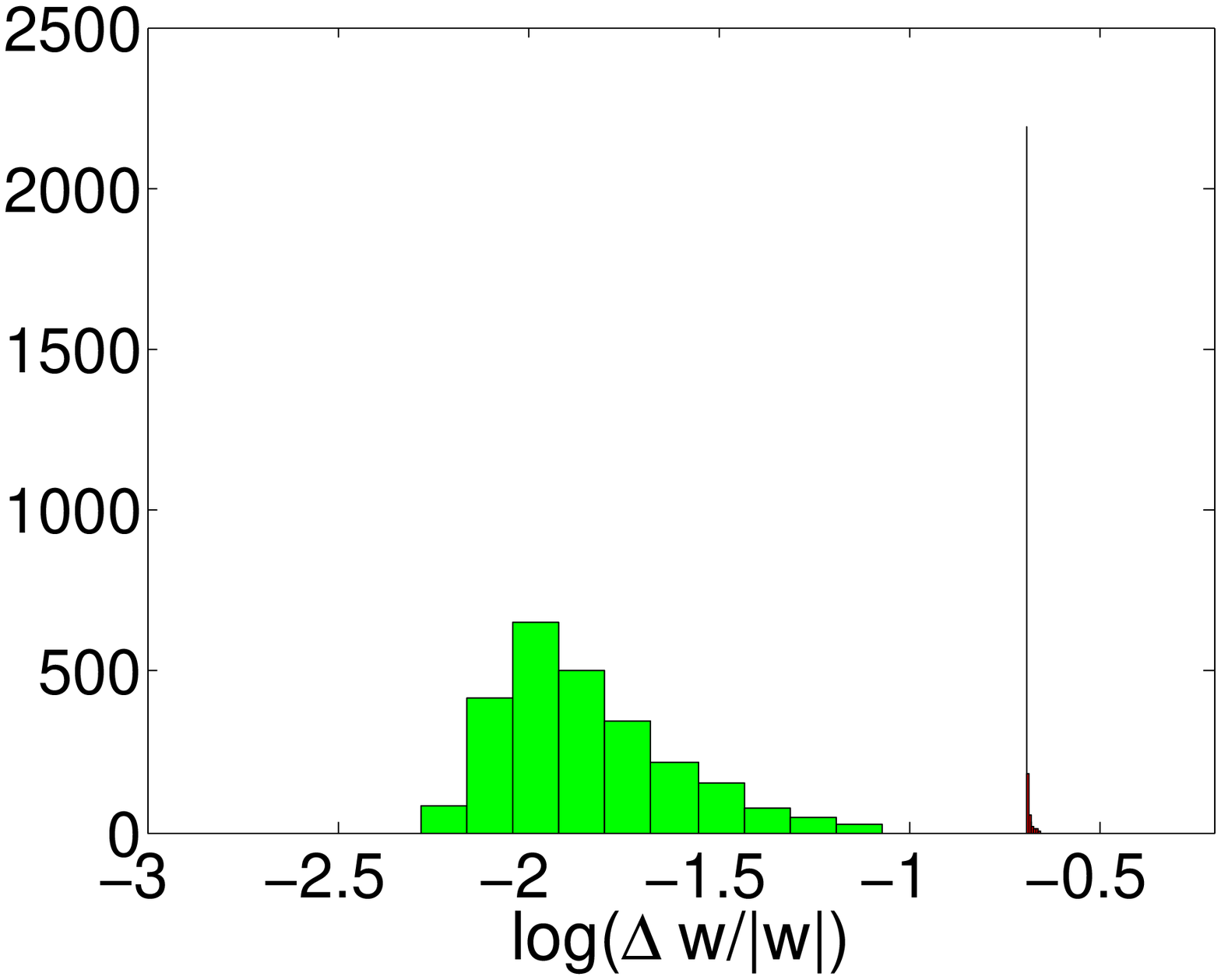} &
\includegraphics[width = 3.5 cm]{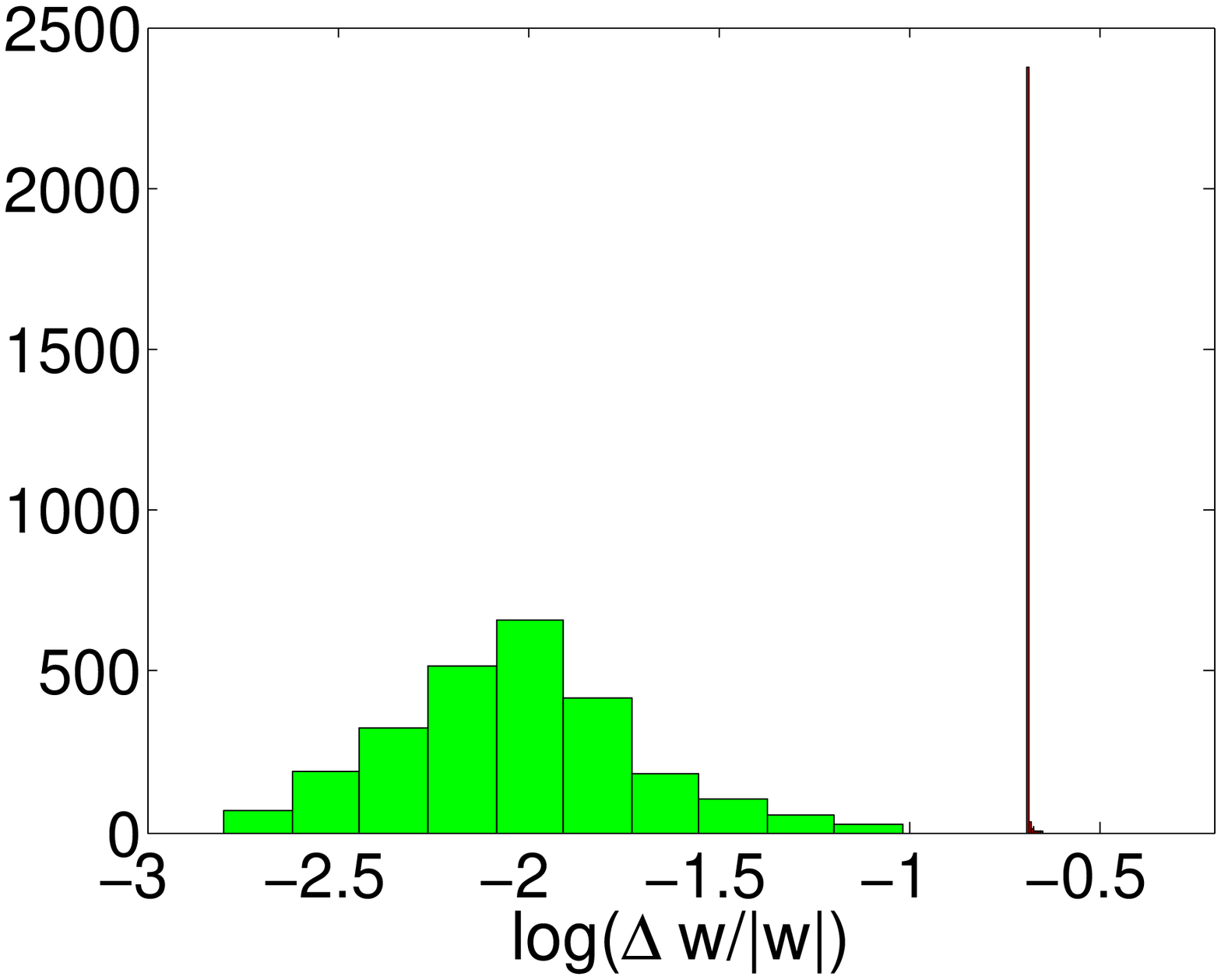} &
\includegraphics[width = 3.5 cm]{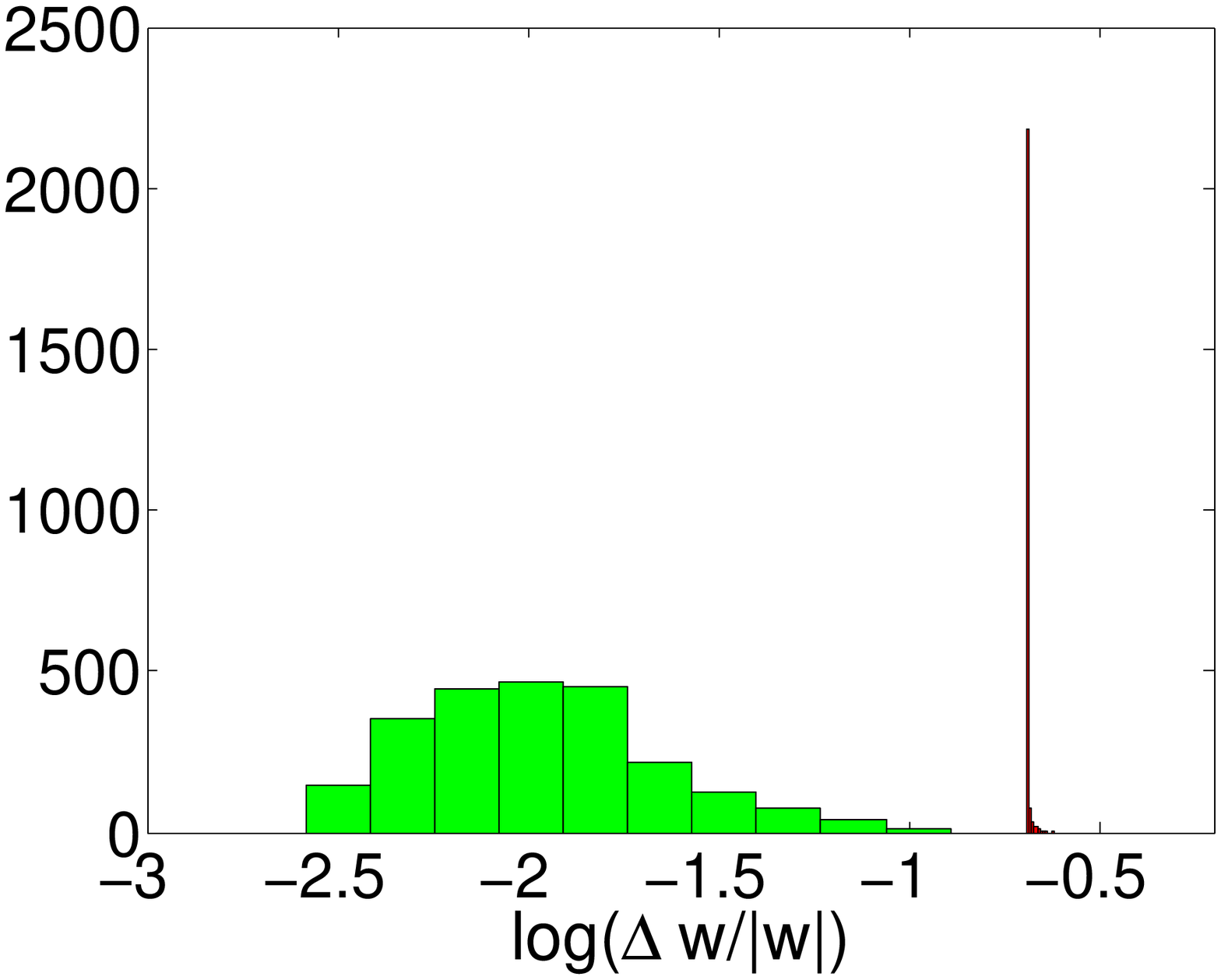} &
\includegraphics[width = 3.5 cm]{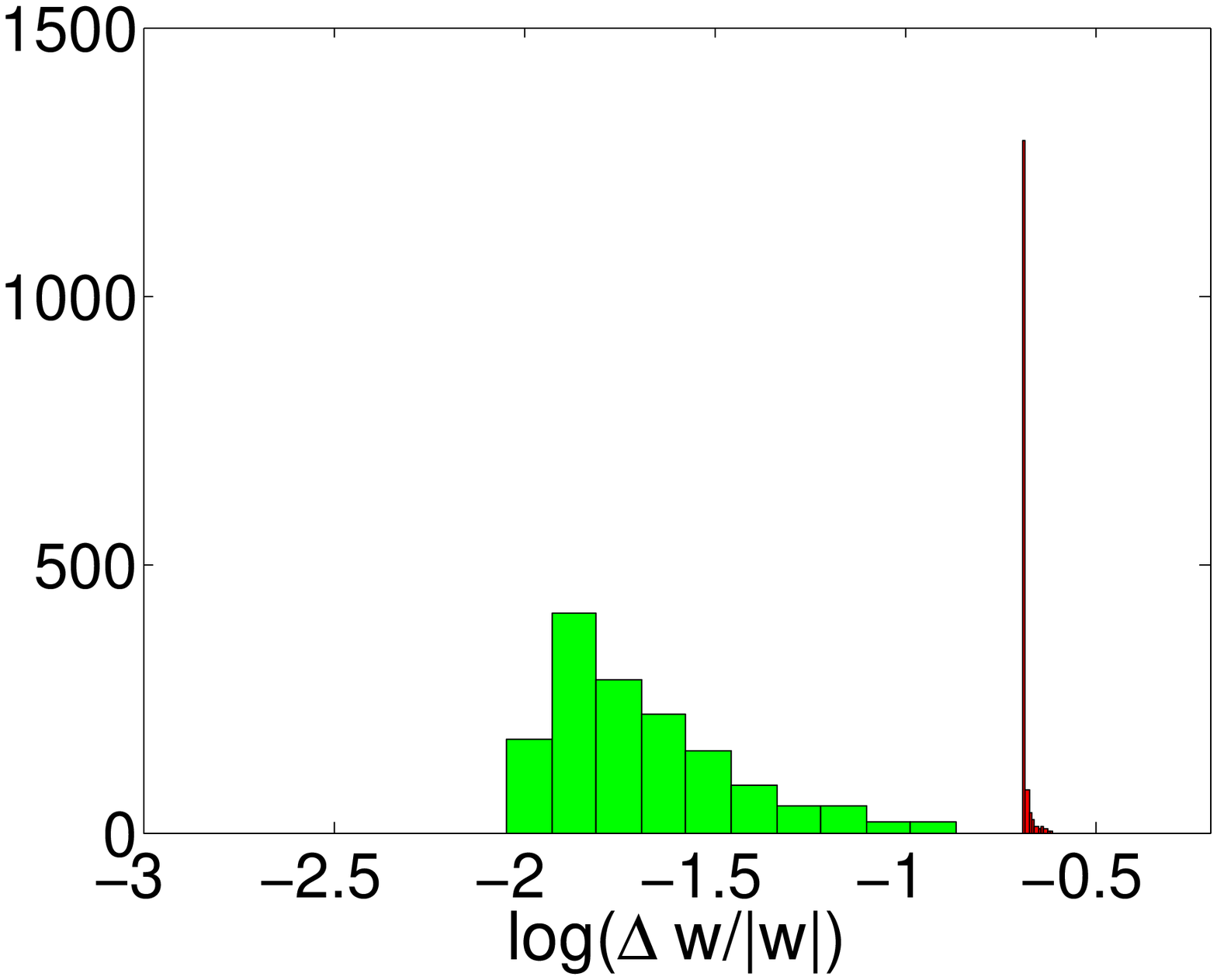}
\\
\includegraphics[width = 3.5 cm]{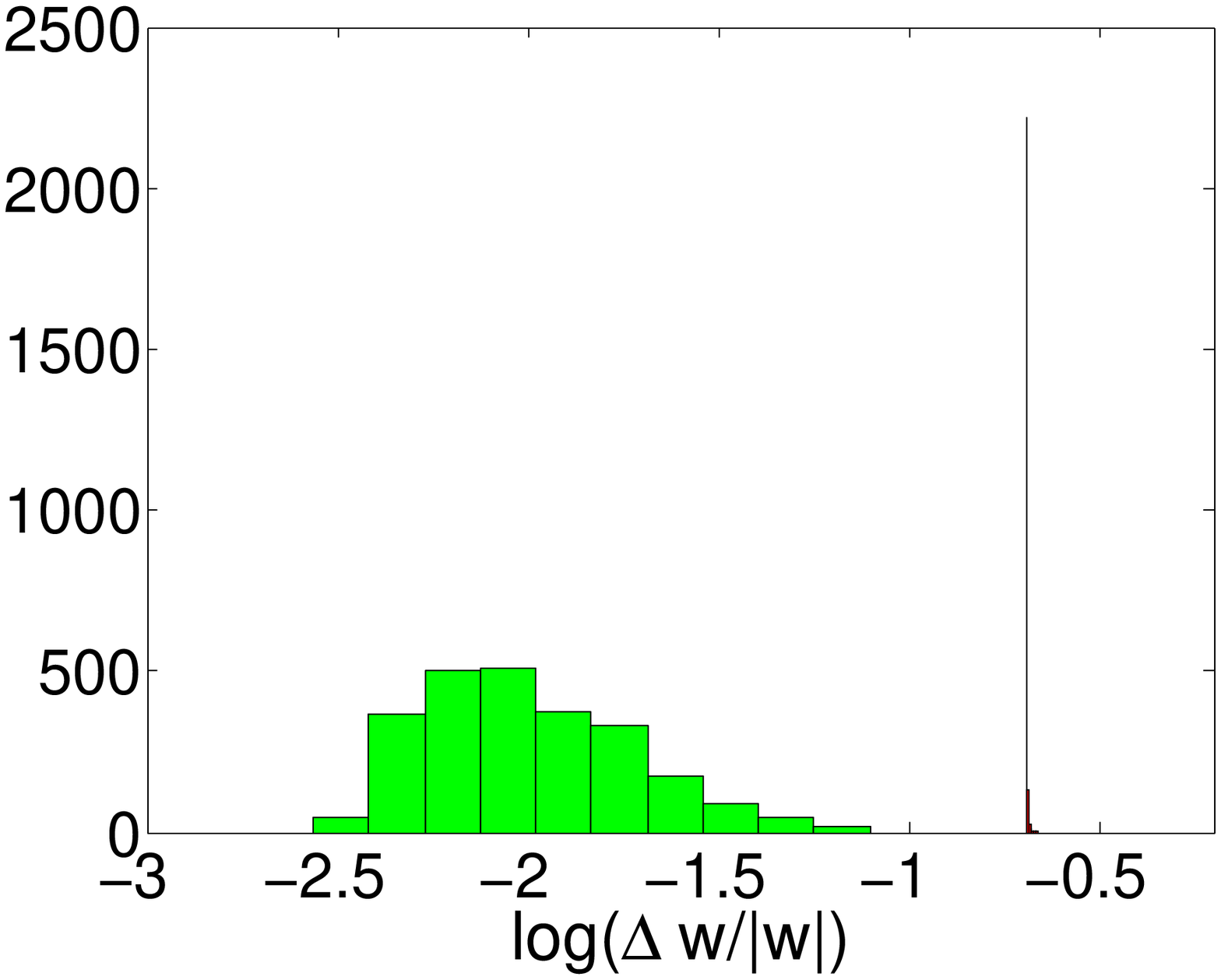} &
\includegraphics[width = 3.5 cm]{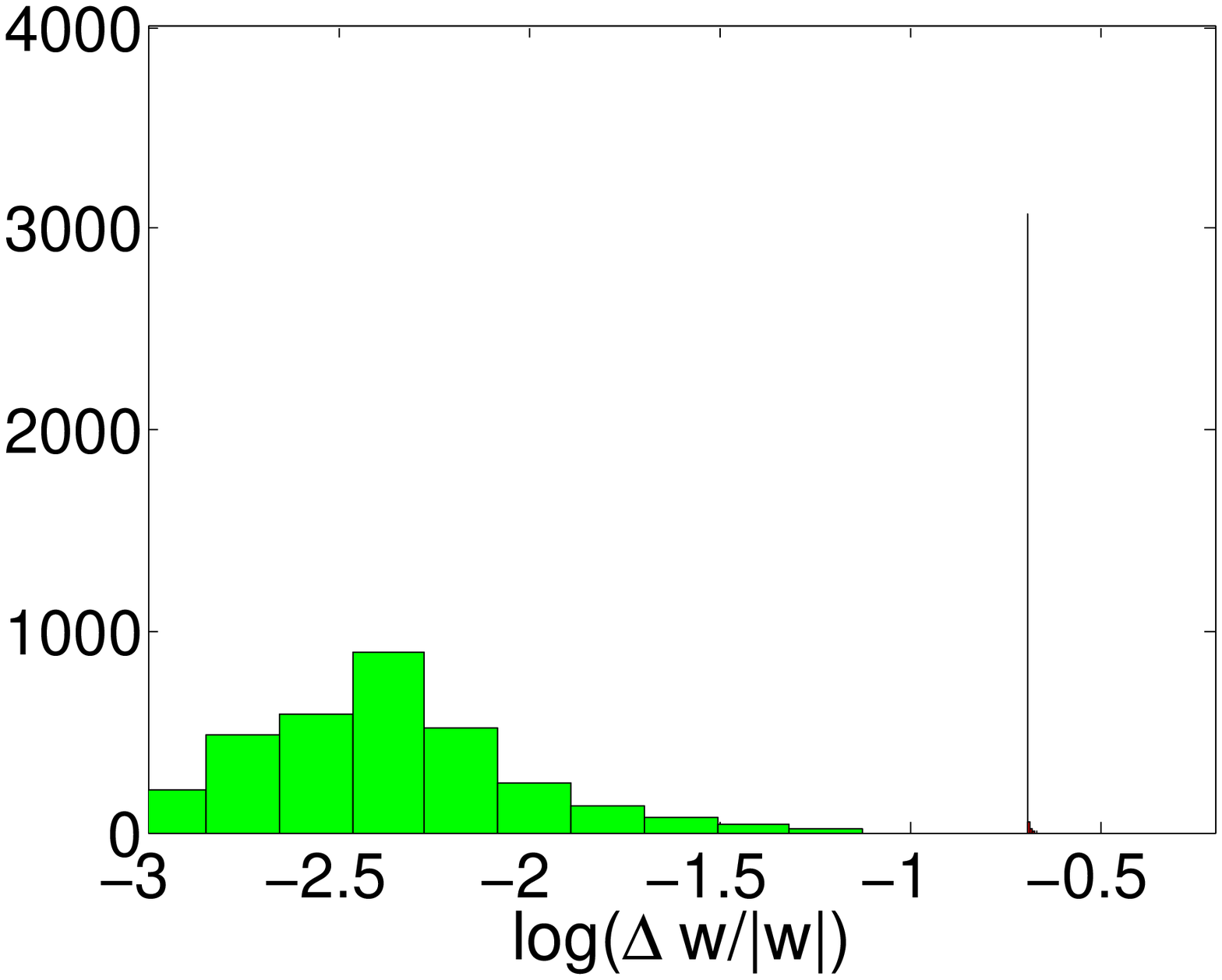} &
\includegraphics[width = 3.5 cm]{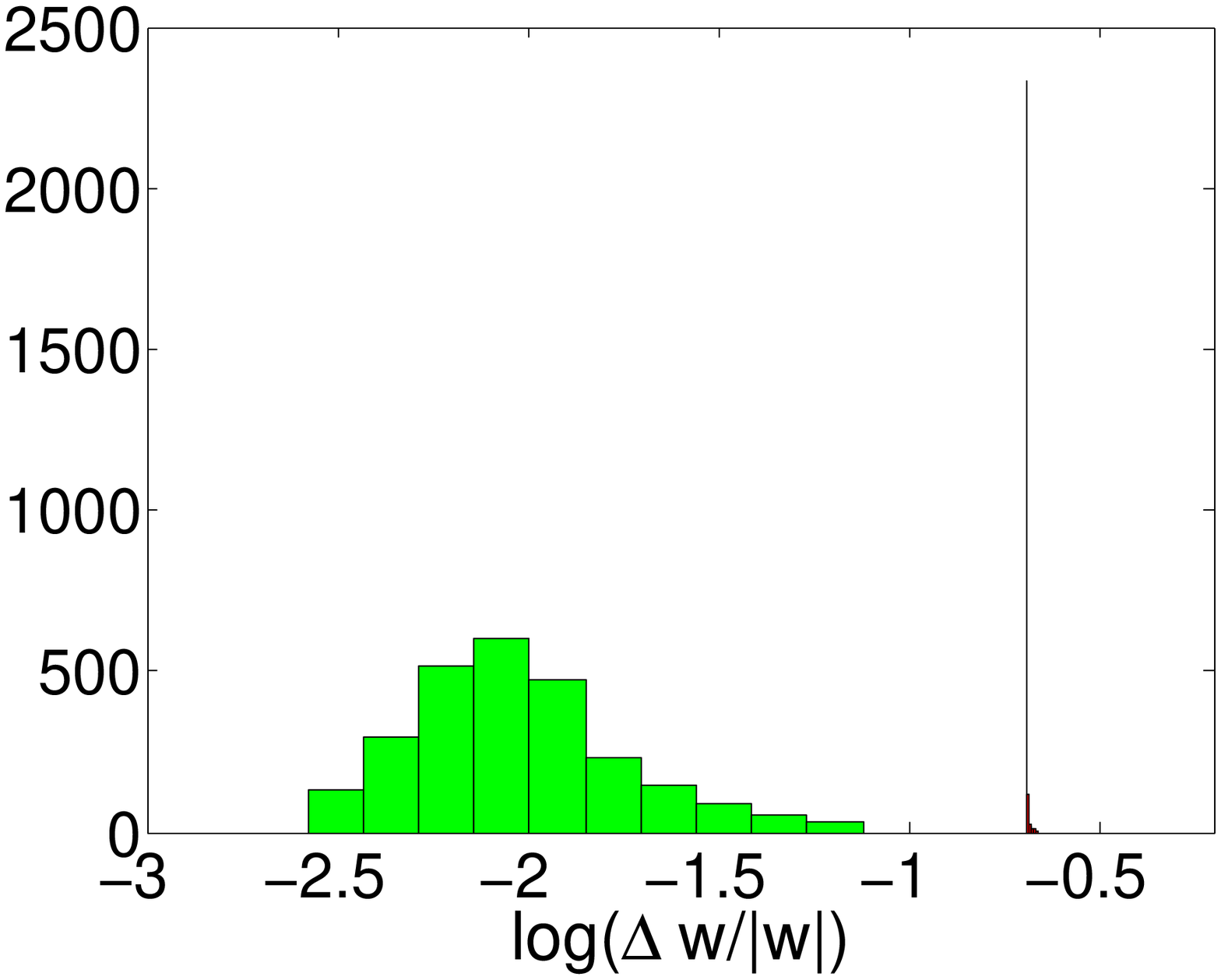} &
\includegraphics[width = 3.5 cm]{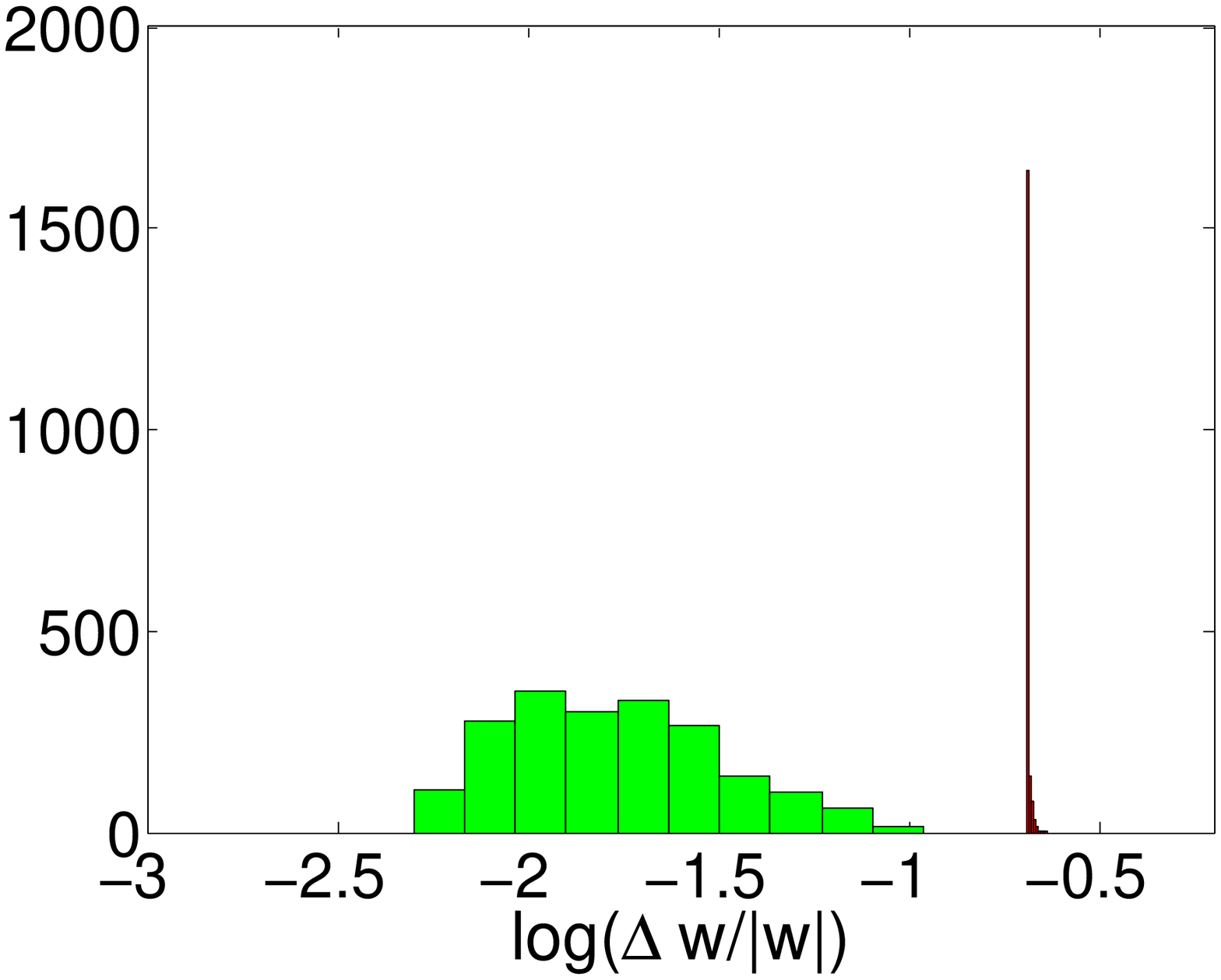} &
\includegraphics[width = 3.5 cm]{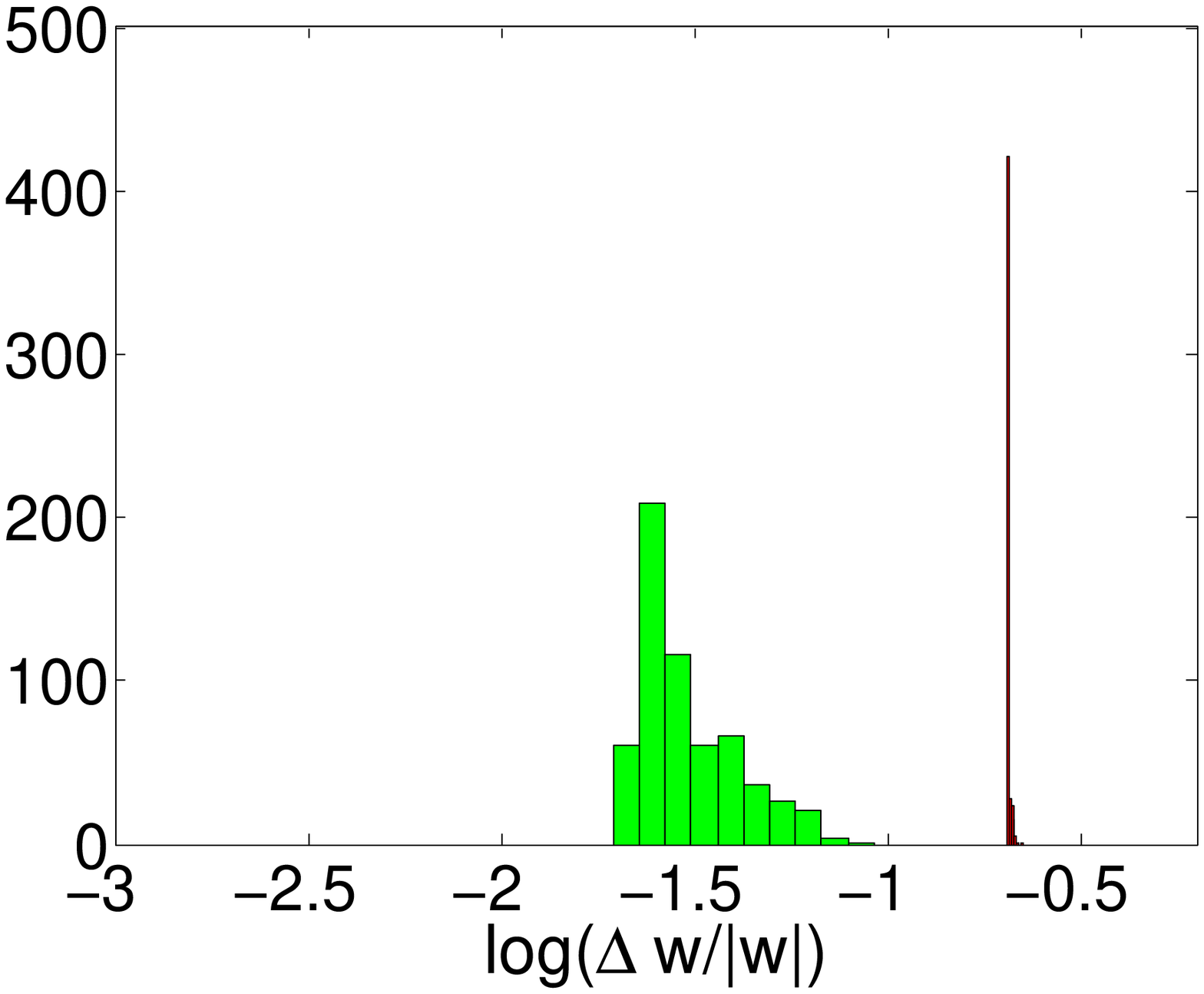}
\end{tabular}
\caption{The same as in Fig.~\ref{Fig.cases} but for $z_0=0.7$ ($D_{\rm L} = 4~\Gpc$).}
\label{Fig.cases_0.7}
\end{figure*}     

\begin{figure*}[t]
\begin{tabular}{ccccc}
 & & & &
\includegraphics[width = 3.5 cm]{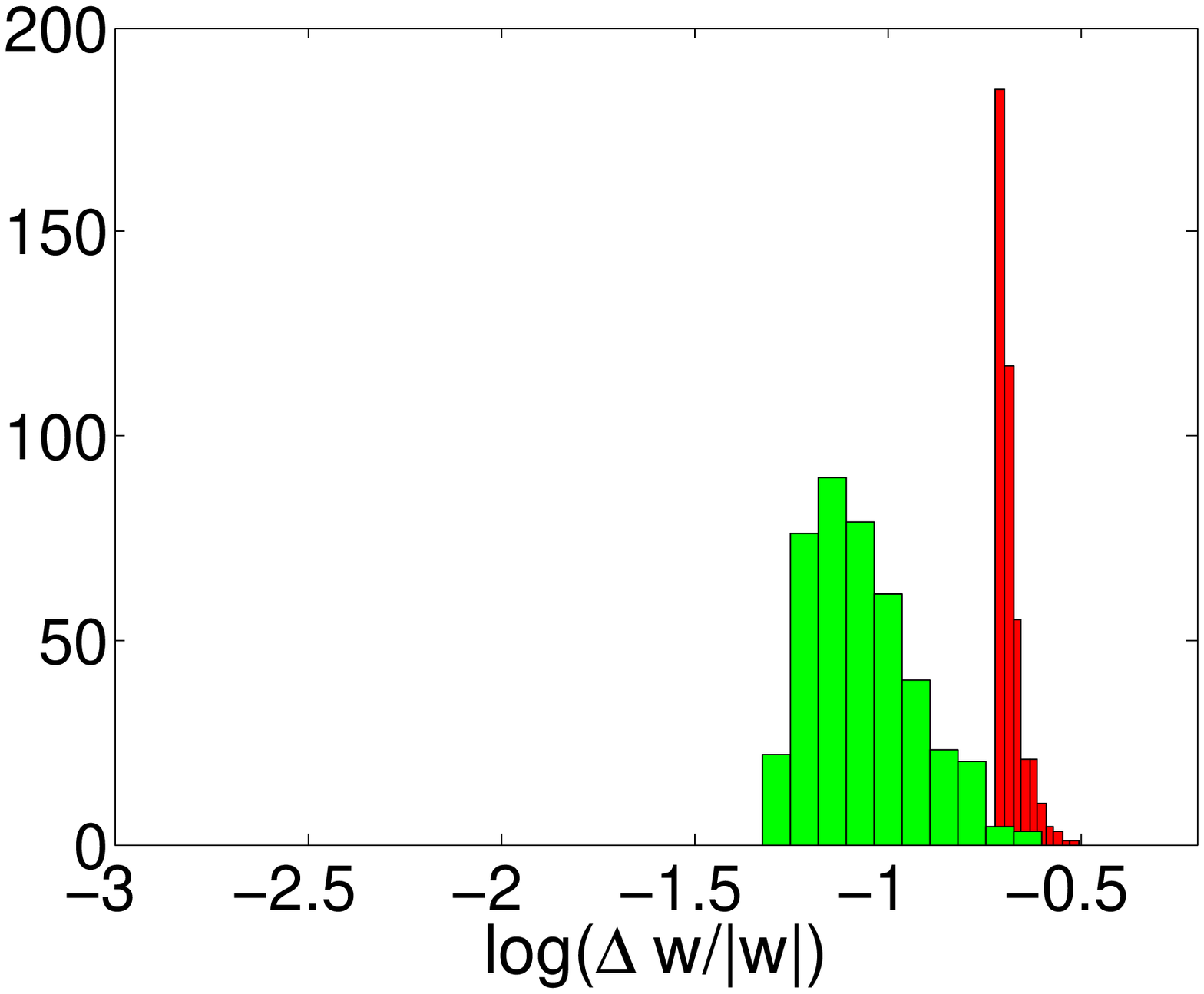}
\\
 & & &
\includegraphics[width = 3.5 cm]{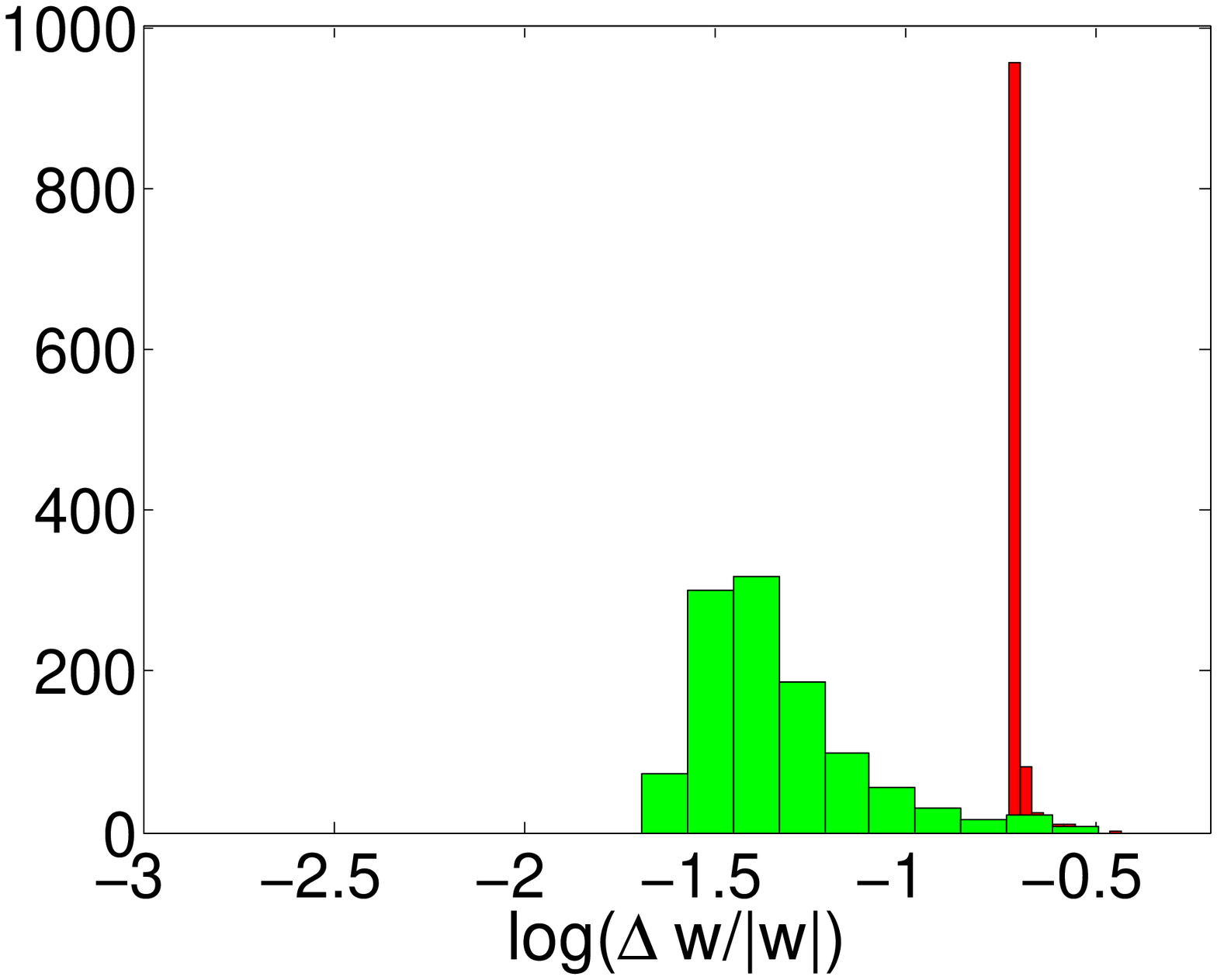} &
\includegraphics[width = 3.5 cm]{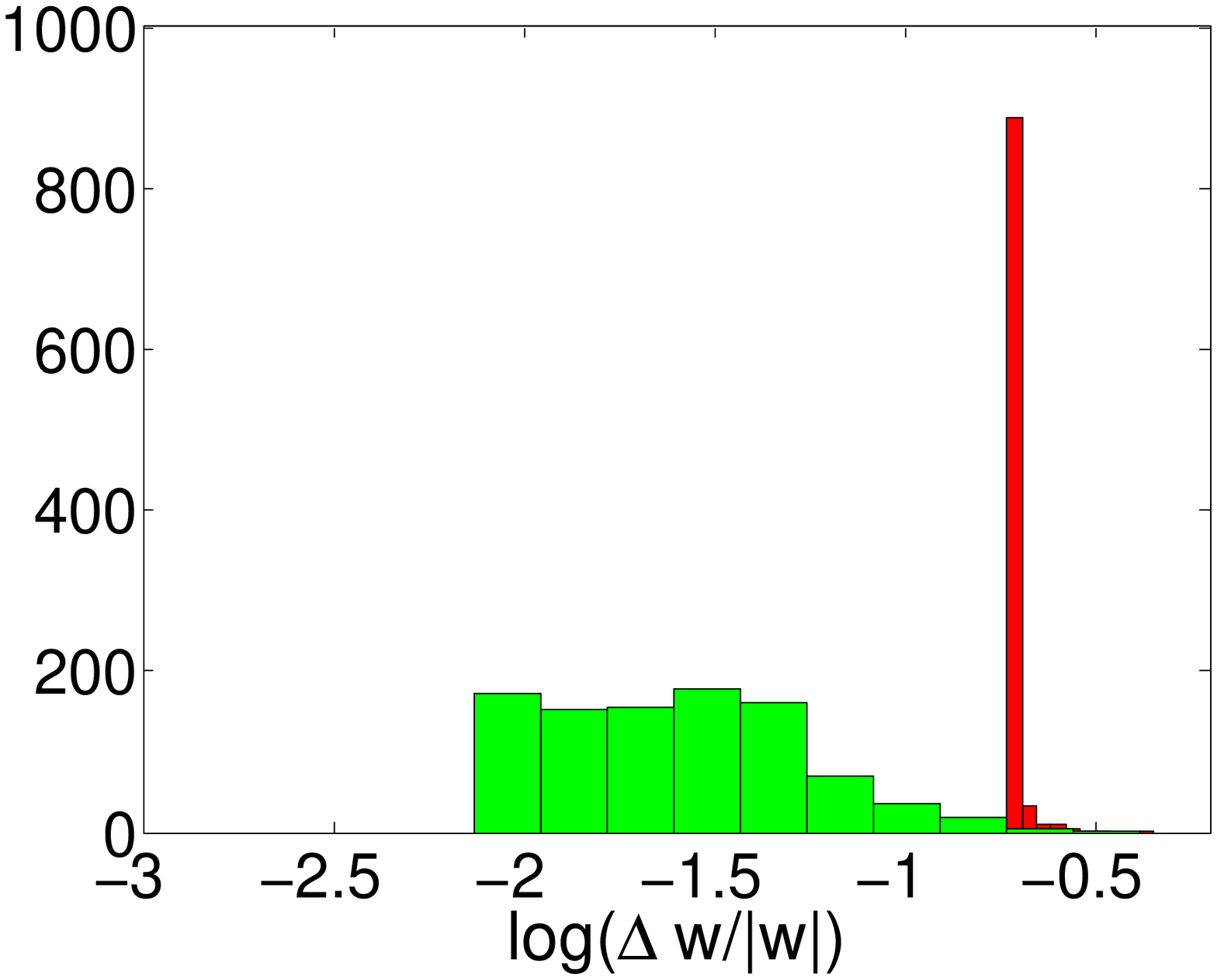}
\\
 & &
\includegraphics[width = 3.5 cm]{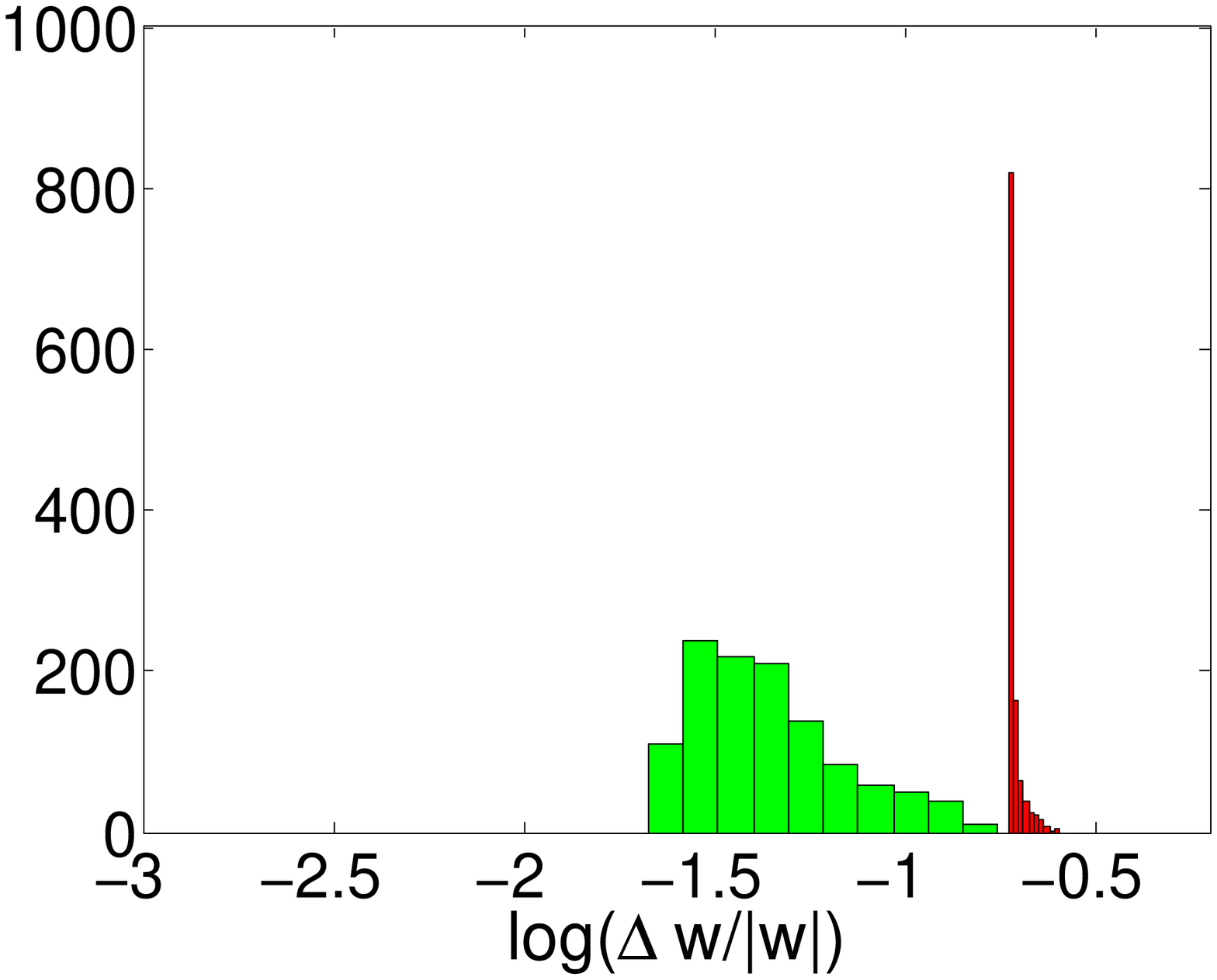} &
\includegraphics[width = 3.5 cm]{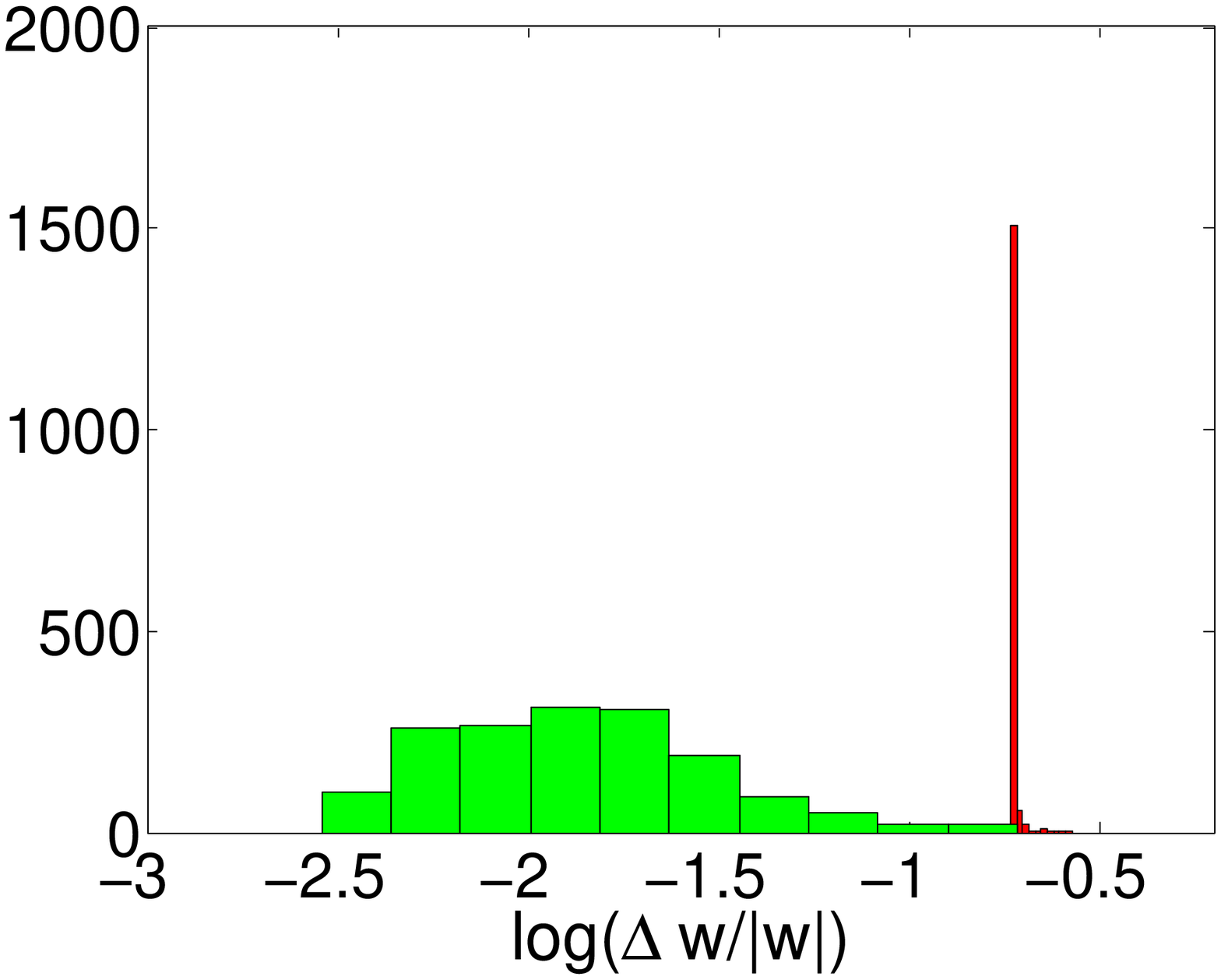} &
\includegraphics[width = 3.5 cm]{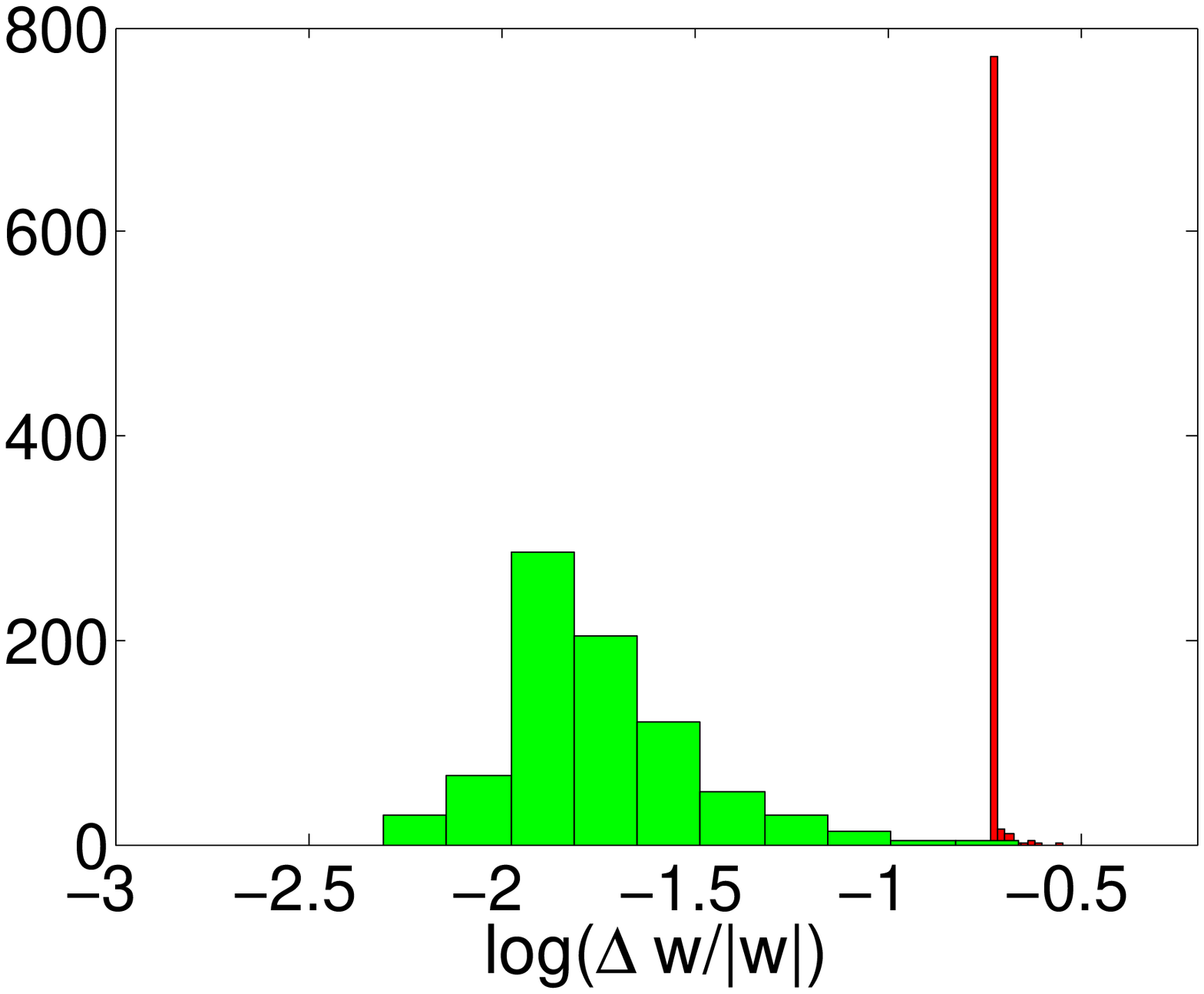}
\\
 &
\includegraphics[width = 3.5 cm]{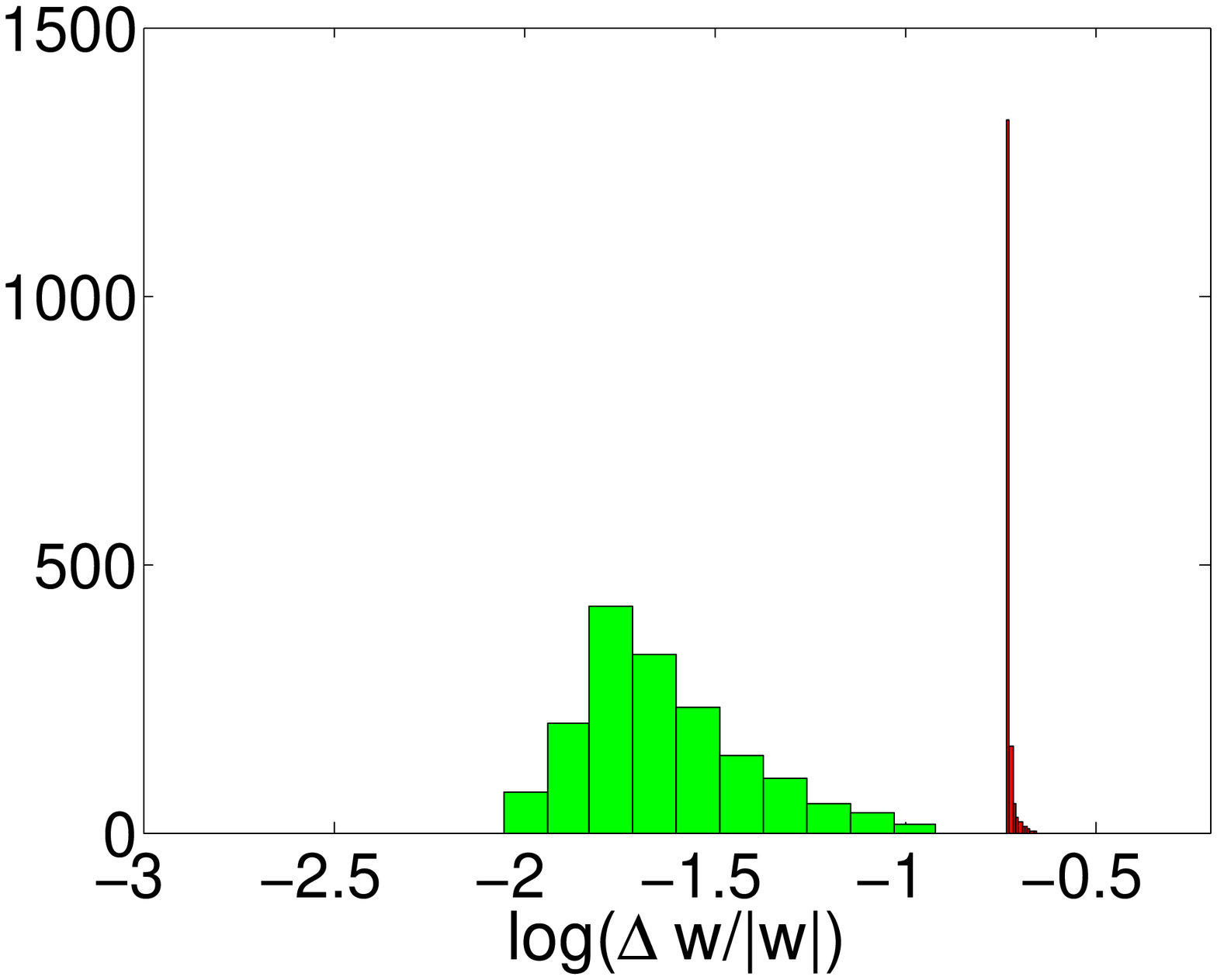} &
\includegraphics[width = 3.5 cm]{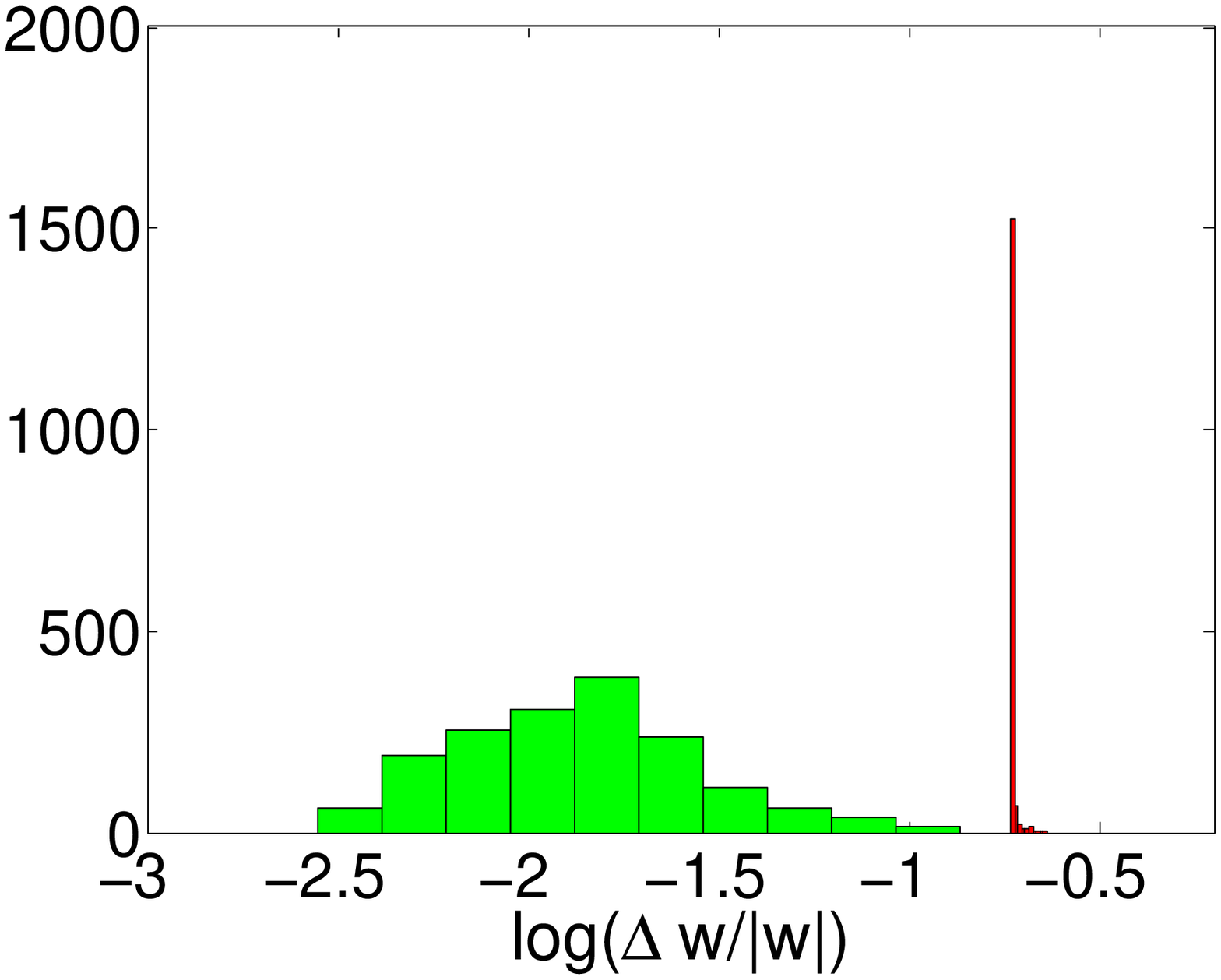} &
\includegraphics[width = 3.5 cm]{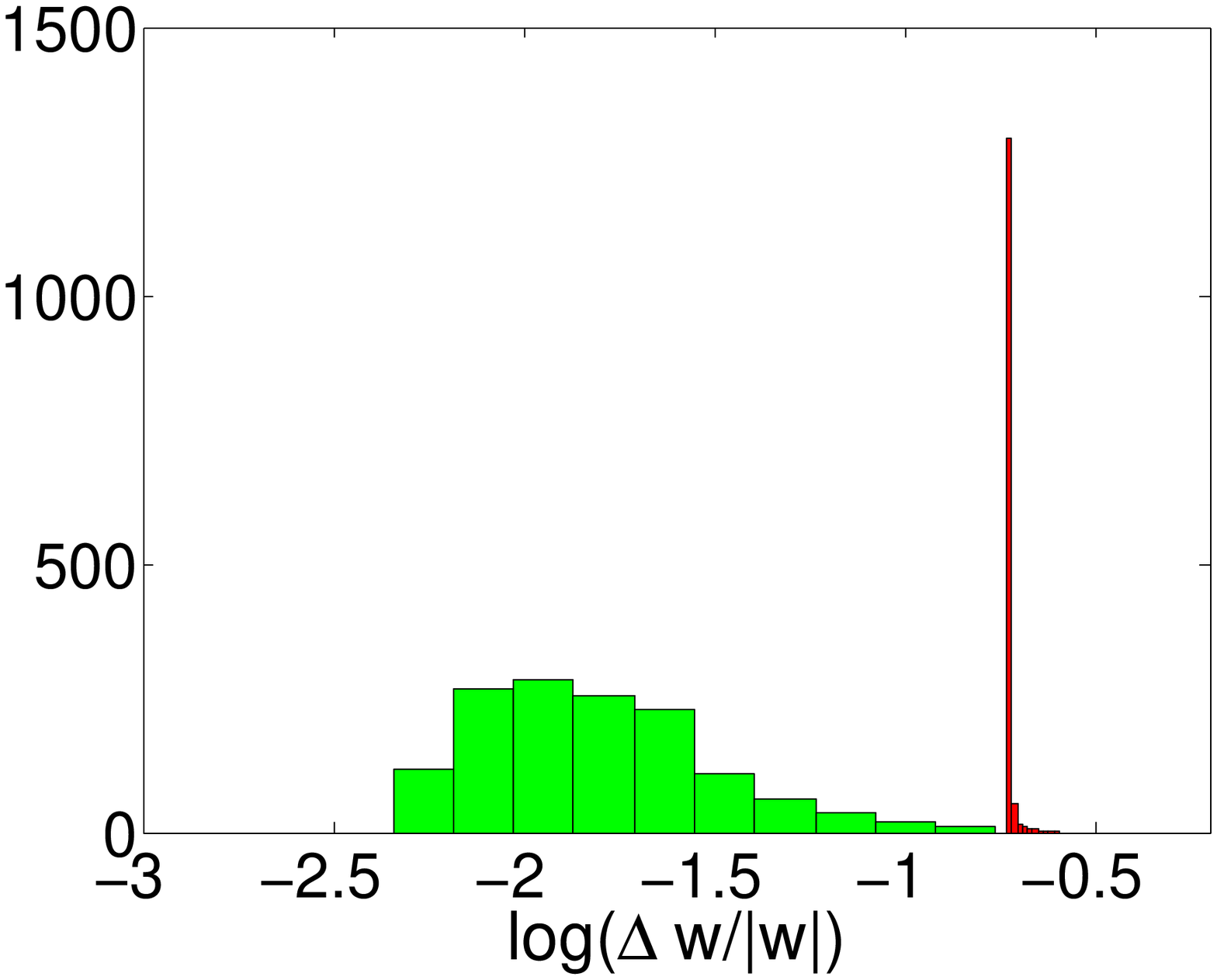} &
\includegraphics[width = 3.5 cm]{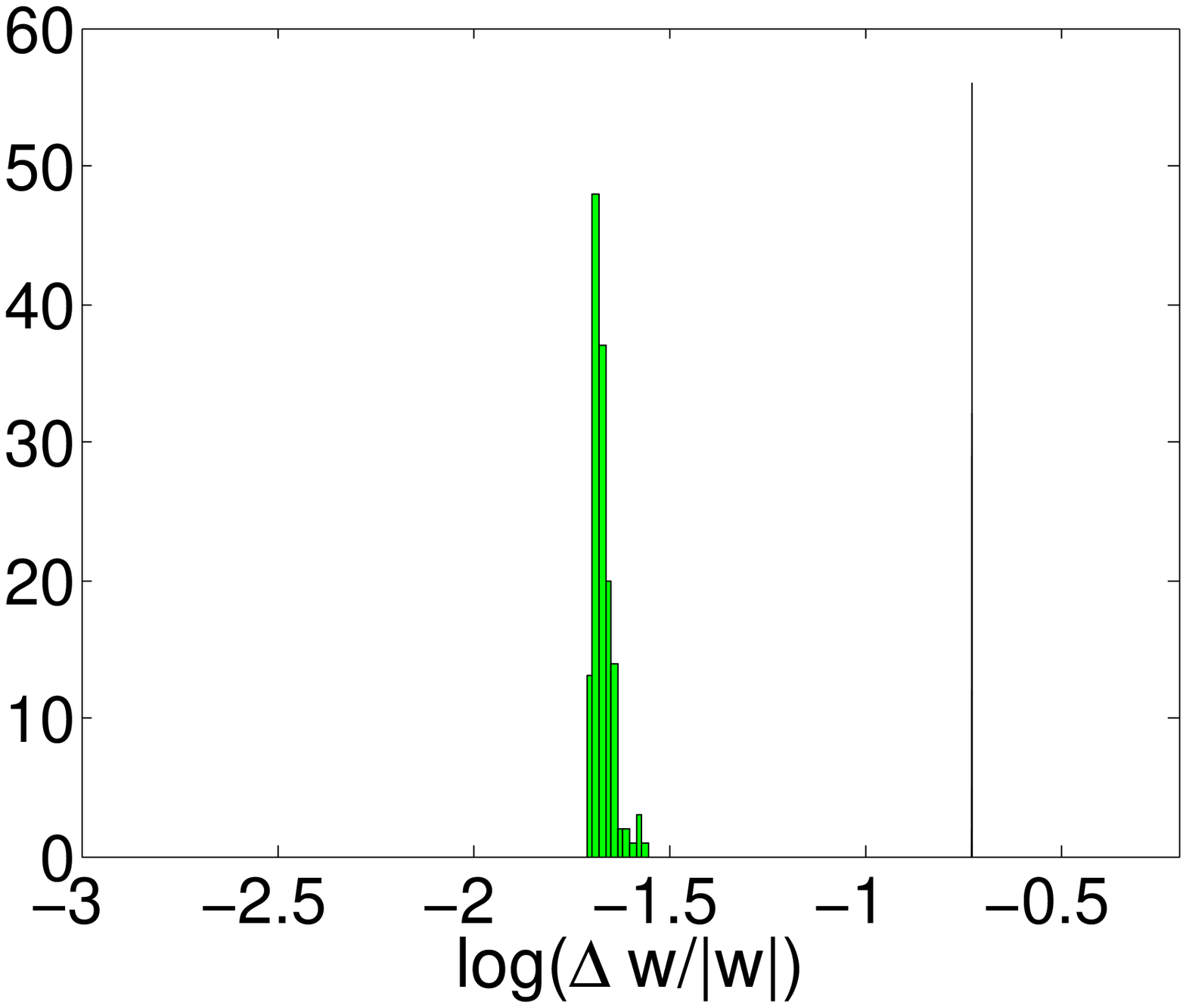}
\\
\includegraphics[width = 3.5 cm]{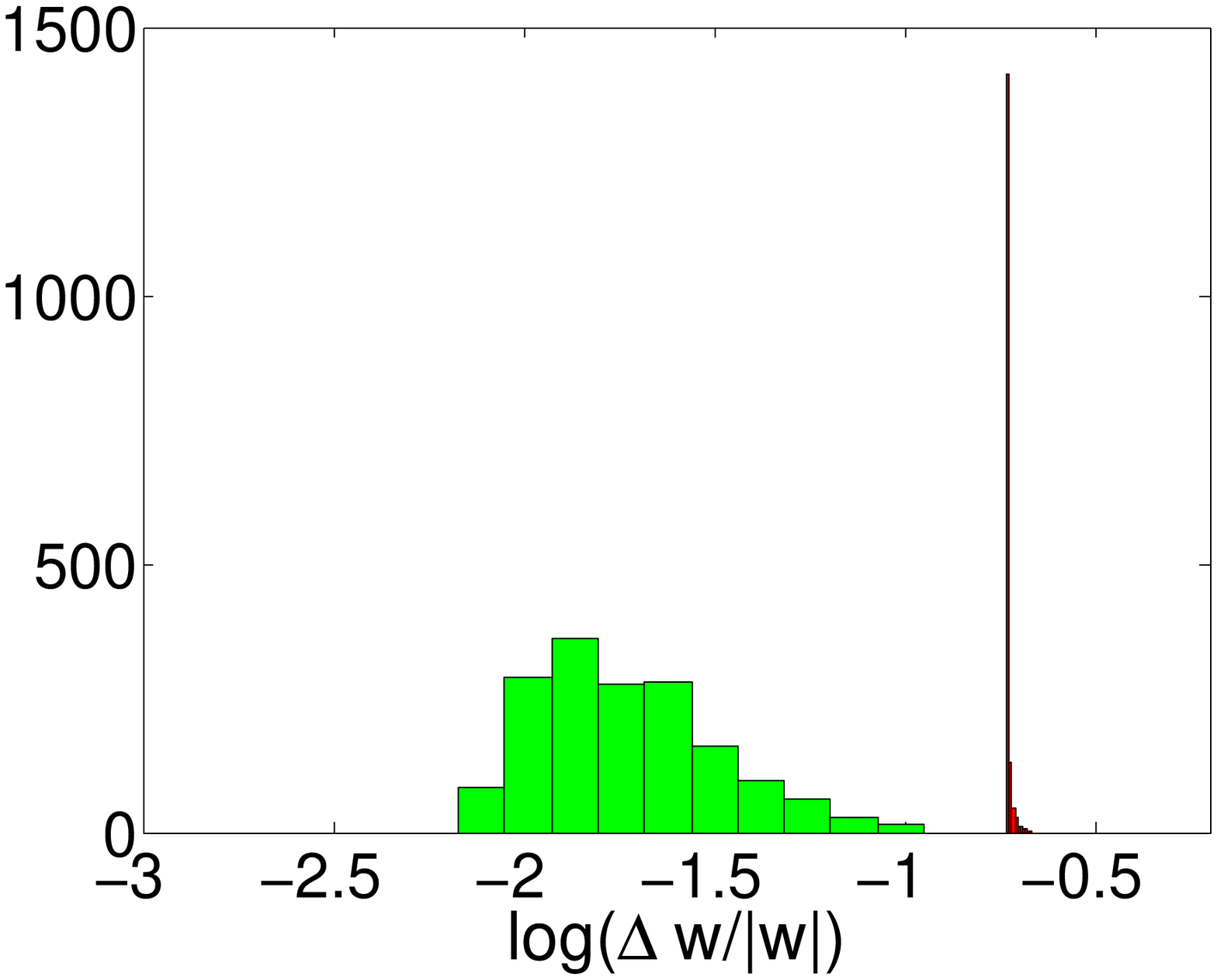} &
\includegraphics[width = 3.5 cm]{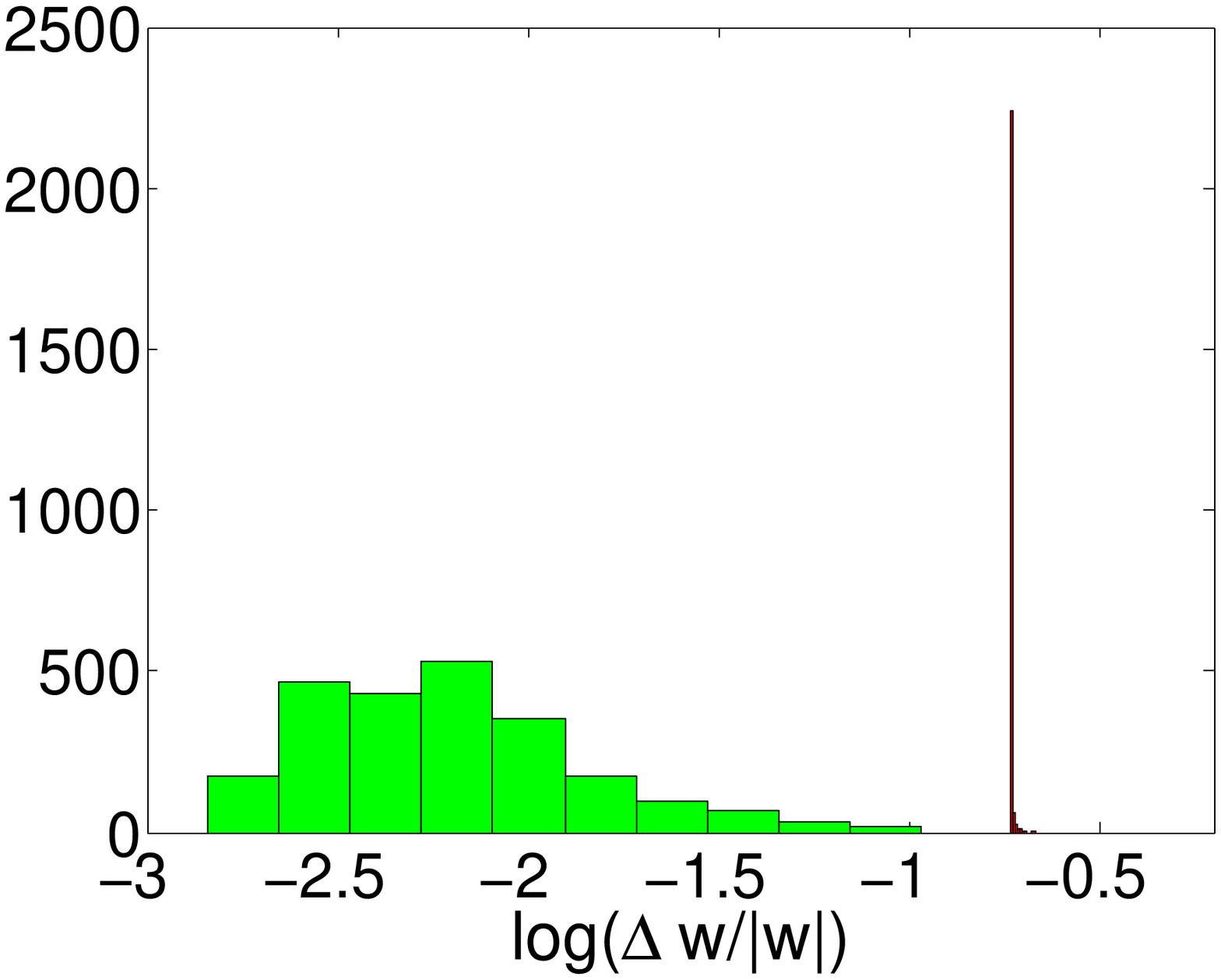} &
\includegraphics[width = 3.5 cm]{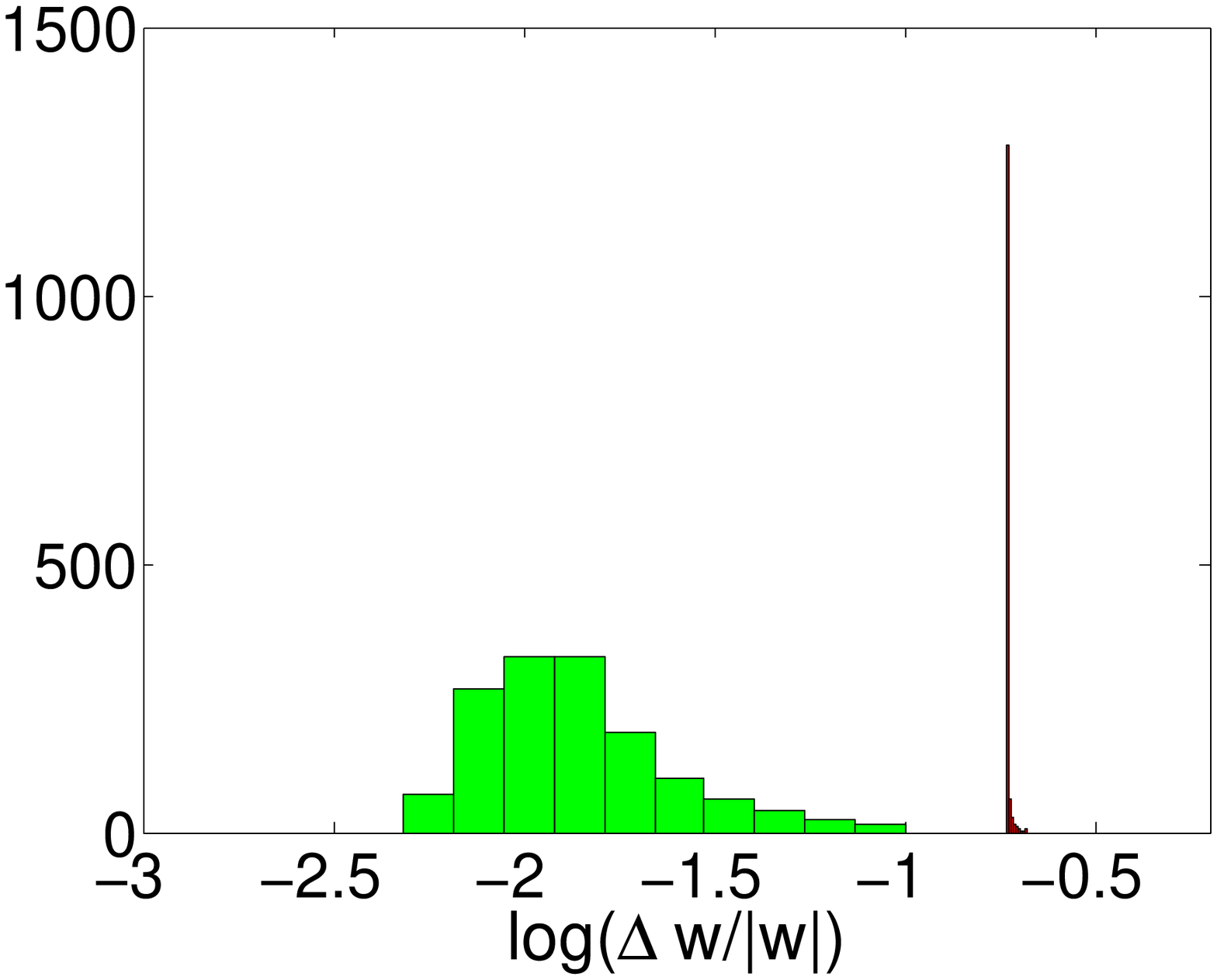} &
\includegraphics[width = 3.5 cm]{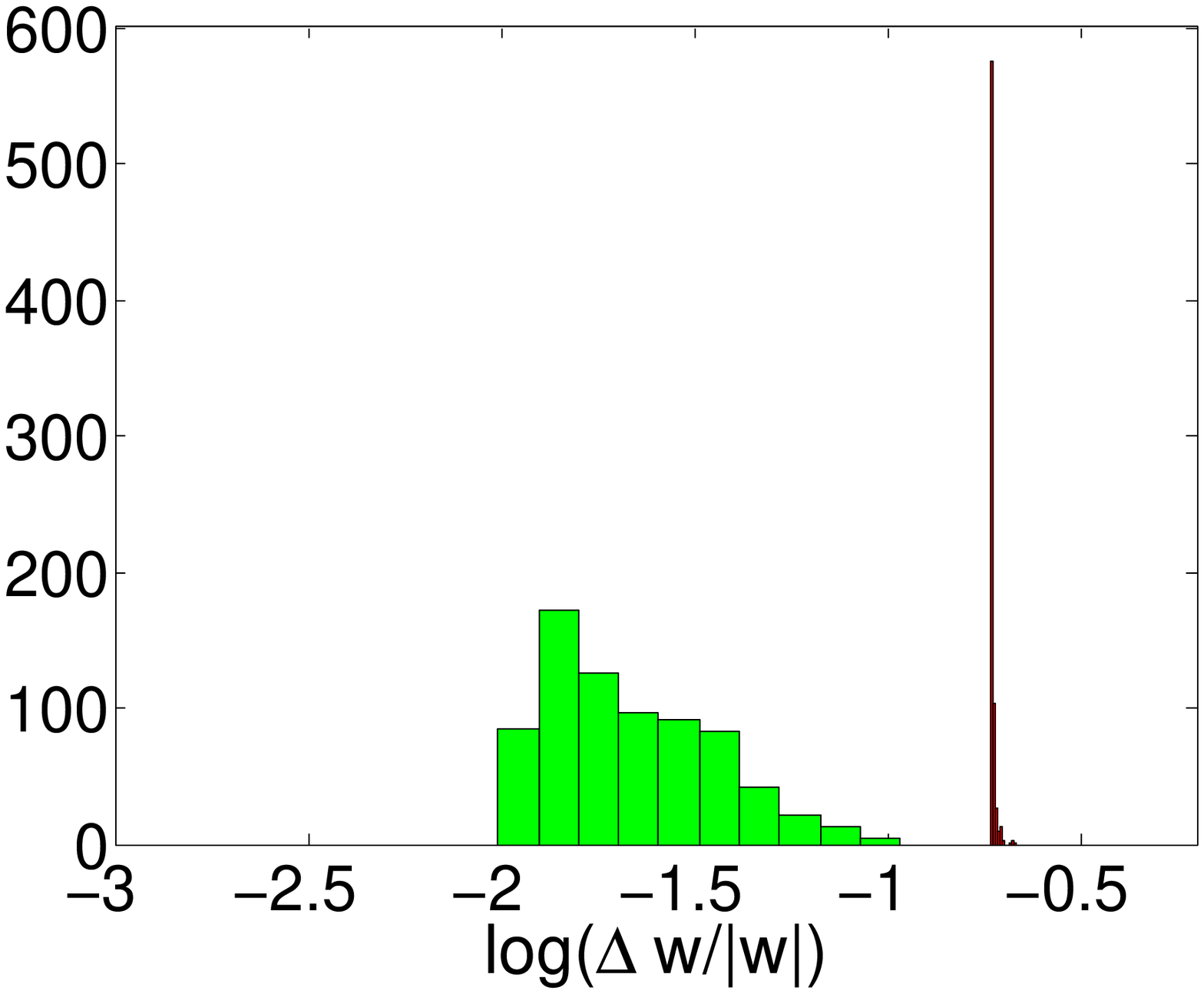} &
\end{tabular}
\caption{The same as in Fig.~\ref{Fig.cases} but for $z_0=1$ ($D_{\rm L} = 6.3~\Gpc$).}
\label{Fig.cases_1}
\end{figure*}   



Assuming that $(H_0, \Omega_{\rm M}, \Omega_{\rm DE})$ are known with 
sufficient accuracy that their uncertainties can be neglected, the 
error on $w$ \cite{Arun:2007hu,AIMSSV} will be given by two contributions: (a) the uncertainty in
$D_{\rm L}$ from LISA observations (GW) and (b) the error on $z$ from the identification of the
host galaxy or galaxy cluster (GC),
\begin{eqnarray}
\Delta w & = & \left|\frac{\partial D_{\rm L}}{\partial w} \right|^{-1} D_{\rm L}
\left[ 
\left(\frac{\Delta D_{\rm L_{,\,GW}}}{D_{\rm L}}\right)^2 +
\left(\frac{1}{D_{\rm L}} \frac{\partial D_{\rm L}}{\partial z} \right)^2 \Delta z_{\rm GC}^2
\right]^{1/2} \nonumber \\
 & = & \left| \frac{\partial D_{\rm L}}{\partial w} \right|^{-1}  D_{\rm L}
\left[ 
\left(\frac{\Delta D_{\rm L_{,\,GW}}}{D_{\rm L}}\right)^2 +
\left(\frac{\Delta D_{\rm L_{,\,GC}}}{D_{\rm L}}\right)^2
\right]^{1/2}
\; .
\label{eq:dw}
\end{eqnarray}
The second contribution to $\Delta w$ could be significant in the cases where the host galaxy
cluster could be identified, but not the individual galaxy. Even then, since the typical radius
of a galaxy cluster is $5$~Mpc \cite{Galaxy_cluster_size}, the contribution from this term
will be $0.17\%$, $0.12\%$ and $0.08\%$ for $z_0=0.55$, $z_0=0.7$ and $z_0=1$, respectively. This is smaller than or 
comparable to the distance error from LISA's noise, so that neglecting it would make a difference
of at most $\sim \sqrt{2}$. In what
follows, we will assume that the host galaxy can be identified (which may be possible with large survey instruments through electromagnetic counterparts \cite{LangHughes08,Kocsisetal08}) and therefore we neglect the
contribution from the size of the cluster; but in any case, the main conclusions of this article would be
the same.

An important problem in determining distances (also in conventional 
astronomy) is that of weak lensing. As the waveform propagates 
through the matter distribution between the source and observer, its 
amplitude will suffer an overall amplification or deamplification, 
leading to an additional random error in the estimation 
of $D_{\rm L}$. For the range of distances considered here, this error 
will be at the level of 3-5\% \cite{Kocsisetal06}. This is substantive: 
at a distance of 3 Gpc (or $z_0 = 0.55$), a distance error of 4\% will 
correspond to an uncertainty in $w$ of 23\%. However, it may be 
possible to largely remove the effect of weak lensing by mapping the 
mass distribution along the line of sight (see, e.g., Refs.\, 
\cite{Gunnarssonetal05,Shapiro:2009ab}). In the next Section we will see 
that, if left uncorrected, weak lensing will completely dominate over 
uncertainties due to LISA's instrumental noise in the determination of $w$.       

\section{Results}
\label{sec:results}

We now discuss the results for the localizability of sources, and 
the values of $\Delta w$ obtained from our Monte-Carlo simulations. 
   
The top panels of Fig.~\ref{Fig.frac_median} show the percentages
of ``useful" systems, i.e., the fraction of simulated inspiral events 
for which the sky position error is sufficiently small that the host 
galaxy could be identified, by our criterion $N_{\rm clusters} < 3$.
Results are given for three choices of redshift: $z_0 = 0.55$, $z_0 = 0.7$, 
and $z_0 = 1$, which, in our fiducial model, correspond to $D_{\rm L} = 3~\Gpc$,
$D_{\rm L} = 4~\Gpc$ and $D_{\rm L} = 6.3~\Gpc$, respectively.
Two trends can be seen: 
\begin{enumerate}
\item When the total mass is high, the 
termination frequency $2 F_{\rm ISCO}$ of the dominant harmonic will 
be low and the signal will have less power in LISA's frequency band, being ``visible"
only during the last orbits before merger. Both the low observed SNR (see Fig.~\ref{Fig.SNR_vs_masses})
and the short time over which the signal is observed will lead to relatively
poor parameter estimation.
\item Parameter estimation will be worse for symmetric systems (i.e., $m_1 = m_2$); 
indeed, all odd harmonics of the orbital frequency are proportional to 
the mass difference $(m_1 - m_2)$ and hence will vanish in the equal 
mass case. The information they could otherwise have carried will then 
not be present. 
\end{enumerate}
For $z_0 = 0.55$, the largest fraction of useful systems is 76.5\%, which 
occurs for light and very asymmetric systems with component masses 
$(1.2 \times 10^6, 3.6 \times 10^5)\,M_\odot$. The smallest fraction 
is 33.5\%, for $(3.6 \times 10^7, 3.6 \times 10^5)\,M_\odot$; although 
these systems are even more asymmetric, they are too heavy to deposit much 
power in LISA's band. For $z_0=0.7$, the largest and smallest fractions of useful systems 
have dropped to 62.5\% and 11.7\%, respectively. For $z_0=1$ the 
analogous numbers are 46.0\% and 0\% respectively. Indeed, for 
$(3.6 \times 10^7, 3.6 \times 10^7)\,M_\odot$, there are no systems in 
our simulated population for which $N_{\rm clusters} < 3$.

In the bottom panels of Fig.~\ref{Fig.frac_median}, the median values of 
$\Delta w/|w|$ are shown, for the ``useful" systems where the host can 
be identified. Binaries with masses $(1.2 \times 10^6, 3.6 \times 
10^5)\,M_\odot$, which gave the largest fraction of useful systems, also 
yield the smallest value for the median error on $w$: for $z_0 = 0.55$ this is
$(\Delta w/|w|)_{\rm median} = 0.3\%$. Still for $z_0 = 0.55$, the largest median error, 
$(\Delta w/|w|)_{\rm median} = 4.1\%$, occurs for very heavy and 
symmetric systems with masses $(3.6 \times 10^7, 3.6 \times 10^7)\,M_\odot$.
At $z_0 = 0.7$, the smallest and largest median errors on $w$ are 0.4\% 
and 5.5\%, respectively. For $z_0=1$, and disregarding the mass pair for 
which there are no useful sources, the smallest and largest median 
errors on $w$ are 0.6\% and 8.2\%, respectively. 

So far we have discussed errors without taking weak lensing into 
account. In Figs.~\ref{Fig.cases}, \ref{Fig.cases_0.7}, and 
\ref{Fig.cases_1}, we show the distributions of errors for each of 
the mass pairs separately, both with and without an additional 4\% 
error folded into the distance error through a sum of quadratures,
to mimic the effect of weak lensing. In all cases, even at $z_0=1$, 
LISA's instrumental errors tends to be far smaller than the combined instrumental and 
weak lensing errors. Indeed, a 4\% error in $D_{\rm L}$ translates
into an 18.5\% to 23\% error in $w$ depending on redshift, and it is around
these high values that the results for $\Delta w/|w|$ are sharply peaked 
when weak lensing is taken into account. This shows that, to make full 
use of LISA's potential, future studies should focus on correcting the weak 
lensing effect.

\section{Conclusions}
\label{sec:conclusions}

LISA has the ability to approximately localize supermassive black hole binary 
coalescence events through modulation of the observed signal due to LISA's motion 
around the Sun.  Electromagnetic follow-up observations could localize the source with negligible 
errors in the source's position in the sky and the redshift $z$ of the host galaxy. 
From the gravitational waveform, the luminosity distance $D_{\rm L}$ can be inferred.
The relationship between $D_{\rm L}$ and $z$ depends sensitively on the
past evolution of the Universe, which affects the gravitational
wave signal as it travels from the source to the detector over cosmological distances. Assuming 
that at sufficiently large scales the Universe is approximated well by a spatially flat 
FLRW model and that the Hubble constant, the density of matter and the density of dark 
energy are sufficiently well known, the observation of a single SMBBH event in LISA can be used 
to measure the equation-of-state parameter of dark energy $w$. Thus, such events can be used
as standard sirens, similar to the standard candles of conventional cosmography, but with no 
need for calibration of distance though a cosmic distance ladder of different kinds of 
sources. 

In \cite{Arun:2007hu,AIMSSV}, it was pointed out that inclusion of higher signal harmonics has
a dramatic effect on source localization, making it more likely that we will be able to find
the host galaxy for relatively close-by ($z \lesssim 1$) SMBBH coalescences; these are the
ones needed for a good estimation of $w$. However, in those papers a relatively small number
of systems were studied. Here we performed large-scale Monte Carlo simulations in order to
exhaustively probe the relevant part of the parameter space. We considered 
15 choices of observed component mass pairs, each being given 5000 possible sky positions 
and orientations, and the results were scaled to three different redshifts. 

For each of the mass and redshift choices, we first computed how many systems would be localizable by the
criterion that within a generous redshift interval there should at most be 3 possible 
host galaxies or galaxy clusters in the sky error ellipse. The fraction of localizable systems
varies widely with total mass and mass ratio (light and asymmetric systems being better), but
at our ``intermediate" redshift of $z_0 = 0.7$ these were between 11.7\% and 62.5\%. We note
that these numbers are likely to be on the conservative side. If a coalescence event is accompanied
by a sufficiently obvious electromagnetic counterpart then our localizability criterion may 
be too strict. Furthermore, a more careful treatment of the coalescence process (inclusion
of spins as well as merger and ringdown) would increase these percentages, as discussed below.

Next we calculated uncertainties on $w$ for those systems which passed our localizability
requirement so that a redshift value would be available. Here too there is considerable 
dependence on the mass parameters; at $z_0 = 0.7$ the median errors came out to be
between 0.4\% and 5.5\%.  

The waveform model used here included higher signal harmonics;
indeed, without these it becomes difficult to even approximately localize the source
in the first place \cite{Arun:2007hu,AIMSSV,Trias:2007fp,Trias:2008pu}. However, spins were 
ignored, which are also known to improve parameter estimation \cite{Vecchio:2003tn,RyanHughes06}. 
Recently Stavridis et al.~have investigated how well $w$ could be measured
with the inclusion of spin-induced precession of the orbital plane, though without
higher harmonics \cite{Stavridisetal09}; they found 1-$\sigma$ uncertainties $\Delta w$ that are 
similar to the ones in the present work. Although higher harmonics and spins break the same degeneracies, 
no doubt some further improvements can be expected by combining the two. We also reiterate
that, as in Ref.\, \cite{Stavridisetal09}, we only looked at the inspiral signal. Inspiral-only 
position estimates can be improved once again if the merger and ringdown signal is taken 
into account. Babak et al.~\cite{Babak:2008bu} and Thorpe et al.~\cite{Thorpe:2008wh} made a start with 
this using waveforms from numerical relativity simulations, and an extensive study on localizability using semi-analytic 
(non-spinning) inspiral-merger-ringdown waveforms was performed by McWilliams et al.~\cite{McWilliamsetal09}. 
It would be interesting to see estimates for sky localizability and luminosity distance measurements 
with waveforms that have both higher harmonics and spins in the inspiral part, and which incorporate
merger and ringdown as well. The results concerning localizability of the source which we presented
here may well be underestimates by factors of several. 

As far as the luminosity distance is concerned,
if we were only limited by LISA's instrumental noise and the confusion background due to Galactic
white dwarf binaries, then here too there would be, with no doubt, room for improvement. 
However, measurements of $D_{\rm L}$ get ``polluted" by weak lensing effects, which in turn affects
the uncertainties on $w$. Recent work indicates that these effects can be substantially reduced by 
exploiting the brightness of galaxies as a tracer of the gravitational fields
of matter along the line of sight \cite{Gunnarssonetal05}, or through mapping shear and flexion of 
galaxy images \cite{Shapiro:2009ab}. With a deep and wide-field image of galaxies, as might be available
with, e.g., Extremely Large Telescope \cite{ELT} and Euclid \cite{Euclid} and with which one could construct a 
flexion map, one may be able to reduce the weak lensing error on $D_{\rm L}$ to about 1.5\% in the redshift 
range $0.5 < z < 1$ \cite{HendryPC}. At $z = 0.55$ this translates into an 
8.5\% uncertainty in $w$, which is competitive with electromagnetic measurements. By utilizing still 
higher-order effects in the apparent deformation of galaxies it may be possible to reduce this number a 
little further. Another idea would be to use high resolution CMB maps to estimate weak lensing effects; it 
is not yet clear, however, how one might infer lensing effects at redshifts of $z \sim 1$ from such maps. 
The clear message of the results we have presented is that in order to use the full potential of LISA as 
a tool for cosmography, further in-depth studies are urgently needed on ways to correct for weak lensing.

\section*{Acknowledgements}
We would like to thank Bernard Schutz for discussions at an early stage of this work, and
Peter Coles and Martin Hendry for discussions on the mitigation of weak lensing effects.
MT and AMS's work was jointly supported by European Union FEDER funds,
the Spanish Ministry of Science and Education (projects FPA2007-60220, HA2007-0042 
and CSD2009-00064) and by the Govern de les Illes Balears, Conselleria
d'Economia, Hisenda i Innovaci\'o. CVDB and BSS were partially supported by
Science and Technology Facilities Council, UK, grant PP/F001096/1. CVDB's work
was also part of the research programme of the Foundation for Fundamental Research on 
Matter (FOM), which is partially supported by the Netherlands Organisation for Scientific 
Research (NWO).


\begin{thebibliography}{99}

\bibitem{Volonteri:2002vz}
  M.~Volonteri, F.~Haardt and P.~Madau,
  Astrophys.\ J.\  {\bf 582} (2003) 559
  [arXiv:astro-ph/0207276].

\bibitem{Begelman:2006db}
  M.~C.~Begelman, M.~Volonteri and M.~J.~Rees,
  Mon.\ Not.\ Roy.\ Astron.\ Soc.\  {\bf 370} (2006) 289
  [arXiv:astro-ph/0602363].

\bibitem{Bardeen70}
  J.M.~Bardeen, Nature {\bf 226} (1970), 64

\bibitem{Thorne74}
  K.S.~Thorne, Astrophys.~J.~{\bf 191} (1974) 507

\bibitem{KingPringle}
  A.R.~King and J.E.~Pringle,
  Mon.~Not.~Roy.~Astron.~Soc.~Lett.~{\bf 373} (2006) L93
  [arXiv:astro-ph/0609598].
  
\bibitem{LISAPE_paper}
  K.~G.~Arun {\it et al.},
  Class.~Quantum Grav.~{\bf 26} (2009) 094027 
  [arXiv:0811.1011 [gr-qc]].

\bibitem{Schutz:1986gp}
  B.~F.~Schutz,
  Nature {\bf 323}, 310 (1986)

\bibitem{HolzHugh03}
D.~E.~Holz and S.~A.~Hughes,
Classical and Quant. Gravity, {\bf 20}, S65 (2005),
[arXiv:astro-ph/02l2218].

\bibitem{HolzHugh05}
D.~E.~Holz and S.~A.~Hughes,
Astrophys.~J {\bf 629}, 15 (2005)
[arXiv:astro-ph/0504616].

\bibitem{Arun:2007hu}
  K.~G.~Arun \etal 
  Phys.\ Rev.\  D {\bf 76} (2007) 104016; 
  [Erratum-ibid.\  D {\bf 76} (2007) 129903 
  [arXiv:0707.3920 [astro-ph]].

\bibitem{AIMSSV}
K.~G.~Arun, C.~K.~Mishra, C.~Van Den Broeck, B.~R.~Iyer, B.~S.~Sathyaprakash, S.~Sinha,
Class.~Quantum Grav.~{\bf 26} (2009) 094021
[arXiv:0810.56727 [gr-qc]] 

\bibitem{Trias:2007fp}
  M.~Trias and A.~M.~Sintes,
  Phys.\ Rev.\  D {\bf 77} (2008) 024030 
  [arXiv:0707.4434 [gr-qc]].
  
\bibitem{Trias:2008pu}
  M.~Trias and A.~M.~Sintes,
  Class.\ Quant.\ Grav.\  {\bf 25} (2008) 184032 
  [arXiv:0804.0492 [gr-qc]].

\bibitem{Porter:2008kn}
  E.~K.~Porter and N.~J.~Cornish,
  Phys.~Rev.~D {\bf 78} (2008) 064005 
  [arXiv:0804.0332 [gr-qc]]

\bibitem{Sintes:1999ch}
  A.~M.~Sintes and A.~Vecchio,
  in \textit{Third Amaldi conference on Gravitational Waves}, 
  edited by S. Meshkov, American Institute of Physics Conference Series 
  (American Institute of Physics, New York, 2000), p. 403.
  [arXiv:gr-qc/0005059].

\bibitem{Sintes:1999cg}
  A.~M.~Sintes and A.~Vecchio,
  in \textit{Rencontres de Moriond:Gravitational Waves and Experimental Gravity}, 
  edited by J. Dumarchez (Frontieres, Paris, 2000).
  [arXiv:gr-qc/0005058].

\bibitem{Moore:1999zw}
  T.~A.~Moore and R.~W.~Hellings,
  Phys.\ Rev.\  D {\bf 65} (2002) 062001 
  [arXiv:gr-qc/9910116].

\bibitem{Schutz:2009a}
  B.~F.~Schutz, Class. Quantum Grav., {\bf 26}, 094020 (2009). 

\bibitem{Kocsisetal06}
B.~Kocsis, Z.~Frei, Z.~Haiman, and K.~Menou,
Astrophys.~J.~{\bf 637} (2006) 27
[arXiv:astro-ph/0505394]

\bibitem{BlanchetLRR}
L.~Blanchet, 
Liv.~Rev.~Rel.~{\bf 5} (2002) 3

\bibitem{Babak:2008bu}
  S.~Babak, M.~Hannam, S.~Husa and B.~F.~Schutz,
  [arXiv:0806.1591 [gr-qc]].

\bibitem{Thorpe:2008wh}
  J.~I.~Thorpe, S.~T.~McWilliams, B.~J.~Kelly, R.~P.~Fahey, K.~Arnaud and J.~G.~Baker,
  Class.\ Quant.\ Grav.\  {\bf 26} (2009) 094026
  [arXiv:0811.0833 [astro-ph]].

\bibitem{McWilliamsetal09}
S.T.~McWilliams, J.I.~Thorpe, J.G.~Baker, and B.J.~Kelly
[arXiv:0911.1078].

\bibitem{Stavridisetal09}
A.~Stavridis, K.G.~Arun and C.M.~Will,
Phys.~Rev.~D {\bf 80} (2009) 067501
[arXiv:0907.4686].

\bibitem{Polarizations3PN}
  L.~Blanchet, G.~Faye, B.R.~Iyer, and S.~Sinha,
  Class.~Quantum Grav.~{\bf 25} (2008) 165003
  [arXiv:0802.1249 [gr-qc]].

\bibitem{Cutler:1997ta}
C.~Cutler,
Phys.\ Rev.\  D {\bf 57} (1998) 7089 
[arXiv:gr-qc/9703068].

\bibitem{Sathyaprakash:1991mt}
B.~S.~Sathyaprakash and S.~V.~Dhurandhar,
Phys.\ Rev.\  D {\bf 44} (1991) 3819.

\bibitem{VanDenBroeck:2006qu}
  C.~Van Den Broeck and A.~S.~Sengupta,
  Class.\ Quant.\ Grav.\  {\bf 24} (2007) 155 
  [arXiv:gr-qc/0607092].

\bibitem{VanDenBroeck:2006ar}
  C.~Van Den Broeck and A.~S.~Sengupta,
  Class.\ Quant.\ Grav.\  {\bf 24} (2007) 1089 
  [arXiv:gr-qc/0610126].

\bibitem{Blanchet:2001ax}
L.~Blanchet, G.~Faye, B.~R.~Iyer and B.~Joguet,
Phys.\ Rev.\  D {\bf 65} (2002) 061501; 
Erratum ibid.\  D {\bf 71} (2005) 129902 
[arXiv:gr-qc/0105099].

\bibitem{Arun:2007qv}
  K.~G.~Arun \etal 
  Phys.\ Rev.\  D {\bf 75} (2007) 124002 
  [arXiv:0704.1086 [gr-qc]].

\bibitem{Finn:1992wt}
  L.~S.~Finn,
  Phys.\ Rev.\  D {\bf 46}, 5236 (1992)
  [arXiv:gr-qc/9209010].

\bibitem{LISAPE_web}
LISA Performance Evaluation wikipage: \texttt{http://www.tapir.caltech.edu/dokuwiki/lisape:home}

\bibitem{Babak:2007zd}
  S.~Babak {\it et al.}  [Mock LISA Data Challenge Task Force Collaboration],
  Class.\ Quant.\ Grav.\  {\bf 25} (2008) 114037
  [arXiv:0711.2667 [gr-qc]].

\bibitem{Arnaud:2007jy}
  K.~A.~Arnaud {\it et al.},
  Class.\ Quant.\ Grav.\  {\bf 24} (2007) S551
  [arXiv:gr-qc/0701170].

\bibitem{Nelemans:2001}
  G.~Nelemans, L.~R.~Yungelson and S.~F.~Portegies Zwart,
  Astron.\ Astrophys.\ {\bf 375} (2001) 890 
  [arXiv:astro-ph/0105221].

\bibitem{Nelemans:2003ha}
  G.~Nelemans, L.~R.~Yungelson and S.~F.~Portegies Zwart,
  Mon.\ Not.\ Roy.\ Astron.\ Soc.\  {\bf 349} (2004) 181 
  [arXiv:astro-ph/0312193].

\bibitem{Cornish:2007if}
  N.~J.~Cornish and T.~B.~Littenberg,
  Phys.\ Rev.\  D {\bf 76} (2007) 083006 
  [arXiv:0704.1808 [gr-qc]].

\bibitem{Sesana:2007sh}
  A.~Sesana, M.~Volonteri and F.~Haardt,
  Mon.\ Not.\ Roy.\ Astron.\ Soc.\  {\bf 377} (2007) 1711
  [arXiv:astro-ph/0701556].

\bibitem{Sesana:2008ur}
  A.~Sesana, M.~Volonteri and F.~Haardt,
  Class.\ Quant.\ Grav.\  {\bf 26} (2009) 094033
  [arXiv:0810.5554 [astro-ph]].

\bibitem{DaHHJ06}
N.~Dalal, D.~E.~Holz, S.~A.~Hughes, and B.~Jain,
Phys.~Rev.~D {\bf 74} (2006) 063006 
[arXiv:astro-ph/0601275]

\bibitem{MacLeod:2008ab}
C.~L.~MacLeod and C.~J.~Hogan,
Phys.~Rev.~D {\bf 77} (2008) 043512
[arXiv:0712.0618]

\bibitem{Nissanke:2009ab}
Samaya Nissanke, Scott A. Hughes, Daniel E. Holz, Neal Dalal, Jonathan L. Sievers,
[arXiv:0904.1017v1 [astro-ph.CO]]

\bibitem{Sathyaprakash:2009ab}
B.~F.~Schutz, B.~S.~Sathyaprakash, and C.~Van Den Broeck,
[arXiv:0906.4151 [gr-qc]]

\bibitem{DE_taskforce}
A.~Albrecht et al.,
[arXiv:astro-ph/0609591]

\bibitem{Bahcalletal}
N.~Bahcall et al.,
Astrophys.~J.~Suppl.~{\bf 148} (2003) 243

\bibitem{LangHughes08}
R.~N.~Lang and S.~A.~Hughes, 
Astrophys.~J.~{\bf 667} (2008) 1184
[arXiv:0710.3795 [astro-ph]]

\bibitem{Kocsisetal08}
B.~Kocsis, Z.~Haiman, and K.~Menou,
Astrophys.~J.~{\bf 684} (2008) 870

\bibitem{Galaxy_cluster_size}
D.W.~Hogg, J.G.~Cohen and R.~Blandford,
Astrophys.~J.~{\bf 545} (2000) 32

\bibitem{Vecchio:2003tn}
A.~Vecchio,
Phys.\ Rev.\  D {\bf 70} (2004) 042001 [arXiv:astro-ph/0304051].

\bibitem{RyanHughes06}
R.N.~Lang and S.A.~Hughes,
Phys.~Rev.~D {\bf 74} (2006) 122001; Erratum ibid.~D {\bf 77} (2008) 
109901
[arXiv:gr-qc/0608062]

\bibitem{Gunnarssonetal05}
C.~Gunnarsson, T.~Dahlen, A.~Goobar, J.~Jonsson, and E.~Mortsell,
Astrophys.~J.~{\bf 640} 417
[arXiv:astro-ph/0506764]

\bibitem{Shapiro:2009ab}
C.~Shapiro, D.~Bacon, M.~Hendry, and B.~Hoyle,
 [arXiv:0907.3635v1 [astro-ph.CO]].

\bibitem{ELT}
ELT Science Working Group, \emph{Report of the ELT Science Working Group}, 2009;
http://www.eso.org/sci/facilities/eelt/science/doc/

\bibitem{Euclid}
R.~Laureijs et al., ESA Science Document DEM-SA-Dc-00001, 2009;
http://sci.esa.int/science-e/www/object/index.cfm?fobjectid=42822

\bibitem{HendryPC}
M.~Hendry, private communication


%
%
%
%
%

  
  

\end{thebibliography}
\end{document}